\documentclass[showpacs,twocolumn,pre,amssymb,amsmath,nobibnotes,aps]{revtex4}
\usepackage{bm}

\newcommand{\nh}{\hat{n}}
\newcommand{\yh}{\hat{y}}
\newcommand{\zh}{\hat{z}}
\newcommand{\xv}{{\bf x}}
\newcommand{\rv}{{\bf r}}
\newcommand{\qv}{{\bf q}}
\newcommand{\hR}{\hat{R}}
\newcommand{\D}{{(\cal{D})}}
\newcommand{\N}{{(\cal{N})}}
\newcommand{\ax}{{(a)}}
\newcommand{\re}{{(r)}}
\newcommand{\bfx}{\mathbf{x}}

\newcommand{\be}{\begin{equation}}
\newcommand{\ee}{\end{equation}}
\newcommand{\bea}{\begin{eqnarray}}
\newcommand{\eea}{\end{eqnarray}}
\newcommand{\oh}{\frac{1}{2}}
\newcommand{\half}{\frac{1}{2}}

\def\rf#1{(\ref{#1})}
\def\rfs#1{Eq.~\rf{#1}}
\usepackage[dvips]{graphicx}
\begin{document}

\title{Stability and distortions of liquid crystal order in a cell
  with a heterogeneous substrate}
\author{Quan Zhang}
\author{Leo Radzihovsky}
\affiliation{Department of Physics, University of Colorado,
   Boulder, Colorado 80309, USA}
\date{\today}

\begin{abstract}
  We study stability and distortions of liquid crystal nematic order
  in a cell with a random heterogeneous substrate. Modeling this
  system as a bulk $xy$ model with quenched disorder confined to a {\em
    surface}, we find that nematic order is marginally unstable to
  such surface pinning. We compute the length scale beyond which
  nematic distortions become large, and calculate orientational
  correlation functions using the functional renormalization group and
  matching methods, finding universal logarithmic and
  double-logarithmic distortions in two and three dimensions,
  respectively.  We extend these results to a finite-thickness
  liquid crystal cell with a second homogeneous substrate, detailing
  crossovers as a function of random pinning strength and cell
  thickness. We conclude with analysis of experimental signatures of
  these distortions in a conventional crossed-polarizer-analyzer light
  microscopy.
\end{abstract}

\maketitle
\section{Introduction}
\label{sec:intro}

\subsection{Motivation and background}
\label{motivation}

Over the past several decades, there has been considerable progress in
understanding the phenomenology of ordered condensed states subject to
random heterogeneities, generically present in real
materials \cite{FisherPhysicsToday,reviewRandom}. These include
``dirty'' charge-density waves \cite{ChargeDensityWave},
superconductors \cite{disorderSC}, and magnets, as well as superfluid
helium \cite{randomHelium} and liquid
crystals \cite{RTaerogelPRL,RTaerogelPRB,BelliniScience} in the random
environment of aerogel and other porous matrices.

Much of the detailed understanding came from the analysis of the $xy$-,
O(N)-, and related smectic \cite{RTaerogelPRL,RTaerogelPRB}
random-field models, in the pioneering works by
Larkin \cite{Larkin,LarkinOvchinnikov}, Fisher \cite{DSFisherFRG} and
Nattermann \cite{Nattermann}, and extensive subsequent studies by Le
Doussal and co-workers \cite
{BalentsFisher,GiamarchiLedoussal,LedoussalWiese,BalentsLedoussal}
using a combination of replica variational and functional
renormalization group (RG) methods. At low temperature and weak
disorder (neglecting enigmatic effects of topological defects that may
become important on much longer scales), the state is characterized by
elastic distortions with universal power-law correlations controlled
by a nontrivial zero-temperature fixed point, the so-called ``Bragg''
(elastic) glass \cite {GiamarchiLedoussal,
  MiddletonBGnumerics,KleinPaperBGexp,LingBGexp,DSFisherBG}.

\begin{figure}
\includegraphics[height=5 cm]{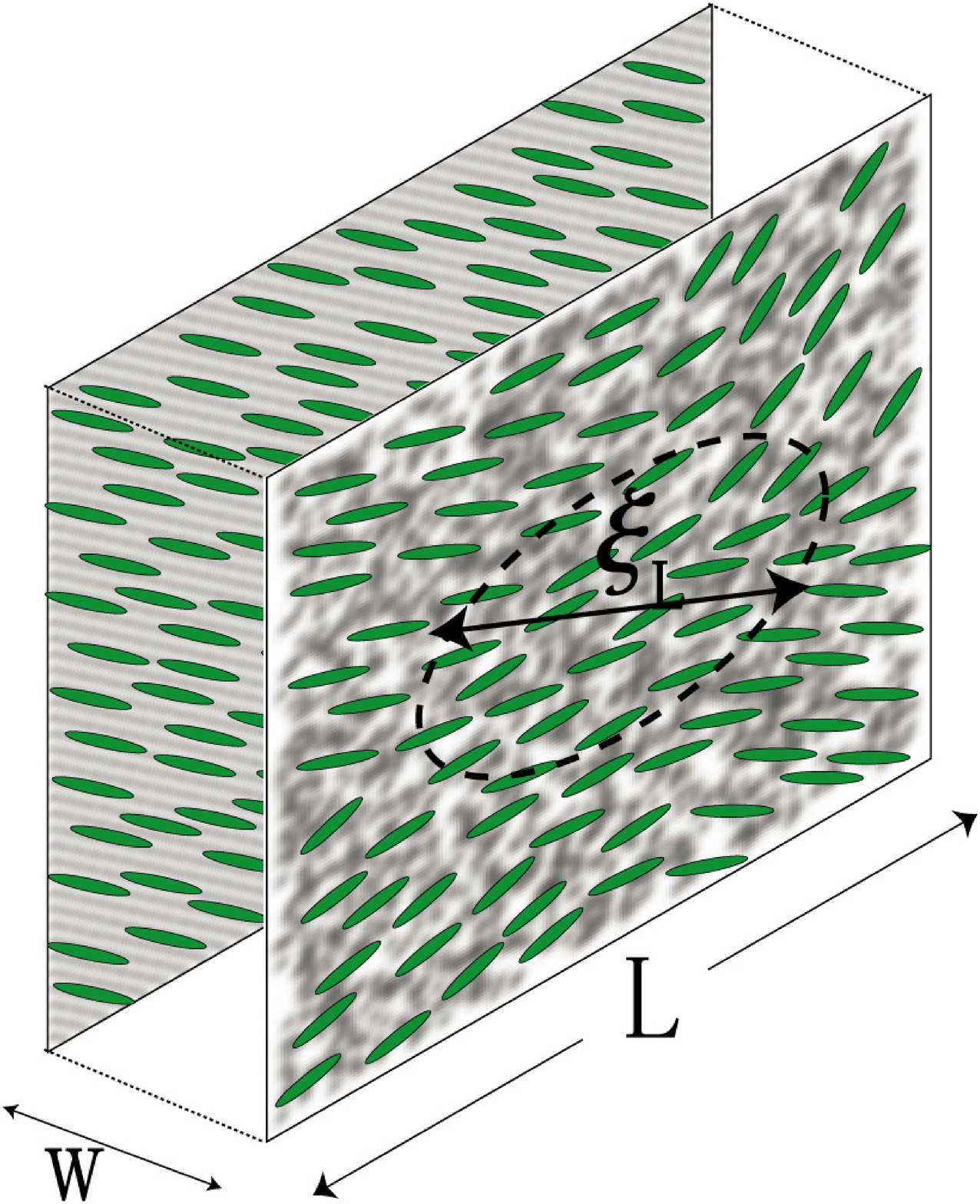}\\   
\includegraphics[height=5 cm]{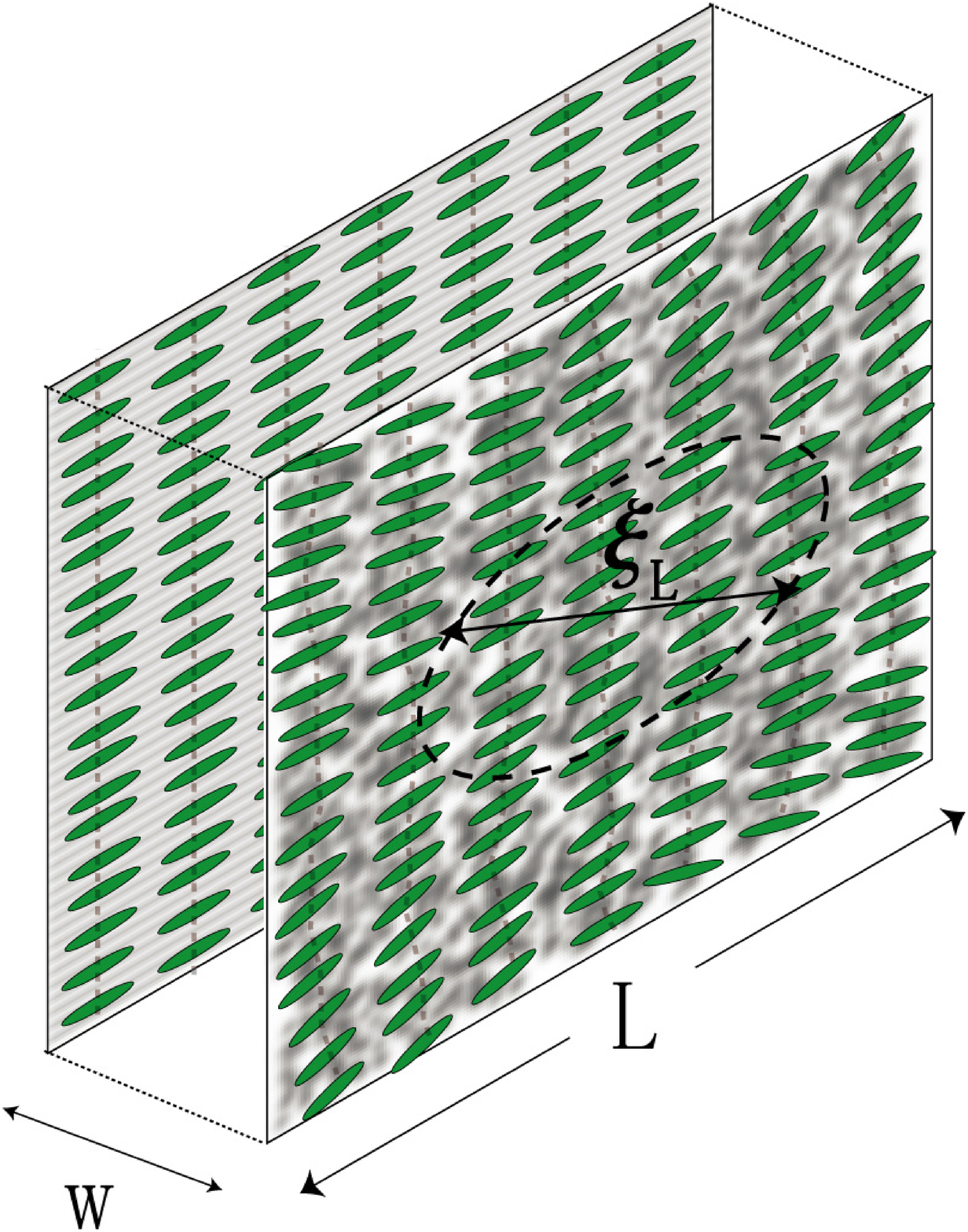}\\   
\caption{(Color online) Nematic (top) and smectic (bottom) liquid crystal cells of
  thickness $w$ with both substrates favoring a planar alignment.  The
  front substrate favors a {\em random heterogeneous} in-plane
  orientation, while the back substrate is rubbed and therefore picks
  out an ordering orientational axis.}
\label{fig:LCcell}
\end{figure}

With the exceptions of the pioneering {\em surface} disorder study by
Feldman and Vinokur \cite{FeldmanVinokurPRL} and our recent extension
to liquid crystal cells \cite{usFRGPRL}, all of the theoretical foci
have been on the {\em bulk} heterogeneity, where the disorder extends
over the full sample. However, there are many realizations in which,
instead, random pinning is confined to a subspace, e.g., a surface of
the sample. A technologically relevant example illustrated in
Fig.~\ref{fig:LCcell} is that of a liquid crystal cell (e.g., of a
laptop display), where a dirty substrate imposes random pinning,
that competes with liquid crystal ordering. In this paper, as a
significant elaboration of our earlier work \cite{usFRGPRL}, we present
a detailed analysis of such a cell.

A commonly observed Schlieren texture \cite{Schlieren_texture} is a
manifestation of such surface pinning in nematic cells. Recent studies
also include photo-alignment and dynamics in self-assembled liquid
crystalline monolayers \cite{ClarkSAMs}, as well as memory effects and
multistability in alignment of nematic cells with heterogeneous
random anchoring substrate \cite{Aryasova}. The existence of the
corresponding phenomena in smectic liquid crystals has been recently
revealed in ferroelectric smectic-$C$ cells in a book-shelf
geometry \cite{ClarkSmC}. This latter system exhibits a long-scale
smectic layers distortion, demonstrated to be driven by collective
random surface pinning, and awaits a detailed description.  Although
here we do not provide an analysis specific to these experiments, we
expect that results presented in this paper are a first step toward
this goal.

We conclude the introduction below by summarizing our main
results. The rest of the paper is organized as follows. In
Sec.~\ref{sec:model}, as a simplest description of nematic liquid
crystal cell, we present a model for an $xy$ system ordered in a
homogeneous bulk, but randomly pinned on a surface. In
Sec.~\ref{sec:Larkin} we analyze this model within the random-torque
Larkin approximation, valid on short length scales. To study the
physics on longer length scales we employ the functional
renormalization group (FRG) in Sec.~\ref{sec:FRG}, combining it with
matching methods in Sec.~\ref{sec:matching}, to compute the
asymptotics of orientational correlation functions. We briefly
consider the limit of strong surface pinning in
Sec.~\ref{sec:strongPinng}.  In Sec.~\ref{sec:NematicSmectic}, we
discuss application and extension of our findings to nematic and
smectic liquid crystals, and analyze experimental signatures of
heterogeneous surface-driven distortions in a
crossed-polarizer-analyzer light microscopy.  We conclude in
Sec.~\ref{sec:conclusion}.

\subsection{Summary of the results}
\label{results}

In this paper, we study stability and distortions of a liquid crystal
order in a cell with a random heterogeneous substrate. We model such
and related systems as a bulk $d$-dimensional $xy$ model with quenched
random-field disorder confined to a $(d-1)$-dimensional {\em surface}
\cite{comment_d}, and analyze its stability to surface pinning and
compute corresponding correlation functions.  As we will show, many of
the tools developed in the context of the {\em bulk} disorder can be
taken over to this ubiquitous surface-pinning problem.

Since the random pinning is confined to a surface, one might a priori
expect its effects to be vanishingly weak compared to the ordering
tendency of the homogeneous bulk, and thus the $xy$ order to be stable
to weak surface disorder in {\em any} dimension.  Our finding
contrasts sharply with this naive intuition \cite{discussionSenthil}.
Namely, our key qualitative observation is that the $xy$ order in such
$d$-dimensional system with $d\leq d_{lc}$, where
\begin{equation}
d_{lc}=3,
\label{dlc}
\end{equation}
is always destabilized by arbitrary weak random surface
pinning \cite{FeldmanVinokurPRL,usFRGPRL}. For $d>3$, the $xy$ order
requires a finite threshold of heterogeneity to be destroyed.  Thus,
while the lower-critical dimension of $d_{lc}=3$ for the system with
surface disorder is indeed reduced relative to that of the {\em bulk}
pinning problem, (characterized by
$d_{lc}^{(bulk)}=4$ \cite{LarkinOvchinnikov,ImryMa}), surface
heterogeneity has a qualitatively strong effect on the bulk ordering
even in the thermodynamic limit.

Above observation can be simply understood from a generalization of
the Imry-Ma argument \cite{ImryMa} to the surface-pinning
problem \cite{FeldmanVinokurPRL,usFRGPRL}. For an ordered region of
size $L$, the interaction with the surface random field can lower the
energy by an order of $E_{pin}\sim
V_p\sqrt{N}_p\sim\Delta_f^{1/2}L^{(d-1)/2}$, where $V_p$ is a typical
pinning strength with zero mean and variance $\Delta_f\approx
V_p^2/\xi_0^{d-1}$ ($\xi_0$ is the pinning correlation length) and
$N_p$ is the number of surface pinning sites. Since a surface
distortion on scale $L$ extends a distance $L$ into the bulk, the
corresponding elastic energy cost scales as $E_e\sim K L^{d-2}$, where
$K$ is the elastic stiffness. By comparing these energies, it is clear
that for $d<3$, on sufficiently long scales, $L >\xi_L\sim
(K^2/\Delta_f)^{\frac{1}{3-d}}$, the surface heterogeneity always
dominates over the elastic energy, and thus on these long scales
always destroys long-range $xy$ order for an arbitrary weak surface
pinning.

A more detailed analysis extends the argument to three dimensions (3D). 
The corresponding {\em surface} Larkin length scale \cite{LarkinOvchinnikov}, 
beyond which the $T=0$ orientational order on the random surface ($z=0$) is
destroyed, is given by
\begin{eqnarray}
\xi_L &=& a e^{c K^2/\Delta_f},\ \ \ \ \ \ \mbox{for $d = 3$},
\label{LarkinLength3}\\
      &=& \left[\frac{(3-d)(2\pi)^2K^2}{C_{d-1}\Delta_f}\right]^{\frac{1}{3-d}},\ \
      \mbox{for $d < 3$}, \label{LarkinLengthl3}
\end{eqnarray}
with $a$ is a microscopic cutoff of order of a few nanometers in the
context of liquid crystals, set by the molecular size, $c=8\pi^3$, and $C_{d-1}=2^{2-d}\pi^{(1-d)/2}/\Gamma(\frac{d-1}{2})$ as in Sec.~\ref{sec:FRG}.

At a distance $z\gg a$ away from the substrate (into the homogeneous
bulk), the orientational order distortions (characterized by
mean-squared fluctuations of $\phi$) across a region of size $\xi_L$
decay according to
\begin{eqnarray}
&&\hspace{-0.5 cm}
\overline{\langle \phi^2(0,z)\rangle}\bigg|_{\xi_L}\nonumber\\
&\sim&\left\{\begin{array}{ll}
1-\frac{2z}{\xi_L}(\ln{\frac{\xi_L}{2z}}+1-\gamma),&2z\ll \xi_L\\
\frac{\xi_L}{2z}e^{-2z/\xi_L},&2z\gg \xi_L
\end{array}\right.\ \mbox{for $d=2$},\nonumber\\
&\sim&\left\{\begin{array}{ll}
1-\Gamma(d-2)(\frac{2z}{\xi_L})^{3-d},&2z \ll \xi_L\\
(3-d)\frac{\xi_L}{2z}e^{-2z/\xi_L},& 2z\gg\xi_L
\end{array}\right.\ \mbox{for $2<d < 3$},\nonumber\\
&\sim& 
\left\{\begin{array}{ll}
1-\frac{\ln(2z/a)}{\ln(\xi_L/a)},&2z \ll \xi_L\\
\frac{\xi_L/2z}{\ln(\xi_L/a)}\ e^{-2z/\xi_L},& 2z\gg\xi_L
\end{array}\right.\ \mbox{for $d = 3$},
\end{eqnarray}
where $\gamma\approx 0.58$ is the Euler's constant. Thus we find that 
the orientational order distortions induced by the heterogeneous surface 
penetrate a distance $\xi_L$ into the bulk.

The physics is more complex on in-plane ($x$) length scales longer
than $\xi_L$, where the random-torque Larkin approximation is invalid
and the $\phi$ nonlinearities of the pinning potential, $V(\phi,\xv)$,
must be taken into account. To access these longer scales, we employed
FRG and matching methods.  On the
substrate ($z=0$), for $d<3$ we find ($q$ is an in-plane wave vector)
\begin{eqnarray}
\overline{|\phi_{q}|^2}\approx\left\{\begin{array}{ll}
\frac{\Delta_f}{K^2}\frac{1}{q^2},&q\gg 1/\xi_L\\
\frac{c(d)}{q^{d-1}},&q\ll 1/\xi_L
\end{array}\right.\ \ \mbox{for $d < 3$},
\label{Correlation_qd}
\end{eqnarray}
where $c(d)$ is a universal number [given in Eq.~\rf{Cqiii}], that in
the physically relevant case of $d=2$ is given by
$c(2)=\frac{\pi^3}{9}$.  At the lower-critical dimension, $d=3$, the
distortion variance is given by
\begin{equation}
\overline{|\phi_{q}|^2}\approx\left\{\begin{array}{ll}
\frac{\Delta_f}{K^2}\frac{1}{q^2},&q\gg 1/\xi_L\\
-\frac{2\pi^3}{9}\frac{1}{q^{2}\ln(q a)},&q\ll 1/\xi_L
\end{array}\right.\ \ \mbox{for $d = 3$}.
\label{Correlation_q3}
\end{equation}

These give the correlation function $C(\xv,z_1,z_2)
=\overline{\langle(\phi(\xv,z_1)-\phi(0,z_2))^2\rangle}$, where
$\langle \cdots\rangle$ is a thermal average and $\overline{\cdots}$
is a quenched disorder realization average, with latter dominating the
former at low temperatures. We report its full spatial dependence in
Secs.~\ref{sec:Larkin} and \ref{sec:matching} and in Figs.~\ref{fig:C2d} 
and \ref{fig:C3d}. On the substrate ($z=0$), its asymptotics in two 
dimensions is given by
\begin{eqnarray}
C_{2D}^{(\infty)}(x,0,0)\approx\left\{\begin{array}{ll}
8\pi^2\frac{\pi x}{2\xi_L},& x\ll \xi_L\\
8\pi^2+\frac{2\pi^2}{9}\ln (x/\xi_L),& x\gg \xi_L,
\end{array}\right.
\end{eqnarray}
for $x\gg \xi_L$ with finite $z$, the correlation 
function is given by
\begin{eqnarray}
\hspace{-0.5cm} C^{(\infty)}_{2D}(x,z,z)&\approx&b_2 e^{-2z/\xi_L} + 
\frac{\pi^2}{9}\ln\left[1+\frac{x^2}{(2z+\xi_L)^2}\right], 
\label{C2d}
\end{eqnarray}
in which $b_2$ is a weak function of $2z/\xi_L$. In the physically most 
relevant three dimensions, on the substrate, it is given
by\cite{FeldmanVinokurPRL,usFRGPRL}
\begin{equation}
C_{3D}^{(\infty)}(\xv,0,0)\approx\left\{\begin{array}{ll}
\frac{8\pi^2}{\ln{(\xi_L/a)}}\ln (x/a),& x\ll \xi_L\\
8\pi^2+\frac{2\pi^2}{9}\ln\left[\frac{\ln (x/a)}{\ln (\xi_L/a)}\right],& x\gg \xi_L.
\end{array}\right.
\end{equation}
At large $x\gg \xi_L$ with finite $z$, the correlation is given as
\begin{eqnarray}
C^{(\infty)}_{3D}(\xv,z,z)
&\approx& b_3 e^{-2z/\xi_L}\nonumber\\
&&+\frac{2\pi^2}{9}\left\{\begin{array}{ll}
\ln\big[\frac{\ln(x/a)}{\ln(\xi_L/a)}\big],&2z \ll \xi_L \ll x\\
\ln\big[\frac{\ln(x/a)}{\ln(2z/a)}\big],&\xi_L\ll 2z \ll x\\ 
\frac{x^2}{16z^2}\frac{1}{\ln{(2z/a)}},&\xi_L \ll x \ll 2z ,
\end{array}\right.\nonumber\\
\label{C3d}
\end{eqnarray}
in which $b_3$ is also a weak function of $2z/\xi_L$. 
The exponentially decaying parts of both two-dimensional (2D) and 3D 
results are contribution from short scales below the Larkin length,
for $z\gg \xi_L$.

With an eye to liquid crystal cell applications, we also analyzed a
nematic cell of finite thickness $w$, with (as above) a heterogeneous
bottom substrate, and a top substrate with a homogeneous Dirichlet
or a homogeneous Neumann boundary conditions. Not surprisingly, this
presents new crossover as a function of the ratio, $w/\xi_L$, of the
cell thickness $w$ to the Larkin length in a (infinitely) thick cell.

For the Larkin length in the Dirichlet cell, illustrated in
Fig.~\ref{fig:xiDN} we find
\begin{eqnarray}
\xi_L^{(\cal{D})}\approx\left\{\begin{array}{ll}
\xi_L,& \xi_L\ll w\\
\frac{c_dw^{\nu_d+1}}{(\xi_L^{*}-\xi_L)^{\nu_d}},&
\xi_L\lesssim\xi_L^{*},
\end{array}\right.
\label{xiD}
\end{eqnarray}
where $\xi_L^{*}=a_d w$ is the crossover ``bulk'' Larkin length beyond 
which effects of the finite cell thickness become important, $c_2=1, 
a_2\approx 1.71, \nu_2=1$ and $c_3\approx 0.79, a_3\approx 1.23, \nu_3=1/2$.

\begin{figure}
\includegraphics[height=5 cm]{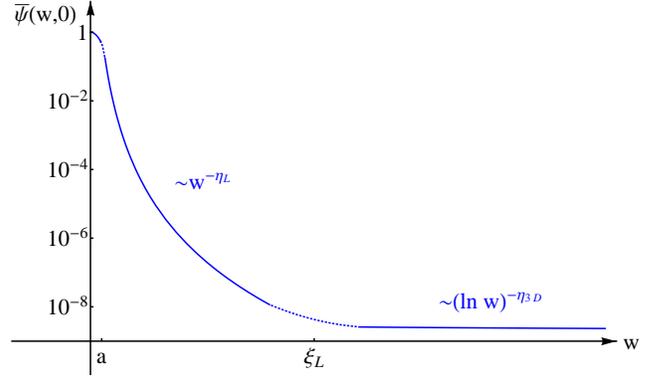}\\  
\caption{(Color online) Orientational order parameter $\overline{\psi}(w,0)$
  (controlling light transmission through a liquid crystal cell) at
  the random pinning substrate of a 3D Dirichlet cell of
  thickness $w$.}
\label{fig:OrderParameter}
\end{figure}

Above behavior in the Dirichlet cell is a manifestation of a crossover
as a function of cell thickness (or equivalently disorder strength )
from a weakly ordered state for a thick cell (and strong disorder) to
a strongly ordered state for a thin cell (and weak disorder).  The
crossover is more clearly reflected in the \textit{surface} $xy$ orientational 
order parameter $\overline{\psi}=\overline{\langle e^{i\phi}\rangle}$, 
that in $d<3$ is given by
\begin{eqnarray}
\overline{\psi}_{d<3}\approx\left\{\begin{array}{ll}
e^{-\alpha(w/\xi_L)^{3-d}},& \mbox{thin cell, $w\ll\xi_L$}\\
e^{-\alpha}\left(\frac{\xi_L}{w}\right)^{\eta_d^*}, & \mbox{thick cell, $w\gg\xi_L$},
\end{array}\right. 
\end{eqnarray}
and in 3D the orientational order parameter, illustrated in Fig.~\ref{fig:OrderParameter},
 is given by
\begin{eqnarray}
\overline{\psi}_{3D}\approx\left\{\begin{array}{ll}
\left(\frac{a}{w}\right)^{\eta_L},& \mbox{thin cell, $w\ll\xi_L$}\\
e^{-\alpha}
\left[\frac{\ln(\xi_L/a)}{\ln(w/a)}\right]^{\eta_{3D}},& \mbox{thick cell, $w\gg\xi_L$},
\end{array}\right.
\end{eqnarray}
where $\eta_d^*=(3-d)\pi^2/18$, $\eta_{3D}=\pi^2/18$ are universal
exponents [given in Eqs.~\rf{eta_d} and \rf{eta_3}] and $\alpha=2\pi^2$,
$\eta_L=2\pi^2/\ln(\xi_L/a)$ are nonuniversal constants.

In contrast, for a cell with a Neumann boundary condition on the top
substrate, the Larkin length, illustrated in Fig.~\ref{fig:xiDN}, is
given by
\begin{eqnarray}
\xi_L^\N\approx\left\{\begin{array}{ll}
\xi_L,& w\gg\xi_L\\
w^{\gamma_d}\left\{\begin{array}{ll}
\sqrt{2\ln(1.2\xi_L/w)},&d=3\\
(\frac{5-d}{3-d})^{\frac{1}{5-d}}(\xi_L)^{\frac{3-d}{5-d}},&d<3
\end{array}\right.&w\ll\xi_L,
\end{array}\right.\nonumber\\
\end{eqnarray}
with $\gamma_d =\frac{2}{5-d}$ and the state is disordered, i.e.,
$\overline{\psi}=0$ for arbitrary weak random pinning.

We expect these crossovers as a function of $w/\xi_L$ to be relevant
to understanding the ordering in real liquid crystal cells. They
should be accessible experimentally in a setup of the type used in
Ref.~[\onlinecite{Aryasova}].

\section{The model}
\label{sec:model}
\begin{figure}
\includegraphics[height=5 cm]{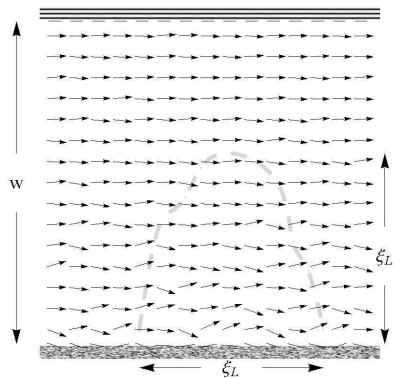}\\   
\caption{Schematic illustration of the orientational distortions in
  an $xy$-model system confined to a thickness $w$ cell with random pinning
  on the bottom substrate and Dirichlet boundary condition (rubbed
  substrate) on the top substrate. A Larkin domain (defined by the faint
  dashed curve) of size $\xi_L$ is also depicted.}
\label{finite_d}
\end{figure}

As a ``toy'' model of a nematic liquid crystal cell with a dirty
substrate and thickness $w$, illustrated in Fig.~\ref{finite_d}, we
employ a $d$-dimensional {\em surface} random-field $xy$ model
characterized by a Hamiltonian
\begin{eqnarray}
H=\int_{-\infty}^{\infty}d^{d-1}x \int_0^wdz
\left[\frac{K}{2}(\nabla \phi(\rv))^2
-V[\phi(\rv),\xv]\delta(z)\right].\nonumber\\
\label{Hbulk}
\end{eqnarray}
In above $\phi(\rv)$ is the $xy$-field distortion at a point
$\rv\equiv(\xv,z)$, the random pinning potential
$V[\phi(\xv,z),\xv]\delta(z)$ with a $2\pi$ periodicity of $\phi$, 
characterized by zero mean and Gaussian distribution with a variance
\begin{eqnarray}
\overline{V(\phi,\xv)V(\phi',\xv')}
=R(\phi-\phi')\delta^{d-1}(\xv-\xv'),
\label{Rdefine}
\end{eqnarray}
is confined to the bottom substrate at $z=0$, and we impose either 
a Dirichlet [$\phi(\xv,z)|_{z=w}=0$] or a Neumann 
[$\partial_z\phi(\xv,z)|_{z=w}=0$] boundary condition on the top
homogeneous substrate at $z=w$ .

A more realistic model must of course include {\em nonplanar} director
distortions, characterized by an additional polar angle, as well as
point and line (strength $1/2$ disclinations) topological defects
allowed by the 3D headless nematic director field, $\hat n$.  We
discuss the effects of these additional ingredients in
Sec. \ref{sec:NematicSmectic}.
The long-scale behavior of the coarse-grained periodic (period $2\pi$)
variance function $R(\phi)$ characterizes low-temperature properties
of our system and will therefore be our main focus.

\subsection{Dimensional reduction}

Because the random pinning potential in the Hamiltonian, \rf{Hbulk}, is
confined to the bottom substrate at $z=0$, no nonlinearities appear in
the bulk ($0<z<w$) of the cell. Consequently, as in other similar
problems \cite{tiltLR,KaneFisherPRB}, it is possible and convenient to
focus on the random substrate and exactly eliminate the bulk degree of
freedom of $\phi(\xv,z)$ in favor of the random substrate field
$\phi_0(\xv)\equiv\phi(\xv,z=0)$. This can be done via a constrained
path-integral by integrating out $\phi(\xv,z)$ with a constraint
$\phi(\xv,z=0)=\phi_0(\xv)$, thereby obtaining an effective
$(d-1)$-dimensional Hamiltonian for $\phi_0(\xv)$ \cite{tiltLR}.
Equivalently \cite{equivalentComment,Aryasova2007} , we can eliminate $\phi(\xv,z)$
by solving the bulk Euler-Langrange equation $\nabla^2\phi(\rv)=0$. To
this end, we Fourier transform $\phi(\xv,z)$ over $\xv$, obtaining an
equation for $\phi(\qv,z)=\int d^{d-1}x\phi(\xv,z)e^{-i\qv\cdot\xv}$,
\begin{equation}
\partial_z^2\phi(\qv,z)-q^2\phi(\qv,z)=0,
\end{equation}
whose solutions for the boundary conditions of interest are obtained
by elementary methods
\begin{eqnarray}
\phi^{(\infty)}(\qv,z)
&=&\phi_0(\qv)e^{-q z},\hspace{1.7cm}w\rightarrow\infty,\label{phi}\\
\phi^{\D}(\qv,z)
&=&\phi_0(\qv)\frac{\sinh{\left[q(w-z)\right]}}{\sinh{(q w)}},
\ \mbox{Dirichlet},\label{phiD}\\
\phi^{\N}(\qv,z)
&=&\phi_0(\qv)\frac{\cosh{\left[q(w-z)\right]}}{\cosh{(q w)}},
\ \mbox{Neumann}\label{phiN}.
\end{eqnarray}
We summarize above three boundary condition cases by
\begin{eqnarray}
\phi^{(a)}(\qv,z)&=&\phi_0(\qv)\varphi^{(a)}(q,z),
\end{eqnarray}
where mode functions $\varphi^{(a)}(q,z)$ are implicitly defined by
Eqs.~(\ref{phi})-(\ref{phiN}).

Substituting these into the Hamiltonian, Eq.~\rf{Hbulk}, we obtain the
$(d-1)$-dimensional (surface) Hamiltonian
\begin{eqnarray}
H_s=\int
\frac{d^{d-1}q}{(2\pi)^{d-1}}\oh\Gamma_q^{(a)}|\phi_0(\qv)|^2
-\int d^{d-1}x V[\phi_0(\xv),\xv],\nonumber\\
\label{Hs}
\end{eqnarray}
which characterizes the behavior of the field $\phi_0(\xv)$ confined
to the random substrate at $z=0$. In the equation above, label $a$ ranges 
over ``free'' ($w\rightarrow\infty$), Dirichlet, and Neumann boundary
conditions on the $z=w$ substrate, with corresponding kernels given by
\begin{eqnarray}
\Gamma_q^{(\infty)}
&=& K q,\hspace{1.5cm}w\rightarrow\infty,\label{GammaI}\\
\Gamma_q^{\D}
&=& K q \coth(q w),
\ \mbox{Dirichlet},\label{GammaD}\\
\Gamma_q^{\N}
&=& K q \tanh(q w),
\ \mbox{Neumann}\label{GammaN}.
\end{eqnarray}
As expected, the finite-thickness Dirichlet ($\Gamma_q^{\D}$) and
Neumann ($\Gamma_q^{\N}$) kernels for $w\rightarrow\infty$ reduce to
the case of the free kernel, $K q$. The $q$ nonanalyticity and long-
wavelength stiffening (relative to the bulk $K q^2$ kernel) of the
latter arises due to a mediation of surface distortions by long-range
deformations in the bulk of the cell. In the opposite limit of a thin
cell and long scales, as expected, the Dirichlet kernel reduces to a
``massive'' one $K/w$, and the Neumann kernel simplifies to $K w q^2$
of an ordinary surface (without a contact with the bulk) $xy$ model.
The advantage of the dimensional reduction above is that formally the
formulation of the random surface problem becomes identical to that of
the extensively studied bulk random
pinning \cite{DSFisherFRG,Nattermann,BalentsFisher,GiamarchiLedoussal}
but in one lower dimension and with a modified long-range
elasticity. 

\subsection{Replicated model}

To compute self-averaging quantities, (e.g., the disorder-averaged
free energy), it is convenient (but not necessary) to employ the
replica ``trick'' \cite{Anderson}, which allows us to work with a
translationally invariant field theory at the expense of introducing
$n$ replica fields (with the $n\rightarrow 0$ limit to be taken at the
end of the calculation). For the free energy, this procedure relies on
the identity for the $\ln(x)$ function
\begin{equation}
{\overline F}=-T\overline{\ln Z}=-T\lim_{n\rightarrow
0}{\overline{Z^n}-1\over n}\;.
\label{replica}
\end{equation}
After replicating and integrating over the random potential
$V[\phi,\xv]$ using Eq.~\rf{Rdefine}, we obtain
\begin{equation}
\overline{Z^n}=\int[d\phi_0^\alpha]e^{-H_s^{(r)}[\phi_0^\alpha]/T}\;.
\end{equation}
The effective translationally invariant replicated Hamiltonian
$H_s^{(r)}[\phi_0^\alpha]$ is given by
\begin{eqnarray}
H_s^{(r)}&=&\sum_{\alpha}^n\int\frac{d^{d-1}q}{(2\pi)^{d-1}}
\oh\Gamma_q^{(a)}|\phi_0^\alpha(\qv)|^2\nonumber\\
&&-\frac{1}{2T}\sum_{\alpha,\beta}^n\int d^{d-1}x R[\phi_0^\alpha(\xv)-\phi_0^\beta(\xv)].
\label{Hsr}
\end{eqnarray}
We will use this Hamiltonian, \rf{Hsr}, in our subsequent
RG analysis of the system.

\section{Larkin analysis}
\label{sec:Larkin}
\subsection{Random torque model}
As with the bulk quenched disorder, the nontrivial nature of the
surface-pinning problem is encoded in the nonlinearity in
$\phi_0(\xv)$ of the random surface potential $V[\phi_0(\xv),\xv]$
in $H_s$, Eq.~\rf{Hs}. However, in an approximation first employed by
Larkin \cite{Larkin} (that now bares his name), for small $\phi_0$
distortions we can Taylor-expand the random potential to linear
order \cite{linearComment} in $\phi_0$
\begin{eqnarray}
H_s^{(L)}&\approx&
\int\frac{d^{d-1}q}{(2\pi)^{d-1}}\oh\Gamma_q^{(a)}|\phi_0(\qv)|^2
-\int d^{d-1}x\ \tau(\xv)\phi_0(\xv)\nonumber\\
&\approx&\int_\qv\left[\oh\Gamma_q^{(a)}|\phi_0(\qv)|^2
-\tau(-\qv)\phi_0(\qv)\right],
\label{HsLarkin}
\end{eqnarray}
obtaining a harmonic Hamiltonian characterized by
$\phi_0$-independent random surface torque
\begin{eqnarray}
\tau(\xv)&=&\partial_{\phi_0}
V[\phi_0(\xv),\xv]\bigg|_{\phi_0=0}\nonumber\\
&=&V'[0,\xv].
\label{torque}
\end{eqnarray}
The random-torque inherits Gaussian statistics from that of the random
potential $V[\phi_0(\xv),\xv]$, with its variance given by
\begin{subequations}
\begin{eqnarray}
\overline{\tau(\xv)\tau(\xv')}
&=&-R''(0)\delta^{d-1}(\xv-\xv')\\
&\equiv&\Delta_f\delta^{d-1}(\xv-\xv').
\label{Delta_f}
\end{eqnarray}
\end{subequations}

\subsection{Correlation functions}
The resulting surface Larkin model (valid on scales shorter than
$\xi_L$) is quadratic in $\phi_0(\xv)$ and can therefore be analyzed
exactly by standard methods. The basic quantity of primary interest is
the Fourier transform of the disorder- and thermally averaged
two-point correlation function
\begin{eqnarray}
\overline{\langle\phi_{\qv}\phi_{\qv'}\rangle}
&=&\overline{\langle\phi_{\qv}\rangle\langle\phi_{\qv'}\rangle}
+ \overline{\langle(\phi_{\qv}-\langle\phi_{\qv}\rangle)
(\phi_{\qv'}-\langle\phi_{\qv'}\rangle)\rangle}\nonumber\\
&=&\left[C_\Delta(q)+C_T(q)\right](2\pi)^{d-1}\delta^{d-1}(\qv+\qv'),
\hspace{1cm}
\end{eqnarray}
with
\begin{eqnarray}
C_{\Delta,Larkin}^{(a)}(q)&=&\frac{\Delta_f}{\left[\Gamma_q^{(a)}\right]^2},\label{C_D}\\
C_{T,Larkin}^{(a)}(q)&=&\frac{T}{\Gamma_q^{(a)}},\label{C_T}
\end{eqnarray}
with the Larkin approximations for the $T=0$ distortions of $\phi_0(\xv)$
and its thermal fluctuations about this pinned ground state,
respectively. Because of the respective structures of $C_\Delta(q)$ and
$C_T(q)$, at long scales (small $q$), clearly the quenched-disorder-driven 
distortions in the ground state dominate over small thermal
fluctuations. Thus, for the remainder of the paper, we will focus on
the system at $T=0$.

\begin{figure}
\includegraphics[height=5 cm]{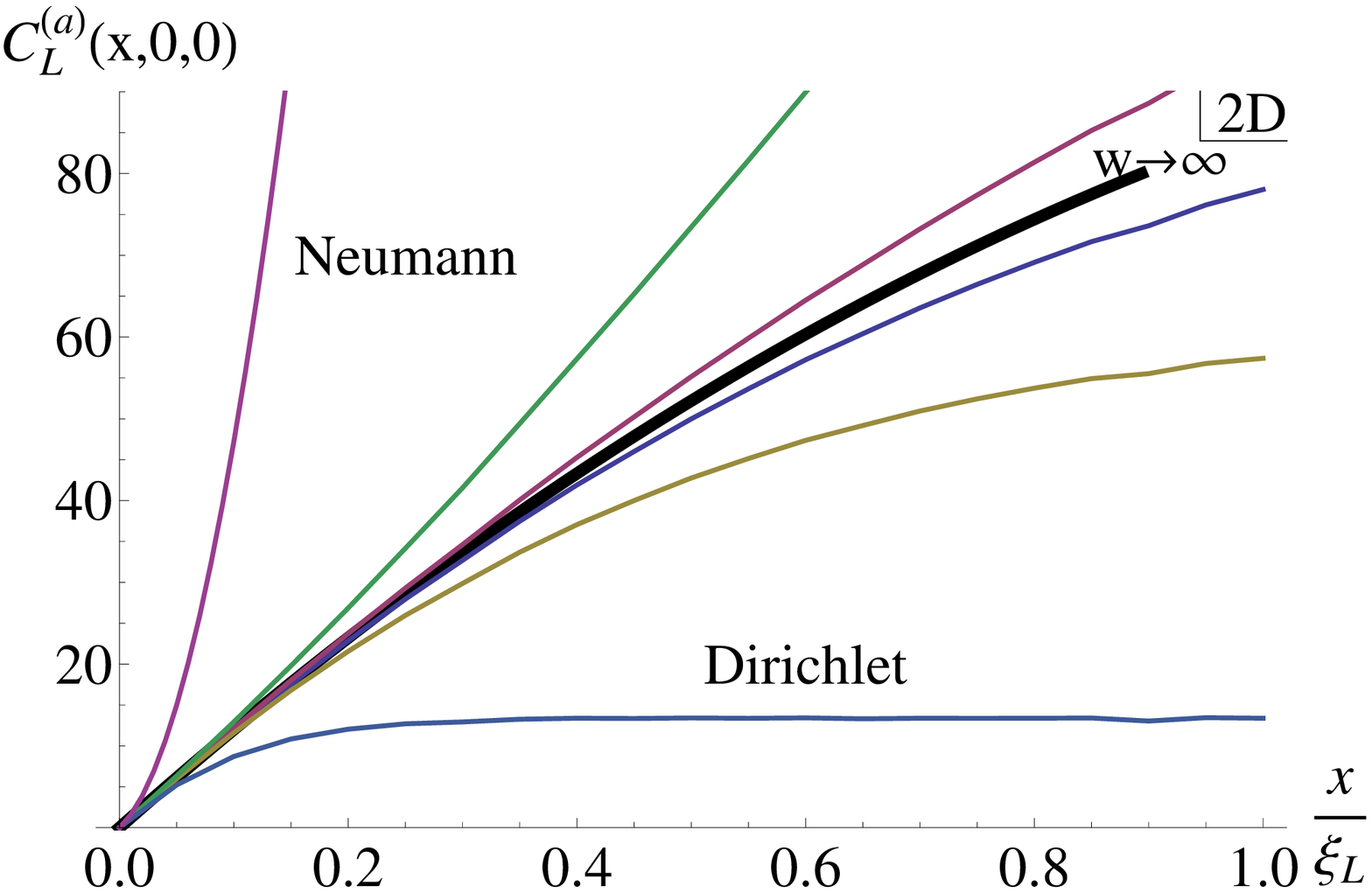}\\   
\includegraphics[height=5 cm]{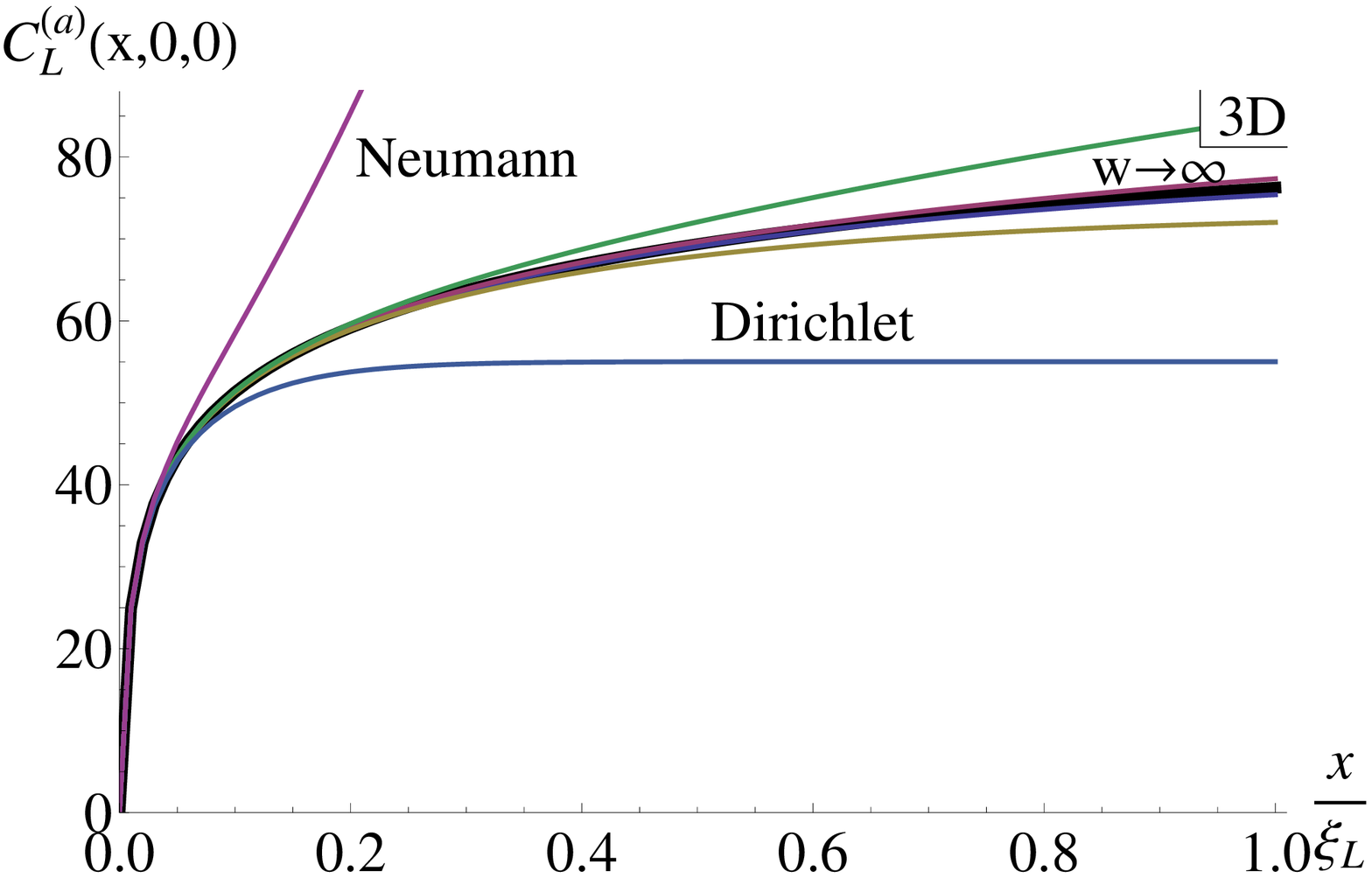}\\   
\caption{(Color online)
  Numerically evaluated Larkin ($x < \xi_L$) correlation
  functions for a finite-thickness cell $w$, with the Dirichlet and
  Neumann top substrates in 2D (top) and 3D (bottom). In both figures,
  the thicker black curve corresponds to the infinite cell thickness,
  above it are the results for the Neumann substrate (with $w/\xi_L=1,
  0.5, 0.1$, bottom to top), and below it are the results for the
  Dirichlet substrate (with $w/\xi_L=1, 0.5, 0.1$, top to bottom). As
  expected, the Dirichlet (Neumann) homogeneous substrate reduces
  (enhances) orientational distortions of $\phi$, with its influence
  growing with reduced cell thickness and weaker disorder, controlled
  by $w/\xi_L$.}
\label{fig:CDN_L}
\end{figure}

The real-space correlation function 
\begin{eqnarray}
C(\xv,z,z')=\overline{\langle(\phi(\xv,z)-\phi(0,z'))^2\rangle},
\end{eqnarray}
which describes surface and bulk distortions that follows via a Fourier
transform of Eq.~\rf{C_D}, combined with Eqs.~(\ref{phi})-(\ref{phiN}). 
Specializing for simplicity to the
case $z=z'$, we compute 
\begin{eqnarray}
\hspace{-0.2cm}C_L^{(a)}(\xv,z,z)&\approx&
2\Delta_f\int\frac{d^{d-1}q}{(2\pi)^{d-1}}(1-\cos\qv\cdot\xv)
\left[\frac{\varphi^{(a)}(q,z)}{\Gamma_q^{(a)}}\right]^2\nonumber\\
\label{CLxzz}
\end{eqnarray}
for the three boundary conditions of interest, with the subscript
``$L$'' denoting that here the validity is limited to the Larkin
approximation. These correlation functions  
are illustrated in Figs.~\ref{fig:CDN_L}, \ref{fig:C_L2d}, and
\ref{fig:C_L3d}.

For the infinitely thick cell ($w\rightarrow\infty$), the asymptotics
(for $a\ll x\ll\xi_L$) is given by (see Appendix \ref{app:LarkinCorr})
\begin{eqnarray}
&&\hspace{-1.2cm}C_L^{(\infty)}(\xv,z,z)\nonumber\\
&\approx&
\frac{2\Delta_f}{K^2}\int_\qv
\frac{(1-\cos\qv\cdot\xv)e^{-2q z}}{q^2},\ \ \mbox{for $x\ll\xi_L$},\nonumber\\
&\sim& \left\{\begin{array}{ll}
x,&z=0\\
\frac{x^2}{2z\xi_L}e^{-2z/\xi_L},&z \gg x
\end{array}\right.,\ \ \ \mbox{for $d=2$},\nonumber\\
&\sim& \left\{\begin{array}{ll}
\left(\frac{x}{\xi_L}\right)^{3-d},&z=0\\
\frac{x^2(2z)^{1-d}}{\xi_L^{3-d}}\Gamma{(d-1,\frac{2z}{\xi_L})},&z \gg x
\end{array}\right.,\ \ \ \mbox{for $d < 3$},\nonumber\\
&\sim&
\left\{\begin{array}{ll}
\frac{\ln{(x/a)}}{\ln{(\xi_L/a)}},&z=0\\ 
\frac{(\frac{x}{2z})^2(1+\frac{2z}{\xi_L})}{\ln(\xi_L/a)}\ e^{-2z/\xi_L},& z\gg x
\end{array}\right.,\ \ \ \mbox{for $d = 3$,}
\label{LarkinCorrelationInfinite}\nonumber\\
\end{eqnarray}
and by definition vanishes as $x\rightarrow 0$. 

For a cell of thickness $w$ with a Dirichlet and Neumann boundary
conditions on the top substrate, the correlation functions are given
by
\begin{eqnarray}
C_L^{\D}(\xv,z,z)&\approx&
\frac{2\Delta_f}{K^2}\int_\qv
\frac{(1-\cos\qv\cdot\xv)\sinh^2q(w-z)}{q^2\cosh^2q w},\nonumber\\
&&\label{LarkinCorrelationIntegralDirichlet} \\ 
C_L^{\N}(\xv,z,z)&\approx&
\frac{2\Delta_f}{K^2}\int_\qv
\frac{(1-\cos\qv\cdot\xv)\cosh^2q(w-z)}{q^2\sinh^2q w}.\nonumber\\ \label{LarkinCorrelationIntegralNeumann}
\end{eqnarray}
For a thick cell ($w>\xi_L$), these can be approximated by the infinite
cell results given above.  For a thin cell ($w < \xi_L$) with a top
Dirichlet substrate, their asymptotics is given by
\begin{eqnarray}
&&\hspace{-1.2cm}C_L^{\D}(\xv,z,z) \nonumber\\
&\sim& \left\{\begin{array}{ll}
(3-d)(\frac{w}{\xi_L})^{3-d},&z \ll w\ll x\\
(3-d)\frac{(z-w)^2}{w^{d-1}\xi_L^{3-d}},&z \lesssim w\ll x
\end{array}\right.\ \ \ \mbox{for $d < 3$}.\label{LarkinCorrelationDirichlet}\nonumber\\
\end{eqnarray}
The asymptotics for $d=3$ and for the Neumann boundary conditions are
more involved and are best evaluated numerically. They are displayed
in Fig.~\ref{fig:CDN_L}.

\subsection{Limits of validity: Larkin length}

An important feature of the Larkin approximation is that it can be
used to estimate its own range of validity. The Larkin model
\rf{HsLarkin} breaks down when the Taylor expansion in $\phi_0$ that
led to it becomes invalid. This can be estimated by looking at the
mean-squared distortions of $\phi_0$, given by
\begin{eqnarray}
\overline{\langle\phi_0^2(\xv)\rangle}
&\approx&\int_\qv C_\Delta^{(a)}(q)\nonumber\\
&\approx&\int_\qv \frac{\Delta_f}{\left[\Gamma^{(a)}_q\right]^2}.
\end{eqnarray}
Focusing first on an infinitely thick cell, characterized by
$\Gamma_q^{(\infty)} = K q$, we note that for $d<d_{lc}=3$,
mean-squared fluctuations of $\phi_0(\xv)$ diverge at long length
scales. Thus, we find that \cite{FeldmanVinokurPRL,usFRGPRL}
\begin{equation}
d_{lc}=3
\end{equation}
is the lower-critical dimension for the stability of the $xy$ order in
the presence of a random {\em surface} quenched pinning. This value of
$d_{lc}=3$ is to be contrasted with the
$d_{lc}^{(bulk)}=4$\cite{Larkin,ImryMa} of a system (spontaneously
breaking continuous symmetry) subjected to {\em bulk} disorder. Thus,
for $d\le 3$, we can define the (so-called) Larkin length, $\xi_L$ as
the scale at which $\phi_0$ distortions grow to order $1$ (that we
take to be $2\pi$, for concreteness). That is,
\begin{equation}
\overline{\langle\phi_0^2(\xv)\rangle}=(2\pi)^2
=\frac{\Delta_f}{K^2}\int_{1/\xi_L}^{1/a}
\frac{d^{d-1}q}{(2\pi)^{d-1}}\frac{1}{q^2},
\label{phi_Lrms}
\end{equation}
which gives (see Appendix \ref{app:LarkinLengths}) the Eqs.~\rf{LarkinLength3} and \rf{LarkinLengthl3} 
\begin{eqnarray}
\xi_L &=& a e^{c K^2/\Delta_f},\ \ \ \ \ \ \mbox{for $d = 3$},\nonumber\\
      &=& \left[\frac{(3-d)(2\pi)^{2}K^2}
                {C_{d-1}\Delta_f}\right]^{\frac{1}{3-d}},\ \
      \mbox{for $d < 3$},\nonumber
\end{eqnarray}
with $a$ a microscopic cutoff of order of a few nanometers in the
context of liquid crystals, set by the molecular size, $c=8\pi^3$, and $C_{d-1}=2^{2-d}\pi^{(1-d)/2}/\Gamma(\frac{d-1}{2})$ as in Sec.~\ref{sec:FRG}.

Since the orientational order parameter 
\begin{eqnarray}
\overline{\psi}&=&\overline{\langle e^{i\phi}\rangle}\nonumber\\
&\approx& e^{-\overline{\langle\phi^2\rangle}/2}
\end{eqnarray}
(somewhat crudely assuming Gaussian correlations in $\phi$ in the
second line above) decays with the growing $\phi_{rms}$, the Larkin
length, $\xi_L$, is the scale beyond which the orientational order
falls off significantly.

As discussed in the Sec.~\ref{sec:intro}, the mean-squared distortions 
of $\phi(\xv,z)$ are suppressed away from the random substrate, 
within Larkin approximation on scale $\xi_L$, for 
$w\rightarrow\infty$ decaying with $z\gg a$
according to (see Appendix \ref{app:msDistortion})
\begin{eqnarray}
&&\hspace{-0.6 cm}\overline{\langle \phi^2(0,z) \rangle}\bigg
   |_{\xi_L}\nonumber\\
&\sim&\left\{\begin{array}{ll}
1-\frac{2z}{\xi_L}(\ln{\frac{\xi_L}{2z}}+1-\gamma),&2z\ll \xi_L\\
\frac{\xi_L}{2z}e^{-2z/\xi_L},&2z\gg \xi_L
\end{array}\right.\ \mbox{for $d=2$},\nonumber\\
&\sim&\left\{\begin{array}{ll}
1-\Gamma(d-2)(\frac{2z}{\xi_L})^{3-d},&2z \ll \xi_L\\
(3-d)\frac{\xi_L}{2z}e^{-2z/\xi_L},& 2z\gg\xi_L
\end{array}\right.\ \mbox{for $2<d < 3$},\nonumber\\
&\sim& 
\left\{\begin{array}{ll}
1-\frac{\ln(2z/a)}{\ln(\xi_L/a)},&2z \ll \xi_L\\
\frac{\xi_L/2z}{\ln(\xi_L/a)}\ e^{-2z/\xi_L},& 2z\gg\xi_L
\end{array}\right.\ \mbox{for $d = 3$}.
\eea
The complete numerically-evaluated behavior is
displayed in Fig.~\ref{SelfCorrelation}. Thus, the orientational order
heals on the scale of the Larkin length, $\xi_L$, into the bulk.

\begin{figure}
\includegraphics[height=5 cm]{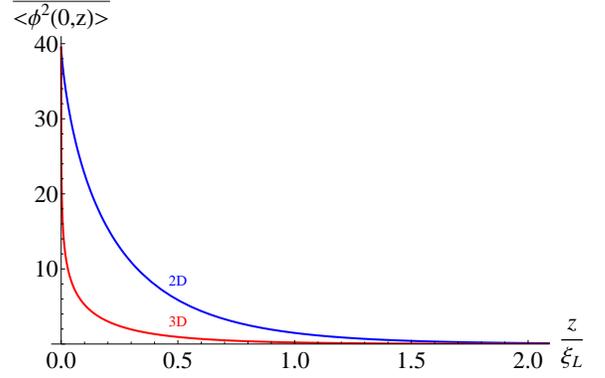}\\   
\caption{(Color online) Mean-squared distortion within the Larkin 
  approximation as a function of $z/\xi_L$ for 2D and 3D systems.}
\label{SelfCorrelation}
\end{figure}

In the interest of applications to liquid crystal cells, we generalize
this analysis to a system with finite thickness, $w$ (along $z$).  For a
homogeneous Dirichlet boundary condition on the top substrate of the
cell, we have
\begin{eqnarray}
\overline{\langle\phi_0^2(\xv)\rangle}^\D=(2\pi)^2
&=&\frac{\Delta_f}{K^2}\int_{1/\xi_L^\D}^{1/a}
\frac{d^{d-1}q}{(2\pi)^{d-1}}\frac{\tanh^2q w}{q^2},\hspace{0.7cm}
\end{eqnarray}
which gives (see Appendix \ref{app:crossover}) Eq.~\rf{xiD} 
\begin{eqnarray}
\xi_L^{(\cal{D})}\approx\left\{\begin{array}{ll}
\xi_L,& \xi_L\ll w\\
\frac{c_d w^{\nu_d+1}}{(\xi_L^{*}-\xi_L)^{\nu_d}},&
\xi_L\lesssim\xi_L^*,
\end{array}\right.\nonumber
\end{eqnarray}
with $\xi_L^{*}=a_d w$ the crossover ``bulk'' Larkin length, $c_2=1,
a_2\approx 1.71, \nu_2=1$, and $c_3\approx 0.79, a_3\approx 1.23,
\nu_3=1/2$. The complete behavior of $\xi_L^{(\cal{D},\cal{N})}$ in 2D and
3D is illustrated in Fig.~\ref{fig:xiDN}.

\begin{figure}
\includegraphics[width=9 cm]{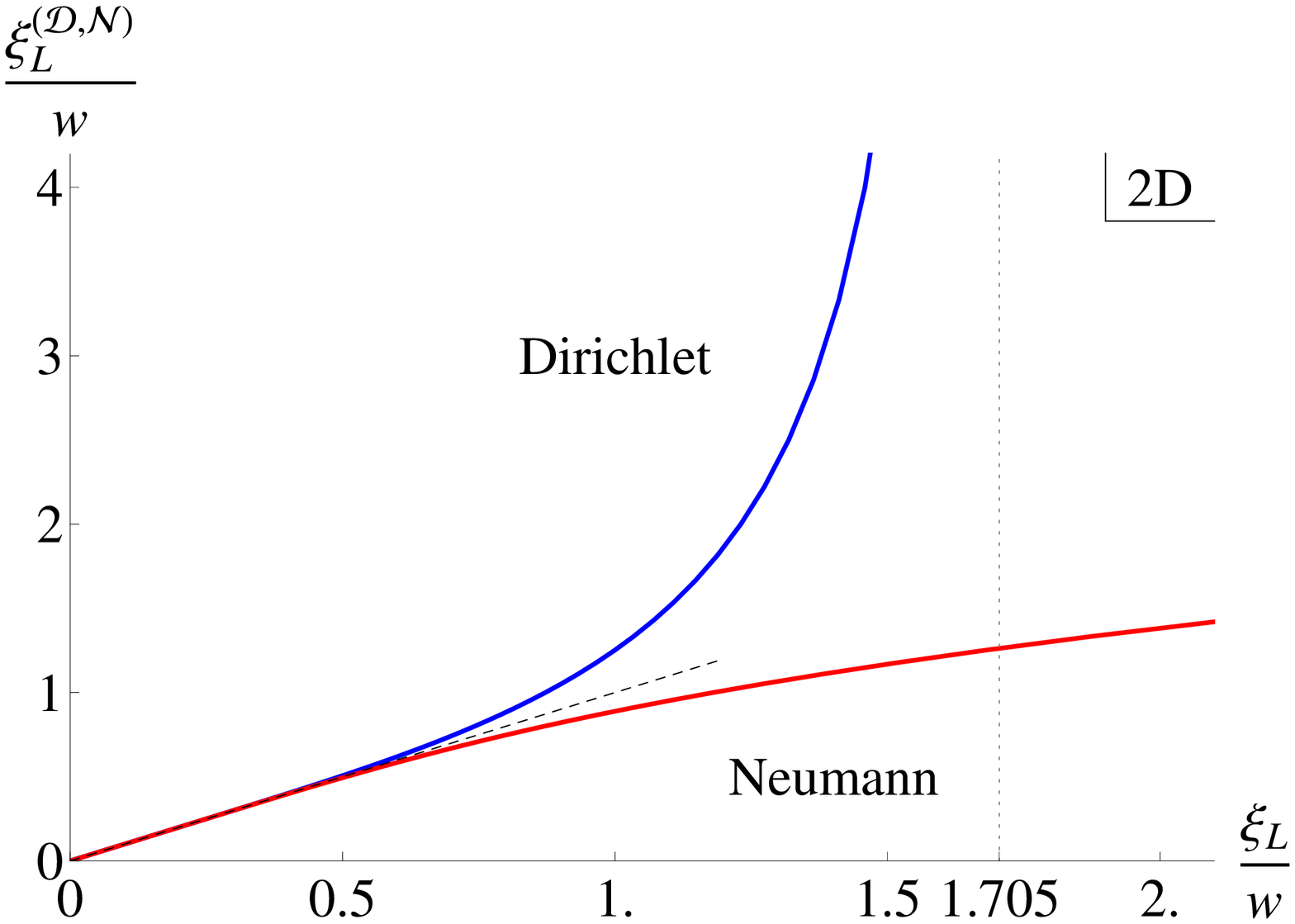}\\ 
\includegraphics[width=9 cm]{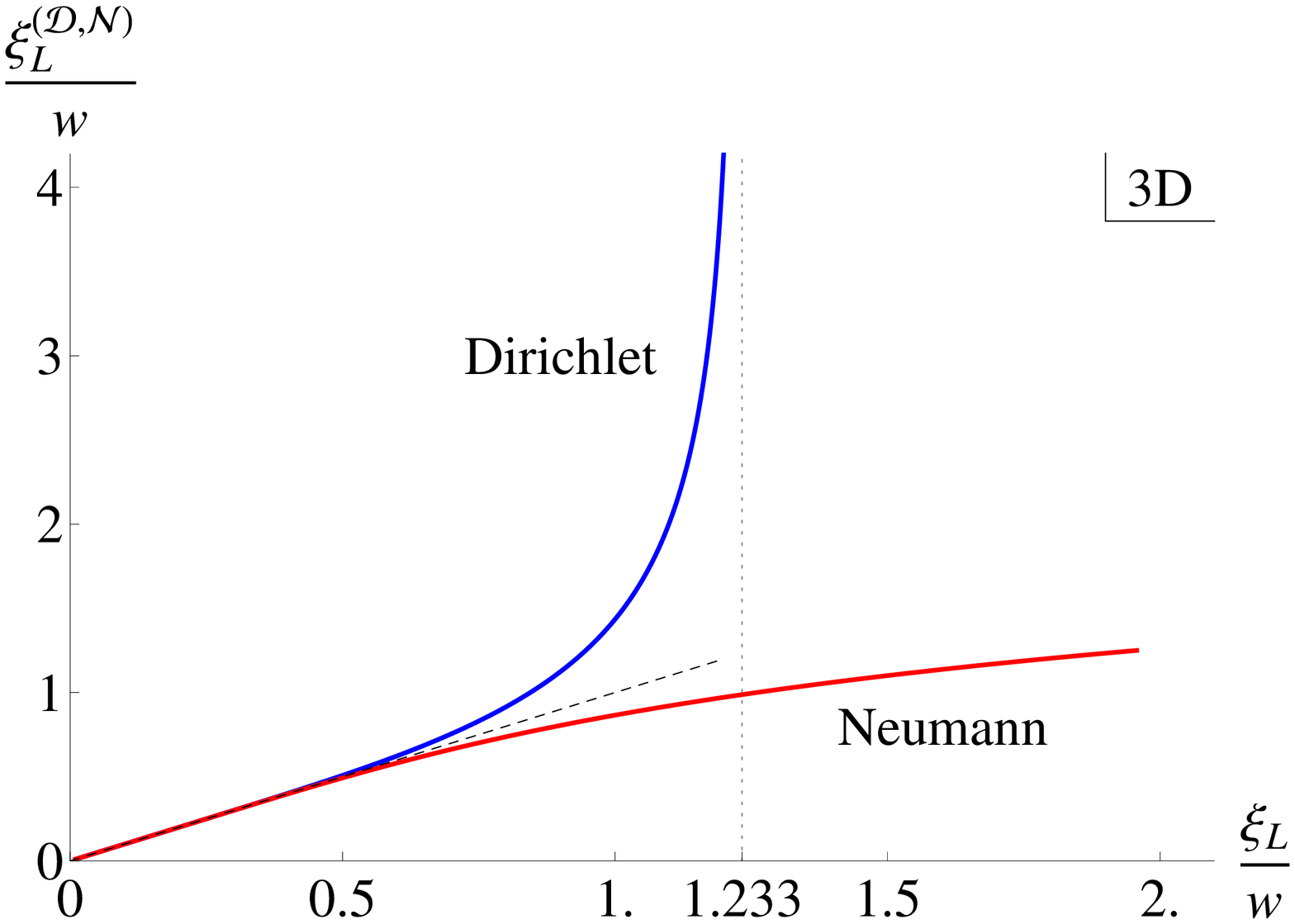}\\ 
\caption{ (Color online)
  Larkin lengths with different boundary conditions on the top
  substrate, for $d=2$ (top) and $d=3$ (bottom). For thick cells 
  ($w\gg\xi_L$), the asymptotic behavior $\xi_L^{(\mathcal{D, N})}\sim \xi_L$ 
  is as expected. For longer $\xi_L$, i.e., weaker surface pinning
  and thinner cell, $\xi_L^{\D}$ ($\xi_L^{\N}$) is always longer
  (shorter) than $\xi_L$ for the infinite-thickness cell, consistent with
  the ordering (disordering) effect of the finite cell thickness.  For
  the Dirichlet homogeneous substrate, the effective Larkin length
  diverges at $\xi_L/w\approx 1.71$ for $d=2$ and $\xi_L/w\approx
  1.23$ for $d=3$, with asymptotics quoted in the
  text.}\label{fig:xiDN}
\end{figure}

This behavior is quite consistent with physical expectations. For a
cell thicker than the bulk Larkin length, $\xi_L$ (thick cell and/or
strong disorder), there is little impact of the top substrate on the
range of the finite $xy$ order, dominated by the random lower
substrate. However, for a thin cell and/or weak disorder, such that
the bulk Larkin length extends across the cell thickness, the Dirichlet
substrate effectively enforces the $xy$-order alignment across the cell,
suppressing $\phi_0^{rms}$ below $2\pi$ and thereby driving the cell
Larkin scale, $\xi_L^\D$, to diverge.

It is important to emphasize that this divergence is {\em not} an
indication of a sharp transition. One signature of this is the fact
that the ``crossover'' scale $\xi_L^{*}$ is a function of a relatively
arbitrary constant [taken here to be $(2\pi)^2$] in the definition of
the Larkin length. Rather, the divergence of $\xi_L^\D$, Eq.~\rf{xiD}, is a
signal of a crossover from a weakly $xy$-ordered state (at strong
disorder and thick cell) for $\xi_L\ll w$ to a strongly $xy$-ordered
state (at weak disorder and thin cell) for $\xi_L\gg w$. In both
limits, the aligning Dirichlet substrate dominates over the random one,
leading to a long-range $xy$ order. We will support this assertion with
a detailed FRG calculation in Sec.~\ref{sec:FRG}.

For a cell with a Neumann boundary condition on the top substrate, the
Larkin length, illustrated in Fig.~\ref{fig:xiDN} is defined by
\begin{eqnarray}
\overline{\langle\phi_0^2(\xv)\rangle}^\N=(2\pi)^2
&=&\frac{\Delta_f}{K^2}\int_{1/\xi_L^\N}^{1/a}
\frac{d^{d-1}q}{(2\pi)^{d-1}}\frac{1}{q^2\tanh^2q w}.\nonumber\\
\end{eqnarray}
This gives
\begin{eqnarray}
\xi_L^\N\approx\left\{\begin{array}{ll}
\xi_L,& w\gg\xi_L\\
w^{\gamma_d}\left\{\begin{array}{ll}
\sqrt{2\ln(1.2\xi_L/w)},&d=3\\
(\frac{5-d}{3-d})^{\frac{1}{5-d}}(\xi_L)^{\frac{3-d}{5-d}},&d<3
\end{array}\right.&w\ll\xi_L,
\end{array}\right.\nonumber\\
\end{eqnarray}
with 
\begin{equation}
\gamma_d =\frac{2}{5-d}.
\end{equation}
Strongly contrasting to the case of the Dirichlet top substrate,
this result is again consistent with our expectations. Namely, that
for a thin cell, $w\ll\xi_L$, such Neumann cell becomes a thin 
($d-1$)-dimensional film, reducing to a ($d-1$)-dimensional $xy$ system pinned by
($d-1$)-dimensional, i.e., bulk disorder, characterized by the expected
lower critical dimension of $5$ (that is, $d_{lc}-1=4$).
Thus on length scales exceeding the Larkin lengths, $\xi_L^{(a)}$,
$\phi_0$ distortions become large and lead to a breakdown of the
Larkin model, \rf{HsLarkin} and of the predictions of correlation
functions [e.g.,
Eqs.~\rf{LarkinCorrelationInfinite} and \rf{LarkinCorrelationDirichlet}]
calculated from it.

\section{Physics beyond $\xi_L$}

On length scales longer than the crossover scale $\xi_L^{(a)}$, large
$\phi_0$ distortions are in the nonlinear regime and the effects of
the random potential $V[\phi_0(\xv),\xv]$ must be treated
nonperturbatively. As with the bulk disorder problems, this can be
done systematically using a renormalization-group
analysis \cite{DSFisherFRG,GiamarchiLedoussal} in an expansion in
$\epsilon=d_{lc}-d=3-d$ about the lower-critical dimension.

\subsection{Functional renormalization-group analysis}
\label{sec:FRG}
It is convenient to work with the translationally invariant replicated
Hamiltonian, $H_s^\re$, \rfs{Hsr}. We employ the standard
momentum-shell RG transformation \cite{Wilson} by separating the $xy$
field into long- and short-scale contributions according to
$\phi_0^\alpha(\xv)=\phi_{0<}^\alpha(\xv) +\phi_{0>}^\alpha(\xv)$ and
perturbatively in nonlinearity
$R[\phi_{0}^\alpha(\xv)-\phi_{0}^\beta(\xv)]$ integrate out the high
wave vector piece $\phi_{0>}^\alpha(\xv)$ that takes support in an
infinitesimal shell $\Lambda/b < q < \Lambda\equiv 1/a$, with
$b=e^{\delta\ell}$. We follow this with a rescaling of lengths and the
long-wavelength part of the field
\begin{subequations}
\begin{eqnarray}
\xv&=&b\, \xv',\label{xb}\\
\qv&=&b^{-1}\qv',\label{qb}\\
\phi_{0<}(b\,\xv')&=&\phi_0'(\xv'),\label{phib}\\
\phi_{0<}(\qv'/b)&=&b^{d-1}\phi_0'(\qv'),\label{phiqb}
\end{eqnarray}
\end{subequations}
so as to restore the UV cutoff back to $\Lambda$. In \rfs{phib}, we
made a convenient choice of a zero scaling dimension for the
real-space field $\phi_0(\xv)$. This is dictated by the convenience of
keeping the periodicity of the disorder variance $R(\phi)$ fixed at
$2\pi$.

\subsubsection{Infinitely thick cell: $w\rightarrow\infty$}
We first focus on an infinitely thick cell, defined by
$w\gg\xi_L$. The above rescaling leads to zeroth order RG flows of the
effective couplings \cite{Delta0Comment} that for a thick cell is
given by
\begin{eqnarray}
K(b)&=&b^{d-2}K\;,\label{Kflow0}\\
R(\phi,b)&=&b^{d-1}R(\phi),\label{Rflow0}
\label{KRflow0}
\end{eqnarray}
indicating that in $d>2$, the effective strengths of both elastic and
pinning energies grow at long scales relative to the thermal energy,
$T$. This is a reflection that in $d>2$, the physics is controlled by
the zero-temperature ground-state competition between elastic and
pinning energies, at long scales both much larger than the thermal
energy.  Equivalently, to emphasize this physics, we can rescale $T$
according to
\begin{eqnarray}
T(b)&=&b^{-(d-2)}T\nonumber\\
&\equiv& b^{-\Theta}T,
\end{eqnarray}
so as to keep the elastic energy fixed at order $1$.  With this
convenient rescaling convention, the measure of the effective pinning
strength grows according to
\begin{equation}
R(\phi,b)=b^{3-d}R(\phi),\label{RT2flow0}
\end{equation}
modified by a factor $(T(b)/T)^2=b^{2(d-2)}$ relative to that in
\rfs{Rflow0} due to the factor of $1/T^2$ in $H_s^\re/T$,
\rfs{Hsr}. Equivalently, without the rescaling of $T$, the
dimensionless combination that arises in the coarse-graining analysis
is given by $R(\phi)/K^2$, and its zeroth order flow is given by
\rfs{RT2flow0}.

In either convention, we find that for $d<3$, the influence of the
random surface pinning grows at long scales relative to the elastic
energy, consistent with the scaling and Larkin analysis that gave
$d_{lc}=3$.

The statistical symmetry\cite{Toner_DiVincenzo_sr} of the bulk
Hamiltonian, $H$ \rf{Hbulk}, under an arbitrary local rotation
$\phi(\rv)\rightarrow \phi(\rv) + \chi(\rv)$ guarantees that the flow
of $K(b)$, \rfs{Kflow0}, and equivalently, the thermal exponent
\begin{equation}
\Theta=d-2,
\end{equation}
are {\em exact}, i.e., do not experience any coarse-graining
corrections. This can equivalently be seen from the replicated
Hamiltonian \rf{Hsr}, where the pinning nonlinearity,
$R[\phi_{0}^\alpha(\xv)-\phi_{0}^\beta(\xv)]$ depends only on the
difference between different replica fields, i.e., independent of the
``center of mass'' field $\sum_{\alpha=1}^n \phi_0^\alpha$. That is,
the only nonlinearity in $H_s^\re$ exhibits a symmetry of a
replica-independent local rotation $\phi_0^\alpha(\rv)\rightarrow
\phi_0^\alpha(\rv) + \chi(\rv)$ and under coarse graining can
therefore only generate terms that also exhibit this symmetry. Thus,
it cannot generate a correction to the elastic term that clearly lacks
this symmetry, implying that $K$ is {\em not} renormalized by the
pinning disorder.

An important consequence of the periodic nonlinearity $R(\phi)$ and
the effective zero-temperature physics, first emphasized by
Fisher \cite{DSFisherFRG}, is that all monomials or (equivalently)
harmonics in the expansion of $R(\phi)$ are equally relevant in
$d<d_{lc}$. Thus, a {\em functional} RG analysis that follows the
coarse-graining flow of the whole function $R(\phi)$ is necessary. The
method is by now quite
standard \cite{DSFisherFRG,BalentsFisher,GiamarchiLedoussal,LedoussalWiese}
and is straightforwardly adapted to the surface-pinning problem,
characterized by $H_s^\re$, \rfs{Hsr}.

We limit the FRG analysis to one-loop order, performing the
momentum-shell integration over the high-wave-vector components
$\phi_{0>}^\alpha$ perturbatively in the nonlinearity
$R[\phi_{0}^\alpha(\xv)-\phi_{0}^\beta(\xv)]$. We find that the change in the
Hamiltonian due to this coarse graining is given by
\begin{equation}
\delta H_s^\re[\phi_{0<}^\alpha]=\langle
H_{p}[\phi_{0<}^\alpha+\phi_{0>}^\alpha]\rangle_>
-\frac{1}{2T}\langle H_{p}^2[\phi_{0<}^\alpha+\phi_{0>}^\alpha]\rangle_>^c\ldots\;,
\label{deltaHsr}
\end{equation}
where $H_{p}[\phi^{\alpha}_0]$ is the nonlinear pinning part of the
Hamiltonian $H_s^\re$, Eq.~(\ref{Hsr}),
\begin{eqnarray}
H_p&=&-\frac{1}{2T}\sum_{\alpha,\beta}^n\int d^{d-1}x R[\phi_0^\alpha(\xv)-\phi_0^\beta(\xv)],
\label{Hp}
\end{eqnarray}
and the averages over short scale fields, $\phi_{0>}^\alpha$, above are
performed with the quadratic (elastic $K$) part of $H_s^\re$. The
superscript $c$ denotes a cumulant average, $\langle H_p^2\rangle^c
=\langle H_p^2\rangle - \langle H_p\rangle^2$.

To lowest order in $R(\phi)$ (dropping a constant term) we find that
$\delta H_s^\re[\phi_{0<}^\alpha]$ is given by
\begin{eqnarray}
\delta H_{s1}^\re&=&-\frac{1}{2T}\sum_{\alpha,\beta}\int_{\xv}
\langle R(\phi_0^\alpha-\phi_0^\beta)\rangle_> \nonumber\\
&\approx&-\frac{1}{4T}\sum_{\alpha,\beta}\int_{\xv}
R''(\phi_{0<}^\alpha-\phi_{0<}^\beta)
\langle(\phi_{0>}^\alpha-\phi_{0>}^\beta)^2\rangle_> \nonumber\\ 
&\approx&-\frac{1}{2T}\int_\qv^>\frac{T}{\Gamma_q^{(\infty)}}\sum_{\alpha,\beta}\int_{\xv}
R''(\phi_{0<}^\alpha-\phi_{0<}^\beta),
\label{deltaH1}
\end{eqnarray}
which when compared to the definition of $H_p$ gives
\begin{equation}
\delta R^{(1)}(\phi)\approx \delta\ell C_{d-1}\Lambda^{d-2}\frac{T}{K}
R''(\phi).
\label{deltaR1}
\end{equation}
In above, the prime indicates a partial derivative with respect to
$\phi$, and $C_d=S_d/(2\pi)^d=1/[\Gamma(d/2)2^{d-1}\pi^{d/2}]$, with
$S_d$ the surface area of a $d$-dimensional unit sphere.

The contribution to second order in $R(\phi)$ is given by
\begin{eqnarray}
\delta H_{s2}^\re&\approx&
-{1\over8 T^3}{1\over2}\sum_{\alpha_1,\beta_1,\alpha_2,\beta_2}\int_{\xv_1,\xv_2}
R''[\phi_{0<}^{\alpha_1}(\xv_1)-\phi_{0<}^{\beta_1}(\xv_1)]\nonumber\\
&&\times 
R''[\phi_{0<}^{\alpha_2}(\xv_2)-\phi_{0<}^{\beta_2}(\xv_2)]
I^{\alpha_2\beta_2}_{\alpha_1\beta_1}(\xv_1,\xv_2)\;,\label{deltaH2}
\end{eqnarray}
where
\begin{eqnarray}
\hspace{-0.2cm}
I^{\alpha_2\beta_2}_{\alpha_1\beta_1}
&=&\frac{1}{2}
\Big\langle\big(\phi_{0>}^{\alpha_1}(\xv_1)-
\phi_{0>}^{\beta_1}(\xv_1)\big)^2
\big(\phi_{0>}^{\alpha_2}(\xv_2)-
\phi_{0>}^{\beta_2}(\xv_2)\big)^2\Big\rangle_>^c.\nonumber\\
&&
\label{Iab}
\end{eqnarray}
Keeping only the most relevant (two-replica) terms and
comparing to $H_p$, we obtain
\begin{equation}
\delta R^{(2)}(\phi)\approx \delta\ell g_2
\left({1\over2}R''(\phi)R''(\phi)-R''(\phi)R''(0)\right)
\label{deltaR2}\;,\ \ \ \ 
\end{equation}
where the constant $g_2$ is defined by
\begin{eqnarray}
\delta\ell\,g_2&=&\int_{\qv}^>\frac{1}{[\Gamma_q^{(\infty)}]^2}\nonumber\\
&\approx&\delta\ell\frac{C_{d-1}\Lambda^{d-3}}{K^2}.
\label{G2}
\end{eqnarray}

Combining the first and second-order contributions to $R(\phi)$,
Eqs.~\rf{deltaR1} and \rf{deltaR2}, with the length and field
rescalings, Eqs.~\rf{Kflow0} and \rf{Rflow0}, we obtain the FRG flow
equation
\begin{eqnarray}
\partial_\ell\hR(\phi)&=&\epsilon\hR(\phi) 
+  C_{d-1}\Lambda^{d-2}\frac{T}{K} \hR''(\phi)\nonumber\\
&&+ \frac{1}{2} \hR''(\phi) \hR''(\phi)-\hR''(\phi)\hR''(0),
\label{FRGflowR}
\end{eqnarray}
where 
\begin{equation}
\hR(\phi)\equiv \frac{C_{d-1}\Lambda^{d-3}}{K^2} R(\phi)
\label{hR}
\end{equation}
is the dimensionless measure of surface disorder.

(i) $2<d< 3$. 
Because as noted above $T/K$ flows to zero as $b^{2-d}=b^{-\Theta}$,
for $d>2$ (independent of the rescaling convention), i.e., the system
is described by the zero-temperature fixed point, the second term on
the right-hand side in \rfs{FRGflowR} can be neglected near $d=3$, and 
FRG equation reduces
to \cite{DSFisherFRG,GiamarchiLedoussal,LedoussalWiese,FeldmanVinokurPRL}
\begin{equation}
\partial_\ell\hR(\phi)=\epsilon\hR(\phi) 
+ \frac{1}{2} \hR''(\phi) \hR''(\phi)-\hR''(\phi)\hR''(0).
\label{FRGflowRT0}
\end{equation}
We note that aside from constant prefactors in the definition of
$\hR(\phi,\ell)$ and the reduced lower-critical dimension giving
$\epsilon=3-d$, the flow equation for the dimensionless disorder
measure $\hR(\phi,\ell)$ is identical to that of the bulk pinning
problem \cite{DSFisherFRG,GiamarchiLedoussal}. Consequently, the
long-scale properties of the low-temperature phase are described by
the same fixed point function,
\begin{equation}
\hR''_*(\phi)=-\epsilon
\left[\frac{1}{6}(\phi-\pi)^2-\frac{\pi^2}{18}\right], 
\label{fixedpointRpp}
\end{equation}
periodically extended (period $2\pi$), with the minimum at the cusp
given by $\hR_*''(0)=-\frac{\epsilon \pi^2}{9}$.

(ii) $d=3$. 
Temperature remains irrelevant at $d=3$ (as for any $d>2$), allowing
us to continue to work at $T=0$. At this lower-critical dimension,
$\epsilon=3-d=0$ and the flow equation \rf{FRGflowRT0} reduces to
\begin{equation}
\partial_\ell\hR(\phi)=
\frac{1}{2} \hR''(\phi) \hR''(\phi)-\hR''(\phi)\hR''(0).
\label{FRGflowRT03d}
\end{equation}
Since it is of the form $\partial_{\ell} f\sim -f^2$, we expect the
solution $\hR(\phi,\ell)$ to decay according to $1/\ell$ and take its
form to be
\begin{equation}
\hR(\phi,\ell)=\frac{\hR_0(\phi,\ell)}{\ell+\ell_0}.
\end{equation}
with the function $\hR_0$ satisfying 
\begin{eqnarray}
\hspace{-0.3cm}(\ell+\ell_0)\partial_\ell\hR_0(\phi)=\hR_0(\phi) 
+ \frac{1}{2} \hR_0''(\phi) \hR_0''(\phi)-\hR_0''(\phi)\hR_0''(0).\nonumber\\
\label{FRGflowR0T0}
\end{eqnarray}

In terms of a new flow variable $t=\ln{(\ell+\ell_0)}$, the equation
for $\hR_0(\phi,t)$ is identical to that for $\hR(\phi,\ell)$,
\rfs{FRGflowRT0}, with $\epsilon=1$. Thus, on the scale beyond the
Larkin length (when $\hR_0$ has crossed away from Gaussian fixed point
toward the nontrivial fixed point), we find that for large $\ell$,
\begin{equation}
\hR''(\phi,\ell)=\frac{1}{\ell}
\left[-\frac{1}{6}(\phi-\pi)^2+\frac{\pi^2}{18}\right], 
\label{R3d}
\end{equation}
with $\hR''(0,\ell)=-\frac{\pi^2}{9\ell}$ also decaying with
$\ell$. This is the same as the result obtained by Chitra 
\textit{et al.} \cite{ChitraGiamarchiDoussal} at the lower-critical
dimension of $d=4$ for the $xy$ model with bulk random-field
disorder. For this special case of an infinitely thick
($w\rightarrow\infty$) 3D cell, the above result also reproduces the
earlier finding in Ref.~[\onlinecite{FeldmanVinokurPRL}].

(iii) $d=2$. 
In two (bulk) dimensions ($d=2$), $\Theta=0$ leads to $\eta=T/(\pi K)$
that is fixed under the RG flow  and the long-scale behavior is no
longer controlled by a zero-temperature fixed point. Instead, the
finite temperature selects eigenfunctions of the RG flow,
\rfs{FRGflowR}, that near the Gaussian fixed point are harmonics
$\cos(n\phi)$ ($n$ integers), with eigenvalues
\begin{equation}
\lambda_n = 1-n^2 \frac{T}{\pi K}.
\end{equation}
Focusing on the eigenfunction $\cos\phi$ with the largest
eigenvalue, $\lambda_1$, the FRG flow reduces to a standard RG flow
equation
\begin{equation}
\partial_\ell g = \Big(1-\frac{T}{\pi K}\Big)g - g^2
\end{equation}
for a single amplitude of this lowest harmonic of $R(\phi)$.

\begin{figure}

\includegraphics[height=5 cm]{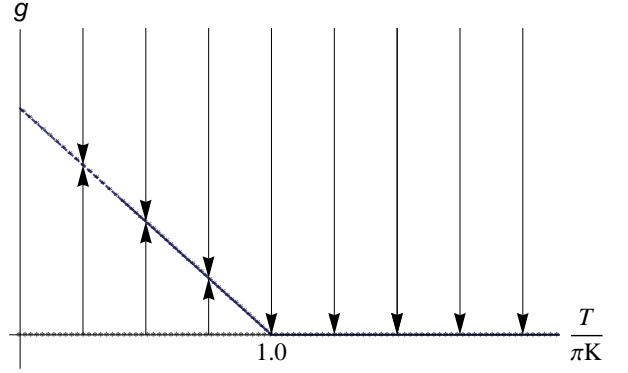}\\   
\caption{(Color online)
  Flow of $g$ at $d=2$: for $T<\pi K$ it flows to the
  $g_*=1-\frac{T}{\pi K}$ fixed line, and for $T>\pi K$, it flows to
  $g_*=0$ fixed line, corresponding to a glass transition at $T_g=\pi
  K$.}\label{flowmap2}
\end{figure}

As illustrated in Fig.~\ref{flowmap2}, $g$ decays to the $g=0$ fixed
line for $T>\pi K\equiv T_g$ and grows to a
\begin{eqnarray}
g_*=1-\frac{T}{\pi K}
\end{eqnarray}
fixed line for $T<\pi K\equiv T_g$. Thus, we find that the 2D cell
exhibits a Cardy-Ostlund-like \cite{CardyOstlund} phase transition at
\begin{eqnarray}
T_g=\pi K
\end{eqnarray}
between a high-temperature phase, where at long scales surface pinning
is smoothed away by thermal fluctuations and a low-temperature glassy
pinned phase, controlled by the nontrivial $g_*(T)$ fixed line.  This
surface-pinned state (and the associated transition) is quite similar
to the super-rough phase of a crystal surface grown on a random
substrate \cite{Toner_DiVincenzo_sr}, to the 1+1 vortex glass phase of
flux-line vortices (confined to a plane) in type-II
superconductors \cite{disorderSC,DSFisherBG,VortexGlass}, and to the 3D smectic
liquid crystals pinned by a random porous environment of e.g.,
aerogel \cite{RTaerogelPRB}.

One key distinction here is the irrelevance in 2D of the tilt pinning
potential \cite{Delta0Comment}
\begin{eqnarray}
H_{\text{tilt}}= \frac{\Delta_h}{T}
\sum_{\alpha,\beta}^n\int d^{d-1}x (\nabla\phi_{0}^\alpha - \nabla\phi_{0}^\beta)^2,
\end{eqnarray}
in contrast to its relevance in these other bulk pinning
systems \cite{disorderSC,Toner_DiVincenzo_sr,DSFisherBG,VortexGlass,RTaerogelPRB},
where it leads to a super-rough phase characterized by $\ln^2x$
roughness. This difference leads to a distinct behavior of correlation
functions in the pinned phase for the surface-pinning problem. Because
this 2D case is somewhat academic, with our main focus on the
experimentally relevant 3D cell, we do not explore it any further
here.

\subsubsection{Finite-thickness cell}

We now focus on the behavior of a finite thickness cell with Dirichlet or
Neumann boundary conditions on the top homogeneous substrate. Much of
the RG analysis of the previous section extends to this case after
substitutions of $\Gamma_q^\D$ and $\Gamma_q^\N$,
Eqs.~\rf{deltaH1} and \rf{G2}, respectively, for
$\Gamma^{(\infty)}$. With these changes the rescalings
Eqs.~\rf{Kflow0} and \rf{Rflow0} are supplemented with the exact flow
equation for the cell thickness
\begin{eqnarray}
w(b) = b^{-1} w.
\label{wflow0}
\end{eqnarray}

The zero-temperature flow equation takes the same form as for an
infinite cell, with $a\in\D,\N$,
\begin{eqnarray}
\partial_\ell\hR_a(\phi)=\epsilon^{(a)}(\ell)\hR_a(\phi) 
+ \frac{1}{2} \hR_a''(\phi) \hR_a''(\phi)-\hR_a''(\phi)\hR_a''(0),\nonumber\\
\label{FRGflowRT0_DN}
\end{eqnarray}
except that the constant $\epsilon=3-d$ is replaced by
$\ell$-dependent functions $\epsilon^\D(\ell)$, $\epsilon^\N(\ell)$,
given by
\begin{subequations}
\begin{eqnarray}
\epsilon^\D(\ell)&=&\epsilon 
- \frac{4 \Lambda w(\ell)}{\sinh[2\Lambda
  w(\ell)]},\;\;\mbox{Dirichlet},
\label{epsD}\\
\epsilon^\N(\ell)&=&\epsilon 
+ \frac{4\Lambda w(\ell)}{\sinh[2\Lambda w(\ell)]},\;\;\mbox{Neumann},
\label{epsN}
\end{eqnarray}
\end{subequations}
for the two boundary conditions on the top substrate. The
dimensionless disorder variance functions in Eqs.~\rf{epsD} and \rf{epsN} have
been, respectively, defined by
\begin{subequations}
\begin{eqnarray}
\hR_{\cal{D}}(\phi)&\equiv&\frac{C_{d-1}\Lambda^{d-3}}
{K^2\coth^2(\Lambda w)} R(\phi),\label{RD}\\
\hR_{\cal{N}}(\phi)&\equiv&\frac{C_{d-1}\Lambda^{d-3}}
{K^2\tanh^2(\Lambda w)} R(\phi)\label{RN}.
\end{eqnarray}
\end{subequations}

The limiting cases of these flow equations can be easily
understood. For a thick cell $\omega\rightarrow\infty$ equations for
both (top substrate) boundary conditions reduce to an infinite cell
analyzed in the previous section. In the opposite extreme of a
microscopically thin cell, such that $\Lambda w\ll 1$, the
$\epsilon^{(a)}(\ell)$ functions reduce to
\begin{subequations}
\begin{eqnarray}
\epsilon^\D(\ell)&\approx&1-d,\;\;\mbox{for $w\ll a$},\label{epsDwa}\\
\epsilon^\N(\ell)&\approx&5-d,\;\;\mbox{for $w\ll a$},\label{epsNwa}
\end{eqnarray}
\end{subequations}
corresponding to flow equations for a ($d-1$)-dimensional {\em bulk}
random-field $xy$ model, which in the case of the Dirichlet boundary
condition is in a uniform external field. In this Dirichlet case, the
eigenvalue is negative for any physical dimension, showing that random
pinning is always dominated by an ordering field of the rubbed top
substrate. In contrast, in the Neumann case, the flow as expected is
identical to that of a ($d-1$) bulk system with the eigenvalue of
$4-(d-1)=5-d$.

For a more realistic situation of a finite cell thickness $w$, there is
a crossover from an infinite $d$-dimensional cell limit at small
$\ell$ such that $\Lambda w(\ell)\gg 1$ to an effective ($d-1$)-dimensional 
system for $\Lambda w(\ell)\ll 1$. The corresponding
crossover scale is given by $b_w^*=w/a$. An independent crossover
scale encoded in flow equations, \rfs{FRGflowRT0_DN} is set by a scale
$b_L^*$ at which the nonlinear terms become comparable to the linear
ones, where the flow leaves the vicinity of the Gaussian fixed point
and (in the bulk system, i.e., for $w=\infty$) would approach the
nontrivial fixed point \rf{fixedpointRpp}. From the flow equations,
\rfs{FRGflowRT0_DN}, one can see that this latter scale is simply set
by the Larkin length, with $b_L^*=\xi_L/a$.

As we discuss below, the detailed nature of distortions strongly
depends on the relative size of these two crossover scales and on the
type of boundary condition on the homogeneous substrate. We naturally
designate the two cases, $w\ll \xi_L$ and $w\gg \xi_L$, as thin and
thick cells, respectively.

\subsection{Correlation function}
\label{sec:matching}

We now turn to a calculation of correlation functions. As discussed
earlier, on short scales (smaller than the Larkin length), these can be
simply computed using the random-torque model of
Sec.~\ref{sec:Larkin}. However, as we have seen in Sec.~\ref{sec:FRG},
for $d\leq 3$ the effective pinning becomes strong (compared to the
elastic energy) on scales longer than the Larkin length, leading to a
breakdown of the perturbative expansion and of the random-torque
model.  Nevertheless, we can utilize the above FRG, which effectively
allows us to treat pinning nonperturbatively to overcome this
difficulty. To see this, we note that the power of the renormalization
group is that it establishes a connection between a correlation
function at a small wave vector (which is impossible to calculate in
perturbation theory due to the aforementioned infra-red divergences)
to the same correlation function at large wave vectors (short scales),
which can be easily calculated in a controlled perturbation
theory \cite{NelsonRudnick,GiamarchiLedoussal,Toner_DiVincenzo_sr,RTaerogelPRB}.

This relation for the Fourier transform of the surface ($z=0$)
correlation function $C(\qv)$ is given by
\begin{eqnarray}
C[\qv,K,w,\hR_a]
&=&e^{(d-1)\ell}C[\qv e^{\ell},K(\ell),w(\ell),\hR_a(\ell)]\;,\nonumber\\
\label{relationCa}
\end{eqnarray}
where the prefactor on the right-hand side comes from the dimensional
rescalings of Sec.~\ref{sec:FRG}, remembering the momentum-conserving
$\delta$ function in the definition of $C(\qv)$, and for simplicity we
chose to keep $T$ fixed under rescaling. We then choose the rescaling
variable $\ell_*$ such that
\begin{equation}
q e^{\ell_*}=\Lambda\;,\label{matching_q}
\end{equation}
which allows us to reexpress $\ell_*$ on the right hand side of \rfs{relationCa}
in terms of the wave vector $q$
\begin{eqnarray}
C[q,K,w,\hR_a]&=&\left(\frac{\Lambda}{q}\right)^{d-1}
C[\Lambda,K(\ell_*),w(\ell_*),\hR_a(\ell_*)]\;.\nonumber\\
\label{relationCb}
\end{eqnarray}
Because the correlation function on the right-hand side is evaluated
at the large wave vector, it is easily computed perturbatively in a
weak pinning potential $\hR_a(\phi,\ell_*)$ if the latter is indeed
small at long scale $e^{\ell_*} a$, i.e., the pinning is weak. To
lowest order, the computation can be done with the replicated
random-torque (Larkin) surface model
\begin{eqnarray}
H_{s,L}^{(r)}&=&\oh\int_\qv\bigg[\sum_{\alpha}^n
\Gamma_q^{(a)}(\ell_*)|\phi_0^\alpha(\qv)|^2\\
&&-\frac{1}{2T(\ell_*)}\sum_{\alpha,\beta}^n R''_a(0,\ell_*)\phi_0^\alpha(\qv)\phi_0^\beta(-\qv)\bigg],\nonumber
\label{HsrL}
\end{eqnarray}
which gives
\begin{eqnarray}
C^\ax[\Lambda,K(\ell_*),w(\ell_*),\hR_a(\ell_*)]&\approx&
\frac{T(\ell_*)}{\Gamma_\Lambda^{(a)}(\ell_*)}
-\frac{R''_a(0,\ell_*)}{\left[\Gamma_\Lambda^{(a)}(\ell_*)\right]^2}\nonumber\\
&\approx&
-\frac{R''_a(0,\ell_*)}{\left[\Gamma_\Lambda^{(a)}(\ell_*)\right]^2},\nonumber\\
\label{C_Lambda}
\end{eqnarray}
where in the last line we neglected the subdominant thermal part. 
To evaluate the resulting correlation function 
\begin{eqnarray}
C^\ax[\qv,K,w,\hR_a]&\approx&-\left(\frac{\Lambda}{q}\right)^{d-1}
\frac{R''_a(0,\ell_*)}{\left[\Gamma_\Lambda^{(a)}(\ell_*)\right]^2}
\label{Cq}
\end{eqnarray}
explicitly requires an analysis of the flow for specific boundary
conditions. 

\subsubsection{Infinitely thick cell: $w\rightarrow\infty$}
We first focus on an infinitely thick cell, for which the above
correlation function reduces to
\begin{eqnarray}
C[\qv,K,\infty,\hR]&\approx&-\frac{1}{q^{d-1}}
\frac{\Lambda^{d-3}R''(0,\ell_*)}{K^2(\ell_*)}\nonumber\\
&\approx&-\frac{1}{q^{d-1}}
\frac{\hR''(0,\ell_*)}{C_{d-1}},\nonumber\\
\label{Cqii}
\end{eqnarray}
where in the second line we used \rfs{hR} to express the result in terms
of the dimensionless disorder variance. 

(i) $2 < d < 3$. 
As we learned in the previous section, for this range of dimensions,
at large $\ell_*$ [which by \rfs{matching_q} corresponds to small $q$],
the dimensionless pinning variance flows to a fixed point
\rf{fixedpointRpp}, giving
\begin{eqnarray}
C[\qv,K,\infty,\hR]&\approx&\frac{1}{q^{d-1}}\frac{(3-d)\pi^2}{9C_{d-1}},
\label{Cqiii}
\end{eqnarray}
as presented in Sec.~\ref{sec:intro}.

We can now use this result to compute real-space correlations on
in-plane scales $x \gg \xi_L$, characterized for $z=z'$ by
\begin{eqnarray}
C(\xv,z,z)&=&\overline{\langle(\phi(\xv,z)-\phi(0,z))^2\rangle}
\nonumber\\ 
&=&2\int\frac{d^{d-1}q}{(2\pi)^{d-1}}
\left(1-\cos\qv\cdot\xv\right)e^{-2 q z}C(q) \nonumber\\
&\approx&C_L(\xv,z,z)+C_*(\xv,z,z).
\label{ClongScale}
\end{eqnarray}
In above, $C_L(\xv,z,z)$ is a (nearly) $x$-independent contribution to
the correlation function from short scales, $\xi_L^{-1}< q < a^{-1}$,
where Larkin approximation [random-torque model, \rfs{HsLarkin}, from
Sec.~\ref{sec:Larkin}] is valid and is given by (see Appendix \ref{app:LarkinCorr})
\begin{widetext}
\begin{eqnarray}
C_L(\xv,z,z)&\approx&
\frac{2\Delta_f}{K^2}\int_\qv
\frac{(1-\cos\qv\cdot\xv)e^{-2q z}}{q^2}\nonumber\\
&\approx&
(3-d)8\pi^2\left(\frac{2z}{\xi_L}\right)^{3-d}\Gamma(d-3,2z/\xi_L,2z/a),
\ \ \ \mbox{for $x\gg\xi_L$}\nonumber\\
&\approx&8\pi^2\left\{\begin{array}{ll}
1-\frac{2z}{\xi_L}(\ln\frac{\xi_L}{2z}+1-\gamma),&a\ll 2z \ll \xi_L\\ 
\frac{\xi_L}{2z}e^{-2z/\xi_L},&2z\gg \xi_L
\end{array}\right.\ \ \ \mbox{for $d =2$},
\label{C_Llongscale}
\end{eqnarray}
\end{widetext}
with $\Gamma(p,z_1,z_2)=\int_{z_1}^{z_2} t^{p-1} e^{-t} dt$ the
generalized incomplete gamma function and $\gamma\approx 0.58$ is 
the Euler's constant. In contrast to its small $x$ ($x\ll\xi_L$)
behavior, \rfs{LarkinCorrelationInfinite}, $C_L(\xv,z,z)$ is
$x$-independent for $x\gg\xi_L$ and is plotted in
Fig.~\ref{fig:C_L2d}.

\begin{figure}
\includegraphics[height=5 cm]{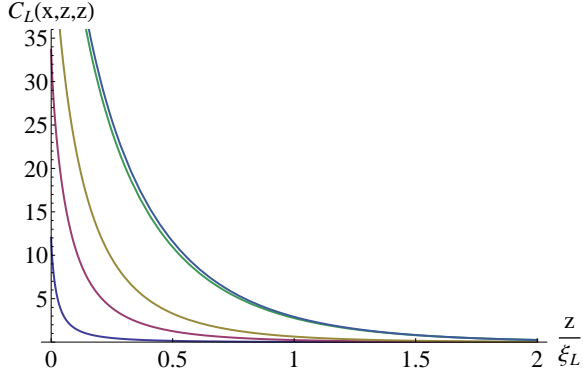}\\    
\caption{(Color online)
  For $d=2$, the contribution of short-scales $1/\xi_L< q <
  a^{-1}$ to the correlation function $C(x,z,z)$ with different $x$
  values (from bottom to top): $0.1$, $0.3$, $0.5$, $10$, $100 \xi_L$.  At
  small $x$ (e.g., $x=0.1\xi_L$, blue curve at the bottom) $C_L(x,z,z)$ is very small but
  $x$-dependent, while at large $x$ values (shown for $x=10,100\xi_L$,
  the nearly overlapping higher curves), the contribution is nearly
  $x$-independent: at $z=0$, it is a constant around $8\pi^2$ and
  decays rapidly to zero.}\label{fig:C_L2d}
\end{figure}

\begin{figure}
\includegraphics[height=5 cm]{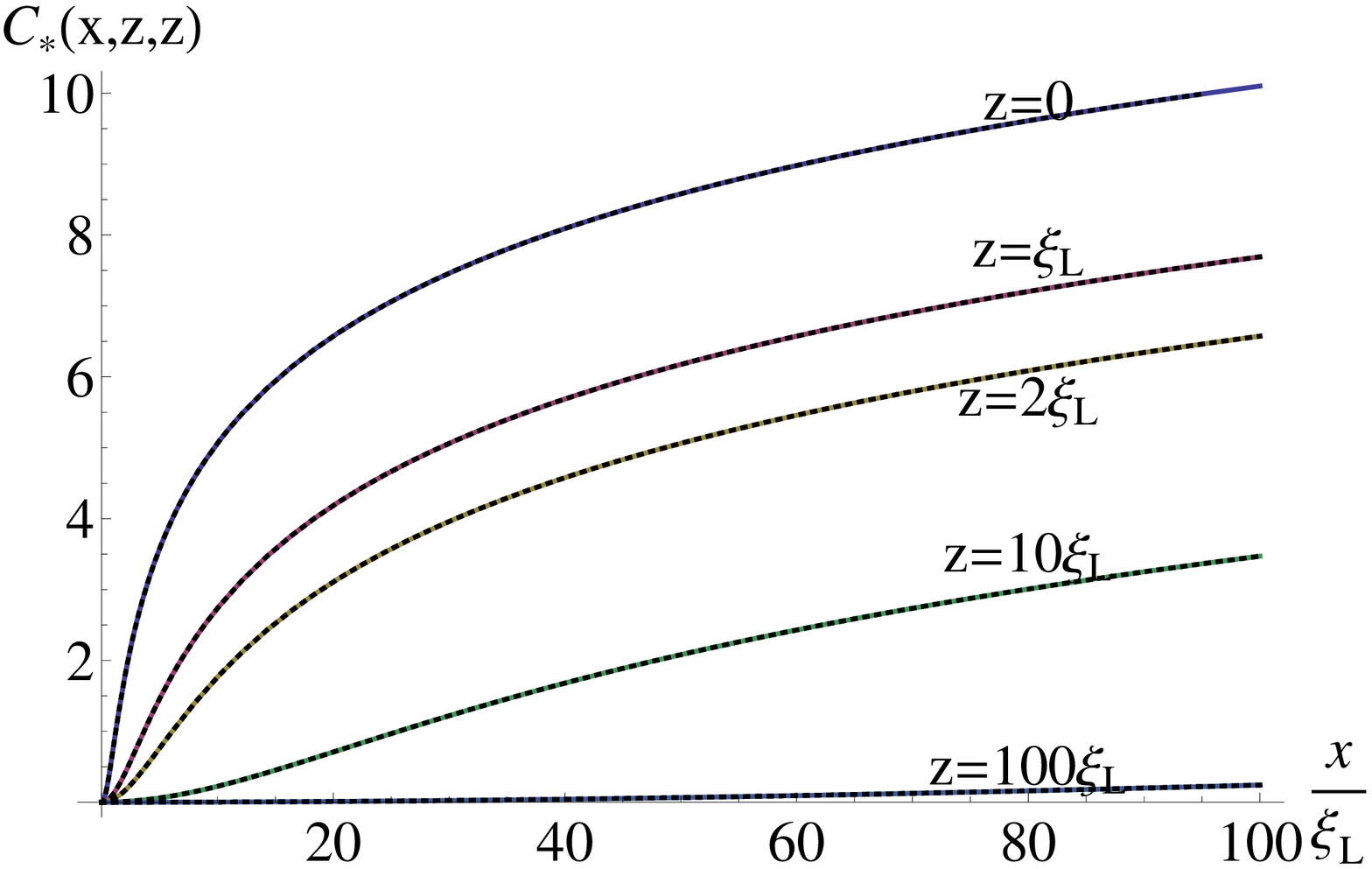}\\
\includegraphics[height=5 cm]{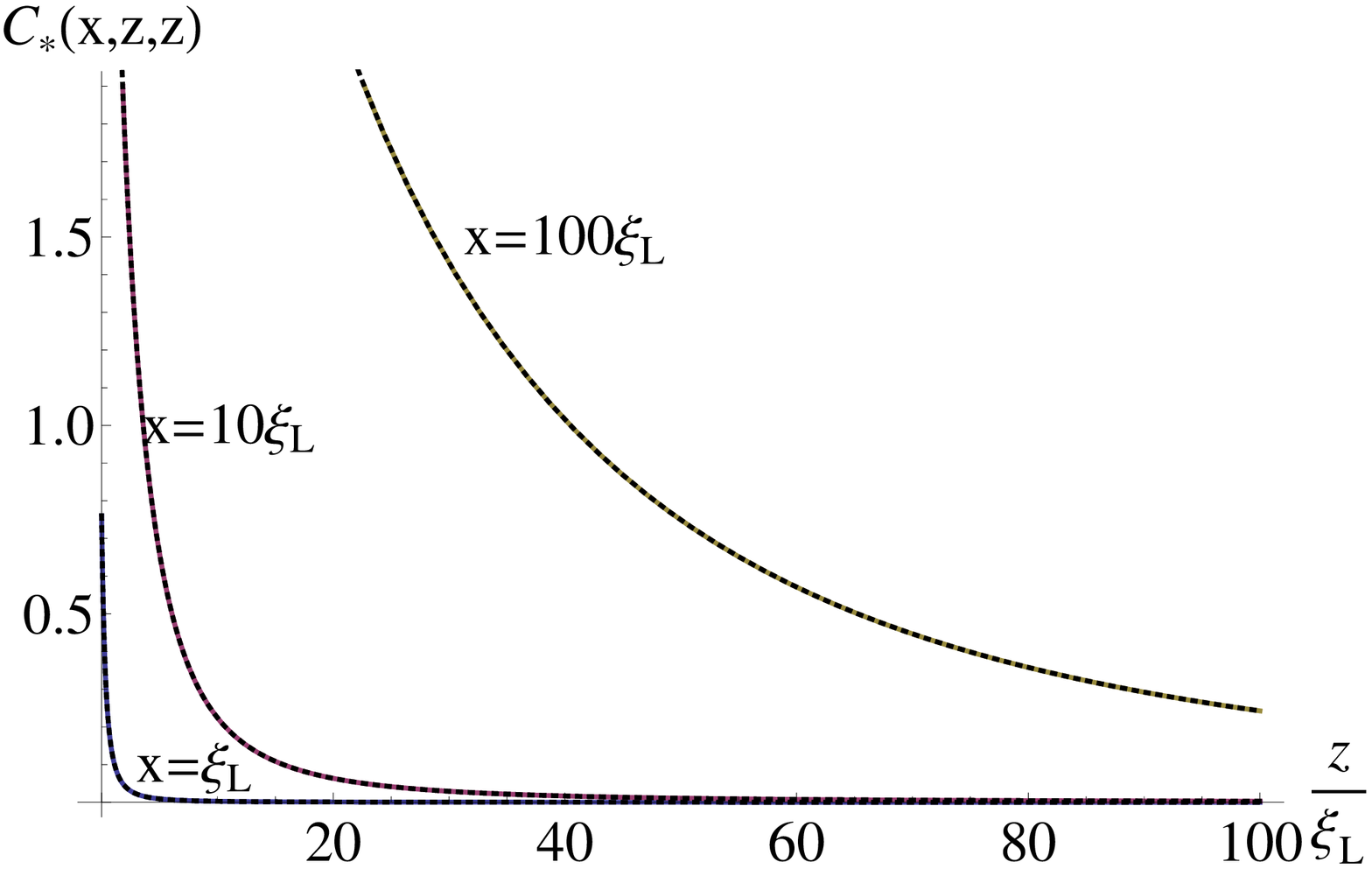}\\    
\caption{(Color online)
  Long-scale contribution, $C_{*}(x,z,z)$, to the 2D
  correlation function as a function of $x,z$. In the top (bottom)
  figure, it is plotted as a function of $x$ ($z$) for a series of $z$
  ($x$) values: $z=0,1,2,10,100 \xi_L$ ($x=1,10,100\xi_L$). The
  approximation
  $\frac{\pi^2}{9}\ln\left[1+\frac{x^2}{(2z+\xi_L)^2}\right]$ 
  (dotted black) provides an excellent interpolation to the overall 
  $x,z$ dependence.}\label{fig:C_s2d}
\end{figure}

\begin{figure}
\includegraphics[height=5 cm]{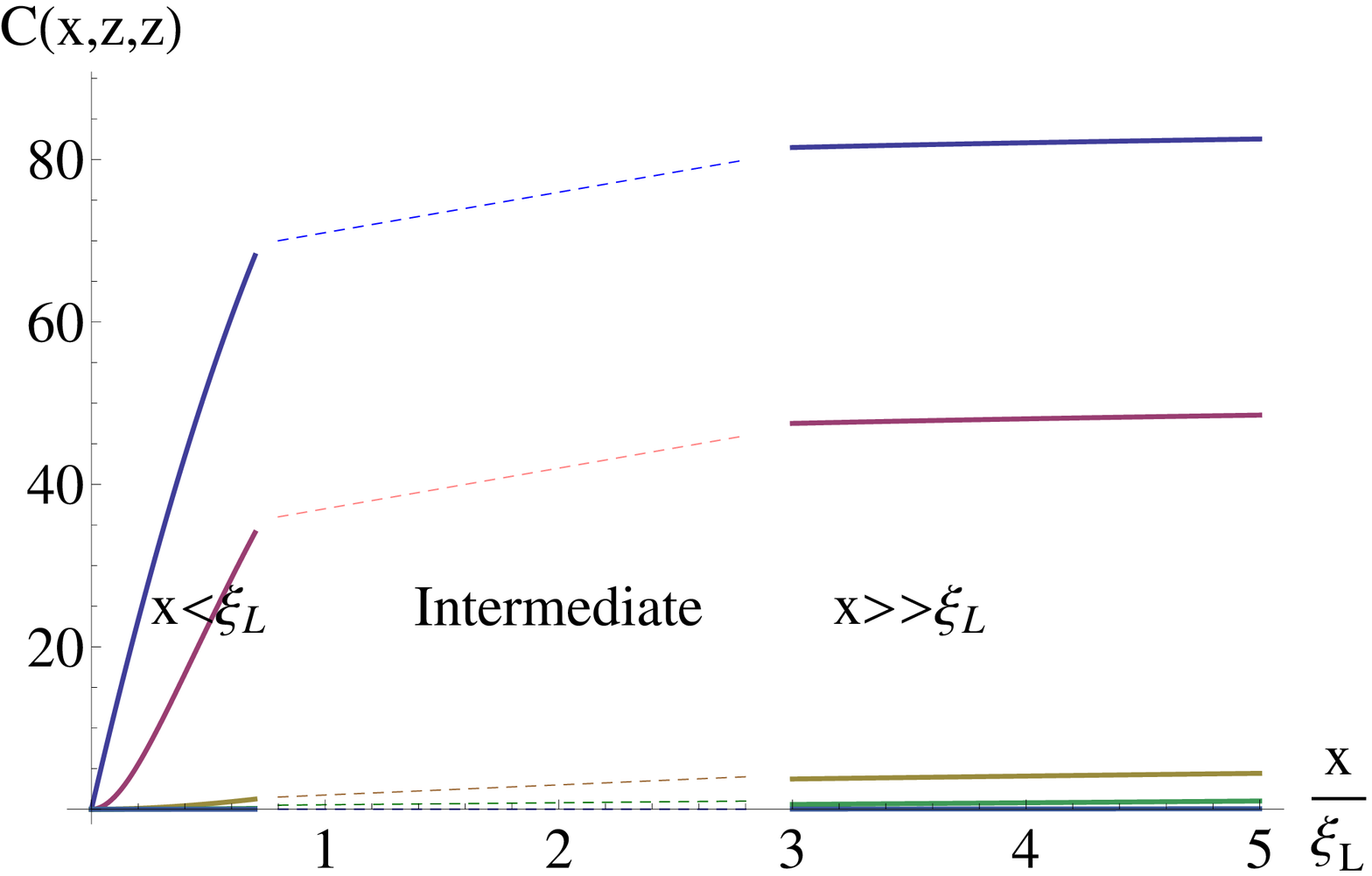}\\
\includegraphics[height=5 cm]{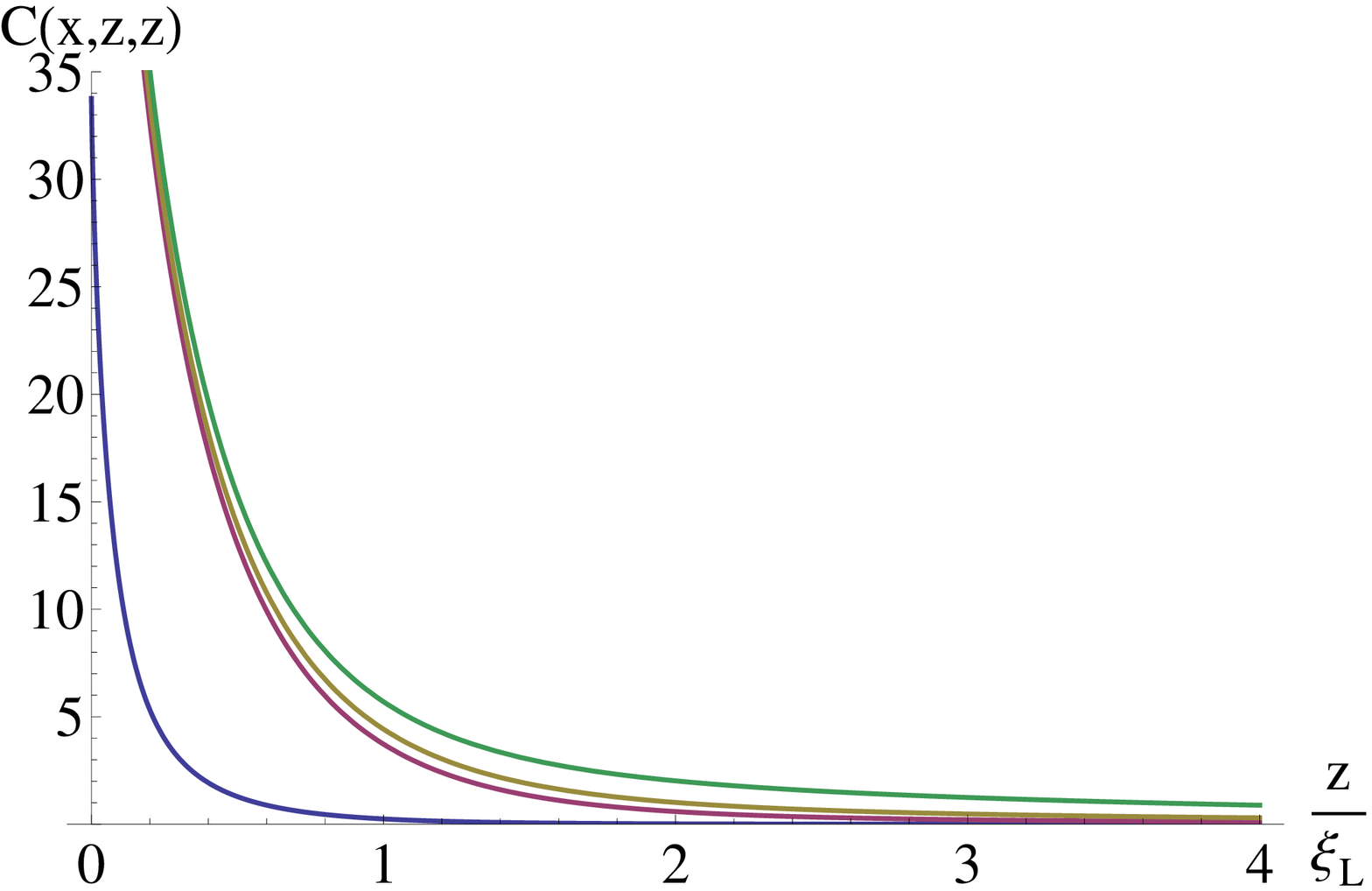}\\     
\caption{(Color online)
  Full 2D correlation function $C(x,z,z)$ in the top (bottom)
  figure plotted as a function $x$ ($z$) for a series of $z$ ($x$)
  values: $z=0, 0.1 ,1, 2, 10\xi_L$ (from top to bottom) 
  [$x=0.3, 3, 5, 10 \xi_L$ (from bottom to top)].  For $z\gg \xi_L$, 
  the correlation is dominated by $C_*(x,z,z)$.}
\label{fig:C2d}
\end{figure}

The second long-scale part, $C_*(\xv,z,z)$ in \rfs{ClongScale} is a
universal contribution [determined by the fixed point function,
\rfs{fixedpointRpp}], that for $d=2$ and low temperatures (when 2D
effects discussed in Sec.~\ref{sec:FRG} can be neglected) can be
straightforwardly computed.  It is compactly given by 
(see Appendix \ref{app:matching})
\begin{widetext}
\begin{eqnarray}
C_*(x,z,z)&\approx&\frac{2(3-d)\pi^2}{9C_{d-1}}\int \frac{d^{d-1}q}{(2\pi)^{d-1}}
\frac{1-\cos\qv\cdot\xv}{q^{d-1}}e^{-2 q z}\nonumber\\
&\approx&\frac{\pi^2}{9}\ln\left[1+\frac{x^2}{(2z+\xi_L)^2}\right],\ \ \mbox{for $x\gg\xi_L$, $d =2$},
\label{C*longscale}
\end{eqnarray}
\end{widetext}
and is plotted in Fig.~\ref{fig:C_s2d}.

The full 2D correlation function $C(x,z,z)$ [defined by
Eqs.\rf{LarkinCorrelationInfinite}, \rf{ClongScale}-\rf{C*longscale}] 
is plotted in
Fig.~\ref{fig:C2d}.

\begin{figure}[t]
\includegraphics[height=5 cm]{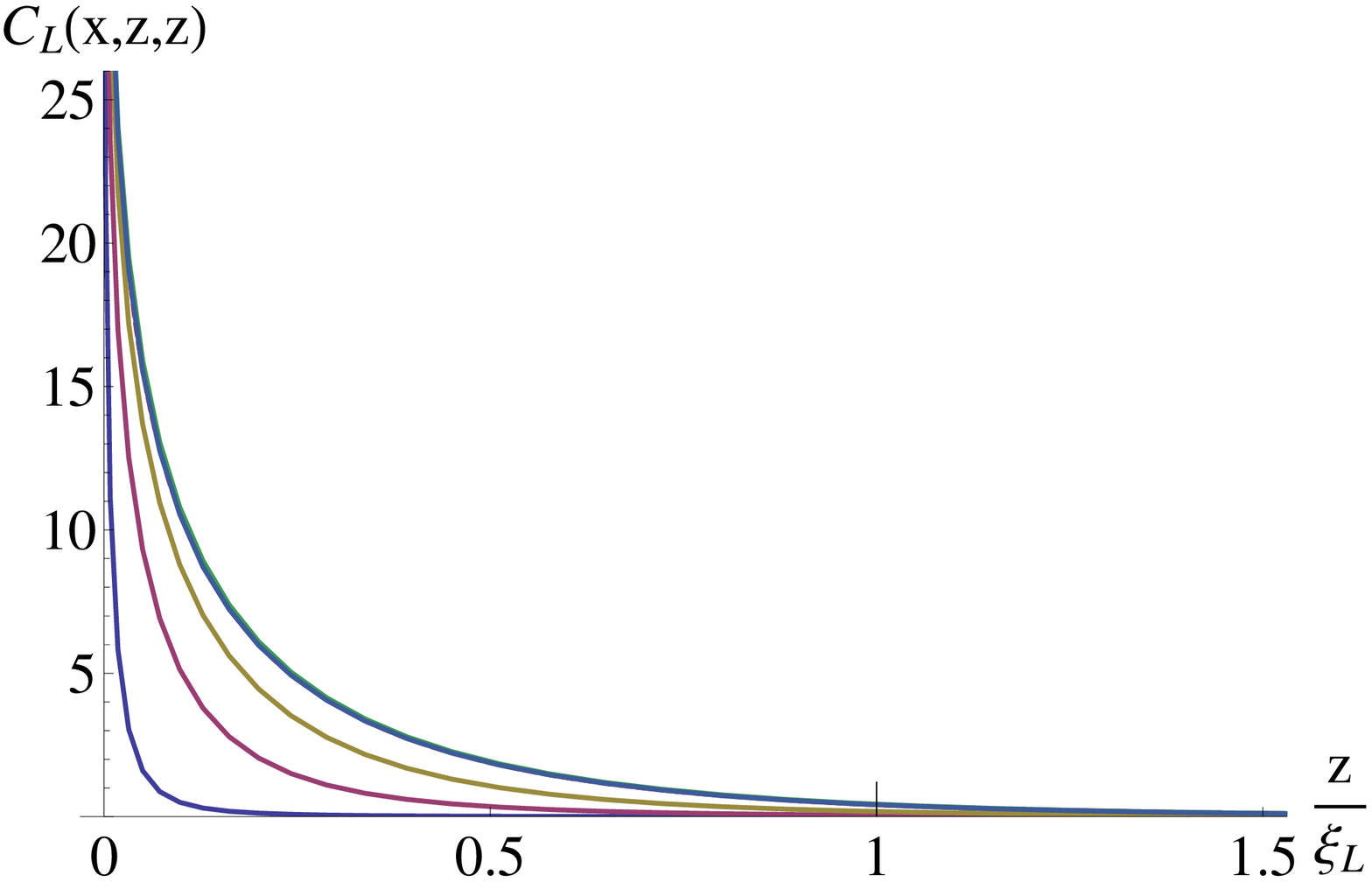}\\    
\caption{(Color online)
  For $d=3$, the contribution of short-scales $1/\xi_L< q <
  a^{-1}$ to the correlation function $C(\xv,z,z)$ with different 
  $x$ values (from bottom to top): $0.1$, $0.5$, $1$, $10$, and 
  $100 \xi_L$.  At small $x$ (e.g., $x=0.1\xi_L$, blue curve at the 
  bottom), $C_L(\xv,z,z)$ decays very fast and is
  $x$-dependent, while at large $x$ values (shown for $x=10,100\xi_L$,
  the nearly overlapping higher curves), the contribution is nearly
  $x$-independent: at $z=0$, it is a constant around $8\pi^2$ and
  decays rapidly to zero.}\label{fig:C_L3d}
\end{figure}
(ii) $d=3$. 
For a three-dimensional (infinite-thick) cell the pinning is
marginally irrelevant, with the large $\ell$ solution given by
\rfs{R3d}. Using this inside \rfs{Cqii}, we obtain
\begin{eqnarray}
\hspace{-1cm} C[\qv,K,\infty,\hR]&\approx&\frac{-1}{q^2\ln(q a)}\frac{\pi^2}{9C_2},
\ \ \mbox{for $d = 3$}.
\label{Cqiii3d}
\end{eqnarray}

The short-scale part, $C_L(\xv,z,z)$, is similar to \rfs{C_Llongscale},
which when evaluated in 3D reduces to (see Appendix \ref{app:LarkinCorr})
\begin{widetext}
\begin{eqnarray}
C_L(\xv,z,z)&\approx&\frac{2\Delta_f}{K^2}\int_{\xi_{L}^{-1}}^{a^{-1}}
\frac{d^2q}{(2\pi)^2}\frac{(1-\cos\qv\cdot\xv)e^{-2q z}}{q^2}
\nonumber\\
&\approx&\frac{8\pi^2}{\ln(\xi_L/a)}\Gamma(0,2z/\xi_L,2z/a),
\ \ \ \mbox{for $x\gg\xi_L$}\nonumber\\
&\approx&8\pi^2\left\{\begin{array}{ll}
1-\frac{\ln(2z/a)}{\ln(\xi_L/a)},&a\ll 2z \ll \xi_L\\
\frac{\xi_L/2z}{\ln(\xi_L/a)}\ e^{-2z/\xi_L},&2z\gg\xi_L
\end{array}\right.\ \ \ \mbox{for $d = 3$},
\label{C_Llongscale3d}
\end{eqnarray}
\end{widetext}
where $C_L(\xv,z,z)$, plotted in Fig.~\ref{fig:C_L3d}, is again nearly
$x$-independent for $x\gg\xi_L$, as the only $x$-dependence enters
through $\cos\qv\cdot\xv$, which averages to zero for these large
wave vectors with $q x\gg 1$.

For $d=3$, the long-scale universal part $C_*(\xv,z,z)$ in
\rfs{ClongScale} is obtained directly from the fixed-point function
\rf{R3d} \cite{FeldmanVinokurPRL,usFRGPRL}, derived in Appendix 
\ref{app:matching}, and it is given by
\begin{widetext}
\begin{eqnarray}
C_*(\xv,z,z)&\approx&-\frac{2\pi^2}{9C_2}\int \frac{d^2q}{(2\pi)^2}
\frac{1-\cos\qv\cdot\xv}{q^2\ln(q a)}e^{-2 q z},\ \ \mbox{for
  $x\gg\xi_L$, $d = 3$},\nonumber\\
&\approx&-\frac{2\pi^2}{9}\left[
  \frac{1}{4}\int_0^1 dk\frac{ke^{-2 k z/x}}{\ln(k a/x)}
  + \int_{1}^{x/\xi_L} dk\frac{e^{-2 k z/x}}{k\ln(k a/x)}
 \right],\nonumber\\
&\sim&
\frac{2\pi^2}{9}\left\{\begin{array}{ll}
\ln\big[\frac{\ln(x/a)}{\ln(\xi_L/a)}\big],&2z \ll \xi_L \ll x\\
\ln\big[\frac{\ln(x/a)}{\ln(2z/a)}\big],&\xi_L\ll 2z \ll x\\ 
\frac{x^2}{16z^2}\frac{1}{\ln{(2z/a)}},&\xi_L \ll x \ll 2z 
\end{array}\right.
\label{C*longscale3d}
\end{eqnarray}
\end{widetext}
and is plotted in Fig.~\ref{fig:C_s3d}.

Thus on the $z=0$ substrate, at long $x\gg\xi_L$ length scales and for
a thick 3D cell, we find
\begin{eqnarray}
C_*(\xv,0,0)&\approx&\frac{2\pi^2}{9}\ln\left[\frac{\ln(x/a)}{\ln(\xi_L/a)}\right],
\nonumber\\
&\sim&\frac{2\pi^2}{9}\ln[\ln(x/a)],
\label{C0_*}
\end{eqnarray}
as claimed in the Sec.~\ref{sec:intro} and first found in
Ref.~[\onlinecite{FeldmanVinokurPRL}].
The complete 3D correlation function $C(\xv,z,z)$ is plotted in
Fig.~\ref{fig:C3d}.

We conclude this section with a computation of the \textit{surface} 
orientational order parameter,
\begin{eqnarray}
\overline{\psi}(w,0)&=&\overline{\langle e^{i\phi(\xv,0)}\rangle}
\nonumber\\
&\approx& e^{-\overline{\langle\phi^2(\xv,0)\rangle/2}},
\end{eqnarray}
where somewhat crudely we approximated it by assuming Gaussian
correlations in $\phi(\xv,0)$. This order parameter is of particular
interest to the application of our results to a finite-thickness cell with
a Dirichlet or a Neumann boundary condition imposed on the homogeneous
substrate (see Fig.~\ref{fig:LCcell}).

Because, as we have seen above [see, e.g., Eqs.\rf{phi_Lrms} and \rf{C0_*}],
in the limit of an infinitely thick cell ($w\rightarrow\infty$)
$\phi_{rms}$ grows without bound with system size $L$, the
orientational order parameter $\overline{\psi}(w,0)$ vanishes in the
thermodynamic limit.  For a more realistic situation of a finite cell,
the decay of orientational order is determined by the cell thickness,
$w$, and the nature of the boundary conditions on the homogeneous
substrate.

\begin{figure}[t]
\includegraphics[height=5 cm]{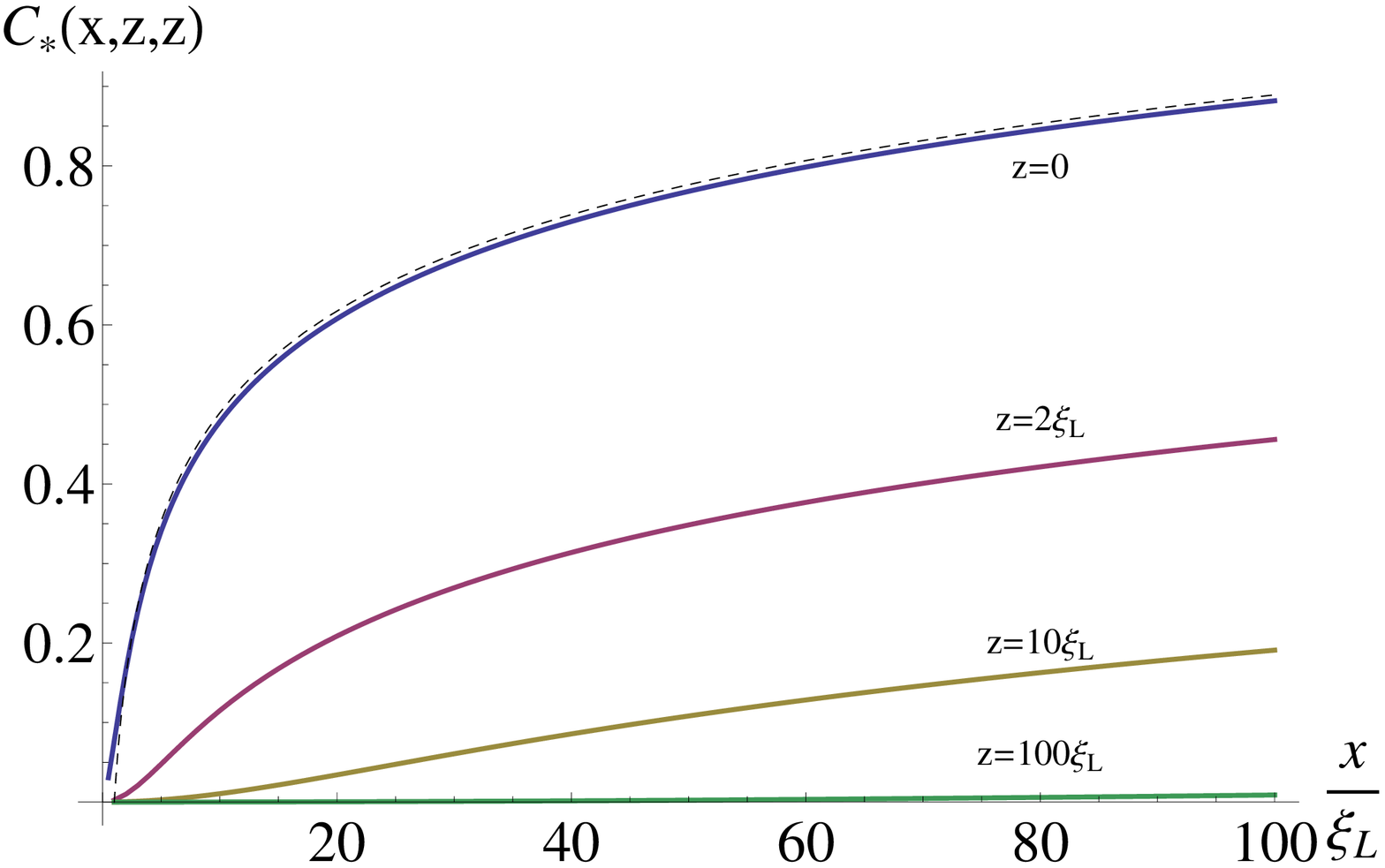}\\    
\includegraphics[height=5 cm]{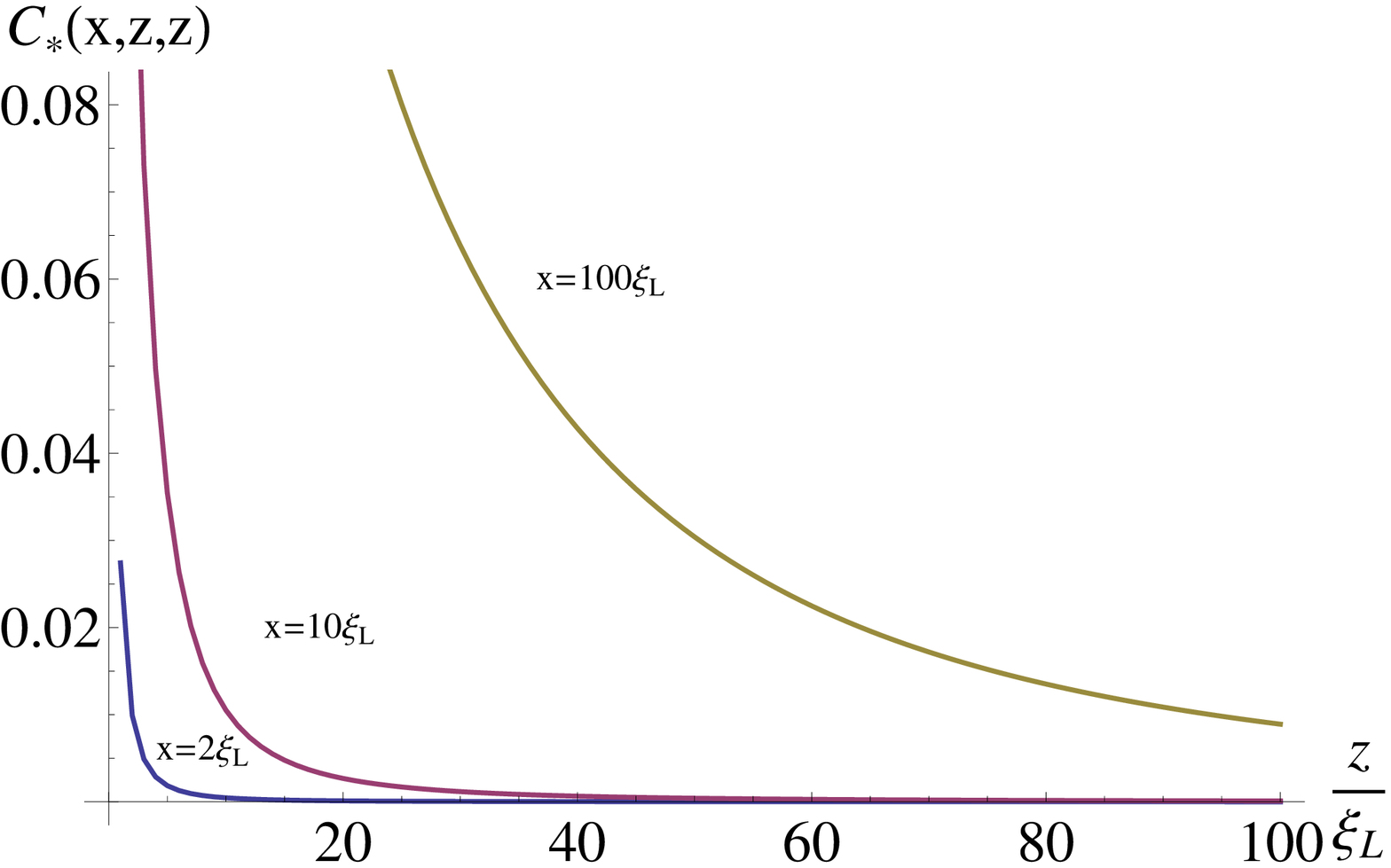}\\     
\caption{(Color online)
  Long-scale part of the 3D correlation function
  $C_{*}(\xv,z,z)$ plotted in the top (bottom) figure as a function of
  $x$ ($z$) for different $z$ ($x$) values: $z=0,2,10,100 \xi_L$ 
  (from top to bottom) [$x=2,10,100\xi_L$ (from bottom to top)]. 
  The dashed curve is the approximation
  $\frac{2\pi^2}{9}\ln{\left[\ln{(x/a)}/\ln{(\xi_L/a)}\right]}$
  summarizing in in-plane correlations on the heterogeneous ($z=0$)
  substrate.}
\label{fig:C_s3d}
\end{figure}

\begin{figure}[t]
\includegraphics[height=5 cm]{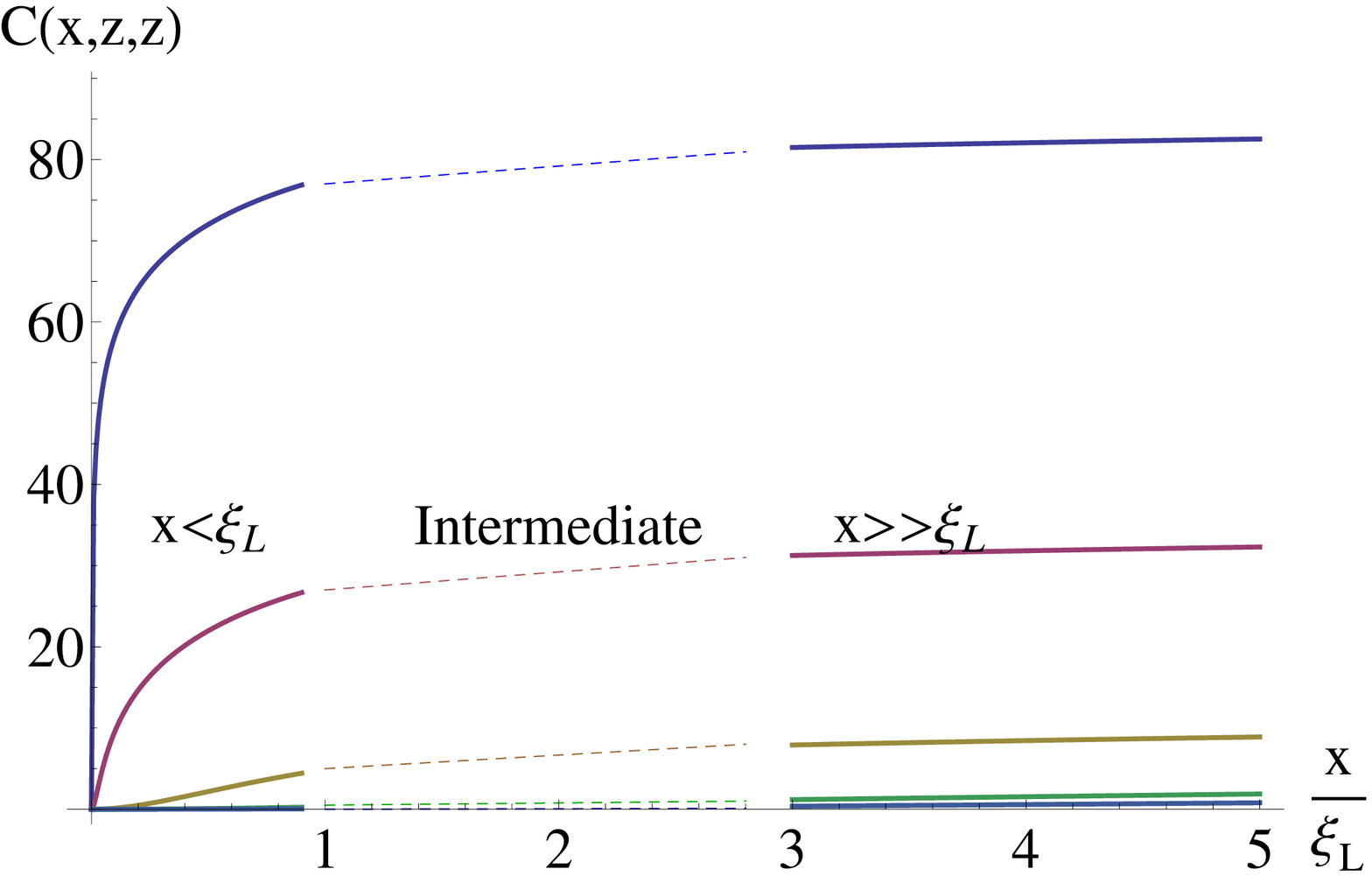}
\includegraphics[height=5 cm]{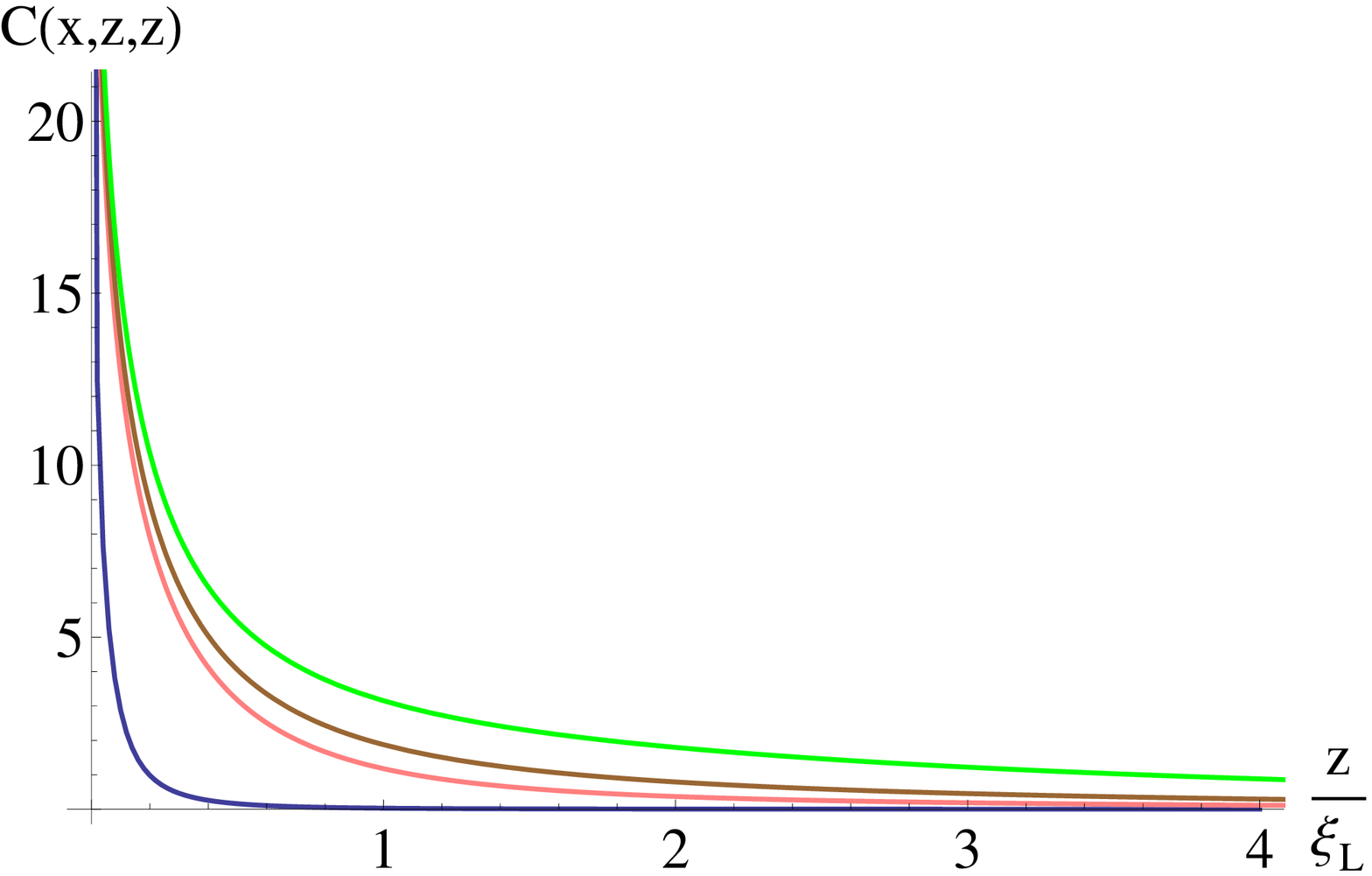}\\     
\caption{(Color online)
  Full 3D correlation function $C(\xv,z,z)$ in the top (bottom)
  figure is plotted as a function of $x$ ($z$) for a series of $z$
  ($x$) values: $z=0, 0.01, 0.2, 1, 2\xi_L$ (from top to bottom) 
  [$x=0.3, 3, 5, 10\xi_L$ (from bottom to top)]. 
  For $z\gg \xi_L$, the correlation is dominated by
  $C_*(\xv,z,z)$.}
\label{fig:C3d}
\end{figure}

\subsubsection{Crossover in a Dirichlet cell}
Contrasting the bulk behavior, for a cell with a Dirichlet
(homogeneous) substrate, we expect the growth of root-mean fluctuations
of $\phi$ to be suppressed by the alignment by this homogeneous
substrate. Thus, in this case, $\overline{\psi}(w,0)$ is nonzero and the
orientational order is stable for an arbitrarily thick (but finite)
cell. 

We can analyze $\overline{\psi}(w,0)$ by estimating $\phi_{rms}(w)$ using the results of FRG found in Sec.~\ref{sec:FRG}. To this end, we examine the
asymptotics of the FRG flow in Eqs.~\rf{wflow0} and \rf{FRGflowRT0_DN}. For
a thin cell (defined by $w\ll\xi_L$), $w(\ell)=e^{-\ell}w$ reaches the
microscopic scale $a$ at $e^{\ell_w^*}=w/a$ and therefore
$\epsilon^\D(\ell>\ell_w^*)\approx \epsilon-2=1-d$ {\em before}
$e^{\ell_L^*}=\xi_L/a$. Since beyond $\ell_w^*$,
$\epsilon^\D(\ell>\ell_w^*)<0$, pinning is irrelevant and the flow is
cut-off at scale $e^{\ell_w^*}$ , scales beyond $\xi_L$ are not probed
(the flow never leaves the vicinity of the Gaussian fixed point), and
$\phi_{rms}$ can be accurately computed within the Larkin
approximation (random-torque model), cut-off by $w$.

In contrast, for a thick cell (defined by $w\gg\xi_L$), the flow
crosses over to the vicinity of the nontrivial fixed point $R_*$
(leaves the Gaussian fixed point) {\em before} it is cut-off by the
finite $w$. In this case, on longer scales, $e^\ell > w/a \gg
e^{\ell_L^*}\equiv\xi_L/a$, the fluctuations are cut-off by $w$ (by the
flow's return to the Gaussian fixed point). In this thick cell regime,
$\phi_{rms}(w)$ is thus given by the matching calculation of
Sec.~\ref{sec:matching} with the diverging $L$ ($>\xi_L$) dependence
cutoff by $w$.

Following this crossover allows us to calculate $\phi_{rms}(w)$. For a
thin cell, $w\ll\xi_L$ and $d<3$, we have
\begin{widetext}
\begin{eqnarray}
\overline{\langle\phi_0^2\rangle}&=&\int\frac{d^{d-1}q}{(2\pi)^{d-1}} C^\D(q)\nonumber\\
&\approx&
\int_{0}^{w^{-1}}\frac{d^{d-1}q}{(2\pi)^{d-1}}\frac{\Delta_f w^2}{K^2}
+\int_{w^{-1}}^{a^{-1}}\frac{d^{d-1}q}{(2\pi)^{d-1}}\frac{\Delta_f}{K^2q^2}\nonumber\\
&\approx&4\pi^2\left(\frac{w}{\xi_L}\right)^{3-d},
\ \ \mbox{for $w\ll\xi_L$, $d < 3$},
\end{eqnarray}
where we used the Larkin approximation \rfs{C_D} together with
\rfs{GammaD}, valid for $w\ll\xi_L$ since the Dirichlet flow in
\rfs{FRGflowRT0_DN} never leaves the vicinity of the Gaussian fixed
point. 

For a thick cell $w\gg\xi_L$ and $d<3$, we have
\begin{eqnarray}
\overline{\langle \phi_0^2\rangle}&=&\int\frac{d^{d-1}q}{(2\pi)^{d-1}} C^\D(q)\nonumber\\
&\approx&
\int_{w^{-1}}^{\xi_L^{-1}}\frac{d^{d-1}q}{(2\pi)^{d-1}}
\frac{\epsilon\pi^2}{9 C_{d-1}}\frac{1}{q^{d-1}}
+\int_{\xi_L^{-1}}^{a^{-1}}\frac{d^{d-1}q}{(2\pi)^{d-1}}\frac{\Delta_f}{K^2q^2}
\nonumber\\
&\approx&\frac{\epsilon\pi^2}{9}\ln(w/\xi_L) + 4\pi^2, 
\ \ \mbox{for $w\gg\xi_L$, $d < 3$},
\label{eta_d}
\end{eqnarray}
\end{widetext}
where we neglected the subdominant contribution from scales longer
than $w$ (where the RG flow for $R(\ell)$ ``turns around'' heading
back toward the Gaussian fixed, i.e., pinning is irrelevant), in the
first term of the second line approximated the $\phi$ correlator
$C(q)$ by its fixed-point value \rf{Cqiii}, valid for
$w/\xi_L\gg 1$ (such that the flow approaches the vicinity of the
nontrivial fixed point), and approximated $C(q)$ in the second
term by its Gaussian fixed point expression (Larkin approximation),
valid for $\xi_L^{-1}<q<a^{-1}$. We also used the definition of
$\xi_L$ to approximate the second term in the last line by $(2\pi)^2$.

Repeating above estimates for $d=3$, for a thin cell ($w\ll\xi_L$), we find
\begin{eqnarray}
\overline{\langle \phi_0^2 \rangle}&\approx&
\int_{w^{-1}}^{a^{-1}}\frac{d^2q}{(2\pi)^2}\frac{\Delta_f}{K^2q^2}\nonumber\\
&\approx&4\pi^2\frac{\ln(w/a)}{\ln(\xi_L/a)},
\ \ \mbox{for $w\ll\xi_L$, $d = 3$},
\end{eqnarray}
and for a thick cell ($w\gg\xi_L$),
\begin{eqnarray}
&&\hspace{-0.5cm}\overline{\langle \phi_0^2\rangle} \nonumber\\
&\approx&
\int_{w^{-1}}^{\xi_L^{-1}}\frac{d^2q}{(2\pi)^2}
\frac{\pi^2}{9 C_2}\frac{-1}{q^2\ln(q a)}
+\int_{\xi_L^{-1}}^{a^{-1}}\frac{d^2q}{(2\pi)^2}\frac{\Delta_f}{K^2q^2}
\nonumber\\
&\approx&\frac{\pi^2}{9}\ln\left[\frac{\ln(w/a)}{\ln(\xi_L/a)}\right] 
+ 4\pi^2,\ \ \mbox{for $w\gg\xi_L$, $d = 3$},\nonumber\\
\label{eta_3}
\end{eqnarray}
where we employed the same asymptotic approximations as for $d<3$. 

Putting these crossovers together, we finally obtain the 
\textit{surface}  orientational order parameter $\overline{\psi}(w,0)$ 
for thin and thick cells in $d<3$,
\begin{eqnarray}
\overline{\psi}_{d<3}\approx\left\{\begin{array}{ll}
e^{-2\pi^2(w/\xi_L)^{3-d}},& \mbox{thin cell, $w\ll\xi_L$}\\
e^{-2\pi^2}\left(\frac{\xi_L}{w}\right)^{\eta_d^*}, & \mbox{thick cell, $w\gg\xi_L$},
\end{array}\right.
\label{psi_d}
\end{eqnarray}
and in 3D
\begin{eqnarray}
\overline{\psi}_{3D}\approx\left\{\begin{array}{ll}
\left(\frac{a}{w}\right)^{\eta_L},& \mbox{thin cell, $w\ll\xi_L$}\\
e^{-2\pi^2}
\left[\frac{\ln(\xi_L/a)}{\ln(w/a)}\right]^{\eta_{3D}},& \mbox{thick cell, $w\gg\xi_L$},
\end{array}\right.\nonumber\\ 
\label{psi3D}
\end{eqnarray}
where $\eta_d^*=(3-d)\pi^2/18$, $\eta_{3D}=\pi^2/18$ are universal
exponents [given in Eqs.~\rf{eta_d} and \rf{eta_3}] and
$\eta_L=2\pi^2/\ln(\xi_L/a)$ is a nonuniversal constant. The 3D
\textit{surface} order parameter for such Dirichlet cell of thickness
$w$ is illustrated in Fig.~\ref{fig:OrderParameter}.

\subsubsection{Crossover in a Neumann cell}
Above analysis straightforwardly extends to a finite-thickness cell
with the Neumann boundary condition on the homogeneous substrate. At
long scales, a finite-thickness Neumann cell reduces to an effective
``film,'' i.e., a $d-1$-dimensional {\em bulk} random-field
$xy$ model. Thus, we expect the disordering effect of the random pinning
to be enhanced compared to the $w\rightarrow\infty$ system, where
additional homogeneous bulk degrees of freedom have a stabilizing
effect against pinning ($d_{lc}$ reduced from $4$ down to $3$).

This is reflected in the behavior of both the correlators in the
random-torque model [given by Eqs.~\rf{GammaN} and \rf{C_D}], and in the
FRG flow that becomes only more divergent on scales
$e^{\ell}>e^{\ell_w^*}=w/a$, as $\epsilon^\N(\ell >
\ell_w^*)\rightarrow\epsilon + 2 = 5-d > \epsilon$. Hence, in contrast
to the Dirichlet cell (where finite $w$ suppresses the effect of the
random potential), in a Neumann cell, finite thickness enhances the
effects of random surface pinning. Consequently, independent of the
Neumann cell thickness, the orientational order parameter, $\overline{\psi}$,
vanishes for $L\rightarrow\infty$.

\section{Strong pinning limit}
\label{sec:strongPinng}
In all of the above analysis, we focused on the most interesting {\em
  weak} surface disorder, where pinning is collective, dominating over
the elastic energy only on the macroscopic length scales, longer than
$\xi_L\gg a$. This assumption is what justified our treatment of the
elastic energy as dominant (at least on short scales, smaller than
$\xi_L$), allowing an expansion about the ordered $\phi=0$ (nematic)
state.  However, it is quite possible that in some (e.g., liquid
crystal) applications, it is the opposite limit of strong pinning that
is of interest.

In the latter strong-disorder limit, the surface-pinning potential (by
definition) dominates over the elastic energy at all, even microscopic
scales, with $V_p > K/a$. To treat this regime, we instead perturb in
the elastic energy about a random ground state, $\phi_0^s(\xv)$, that
exactly minimizes the random pinning potential $V[\phi_0(\xv),\xv]$. That
 is, $(\partial_{\phi_0} V)|_{\phi_0(\xv)=\phi_0^s(\xv)}=0$. We then 
expand about this random ground state, obtaining
\begin{eqnarray}
\delta H_s&\approx&
\int\frac{d^{d-1}q}{(2\pi)^{d-1}}\left[\frac{K}{2} q|\phi_0(\qv)|^2
+\frac{g}{2}|\phi_0(\qv)-\phi_0^s(\qv)|^2\right],\nonumber\\
\label{deltaHs}
\end{eqnarray}
where $g=-\overline{\partial^2_{\phi_0}
  V}|_{\phi_0(\xv)=\phi_0^s(\xv)}\approx V_p/(2\pi)^2$.  A
minimization of the above Hamiltonian then straightforwardly gives
\begin{eqnarray}
\phi_0(\qv)&\approx&
\frac{g}{K q + g}\phi_0^s(\qv).
\label{phi_0}
\end{eqnarray}

From this analysis we can readily identifying a strong-coupling
pinning length
\begin{eqnarray}
\xi_s&=&K/g,\nonumber\\
&\sim& \frac{K}{\Delta_f^{1/2}}\xi_0^{(d-1)/2},\nonumber\\
&\sim& \xi_L^{(3-d)/2}\xi_0^{(d-1)/2},
\end{eqnarray}
which is a scale below which the $xy$-order parameter no longer
faithfully follows spatial variations of the local random potential
and thus it is a strong-coupling version of the Larkin length.  In
above, we restored the pinning potential correlation length $\xi_0$ to
also account for the more realistic case where $\xi_0$ is distinct and
longer than the microscopic molecular cutoff scale $a$.  Hence we
conclude that below the pinning correlation length $\xi_0$, there is a
crossover from weak to strong pinning limit when the (weak-coupling)
Larkin length, $\xi_L$, drops down to $\xi_s$. On these shorter scales,
the collective pinning analysis of previous sections and
corresponding results break down.

\section{Application to liquid crystal cells}
\label{sec:NematicSmectic}

As discussed in the Sec.~\ref{sec:intro}, liquid crystal cells provided a
strong motivation for our study of the orientational order in the
presence of surface random pinning. However, although there is a
qualitative overlap, in detail a model of a surface-pinned liquid
crystal cell, a priori can be quite different from a basic $xy$ model
studied above. Furthermore, the detailed model very much depends on
the specific nature of the liquid crystal phase and thus requires an
extensive study that lies beyond the current paper. However, to
put our above results for an $xy$ model in a physical context, we now
briefly examine a formulation of a surface-pinning problem for real
liquid crystals, focusing on nematic and smectic phases, deferring
their detailed analysis to a future study.

\subsection{Nematic liquid crystal phase}

The key distinction of the nematic liquid crystal phase as compared to
the $xy$ model studied so far is the nature of the Goldstone modes, that
for a nematic is given by a {\em three}-dimensional unit vector
(strictly speaking with opposite ends identified forming an $RP_2$
manifold), the nematic director
\begin{equation}
\hat{n}=(\cos{\theta}\cos{\varphi},\cos{\theta}\sin{\varphi},\sin{\theta}),
\label{n}
\end{equation}
as opposed to its $xy$-model counterpart, where it is a single azimuthal
(planar) angle, $\phi$, with the polar angle $\theta$ fixed at zero by
some easy-plane anisotropy. We note that in above, we chose a somewhat
nonstandard (but here convenient for treating parallel surface
alignment) convention for $\theta$. We also implicitly ignored the
difficult question of topological defects proliferation and the
corresponding stability of the elastic glass. If indeed important, for
weak disorder, we expect their effects to set in on much longer length
scales, thereby providing a wide intermediate range of scales, where
defect-free model is of interest.

In the nematic phase, the bulk energy of a nematic director
$\hat{n}(\rv)$ is described by the well-known Frank-Oseen
expression \cite{deGennes}
 
\bea 
H_{F}&=&\frac{1}{2}\int
d^{d-1}xdz\bigg\{K_s(\nabla\cdot\hat{n})^2+K_t\left[\hat{n}\cdot(\nabla\times\hat{n})\right]^2\nonumber\\
&&+K_b\left[\hat{n}\times(\nabla\times\hat{n})\right]^2\bigg\}.
\eea 
Within a simplifying one-elastic-constant approximation,
$K_s=K_t=K_b\equiv K_n$, and together with the surface-pinning energy
and polar parameterization, \rfs{n}, the above equation gives the elastic
Hamiltonian for the nematic surface-pinned cell

\bea
H_{nematic}&=&\frac{K_n}{2}\int d^{d-1}x\int_0^w dz
\left[(\nabla\theta)^2+\cos^2{\theta}(\nabla\varphi)^2\right]
\nonumber\\
&& + H_{pin},
\label{Hnematic}
\eea
where the surface-pinning energy
\bea
H_{pin}&=&-\int
d^{d-1}x\bigg[\left(W_0+V(2\varphi,\xv)\right)\cos^2{\theta}|_{z=0}
\nonumber\\
&&+W_w\cos^2{\theta}|_{z=w}\bigg]
\label{Hnem_pin}
\eea
is given by a purely homogeneous planar ($\theta=0$) alignment on the
top ($z=w$) substrate and a planar alignment with a heterogeneous
azimuthal component on the bottom ($z=0$) substrate. The latter is
encoded in a random pinning function $V(\phi,\xv)$ with a $2\pi$ 
periodicity of $\phi$, presented in
Sec.~\ref{sec:model}, with $\phi=2\varphi$ capturing the 
$\hat{n}\leftrightarrow -\hat{n}$ symmetry of the nematic liquid
crystal phase.

For a thin cell and weak random pinning on the bottom substrate, such
that the scale of $V(\phi,\xv)$ is much smaller than $W_0$, clearly
the planar alignment, while of random strength (on the bottom
substrate), remains planar on both substrates. We further note that
because $(\nabla\phi)^2$ (computed within the $xy$-model approximation)
remains finite in the physically interesting dimensions (i.e.,
infrared convergent for $d>1$) and small for weak pinning (decaying
into the bulk with $z$), based on Fredericks transition
phenomenology \cite{deGennes}, we do not expect planar surface alignment
by $W_{0,w}$ to be overturned by a weak second term in \rfs{Hnematic}.

Thus, with the exception of small regions (that are rare for weak
disorder), $\theta=0$ is the solution that minimizes the total
energy. With this, the nematic cell model reduces to the surface 
random-field $xy$ model for the director's azimuthal orientation
$\phi=2\varphi$, studied
above \cite{FeldmanVinokurPRL,usFRGPRL}. Hence, for weak pinning, all of
$xy$-model results detailed above apply directly to a nematic liquid
crystal cell.

In contrast, we expect strong pinning to lead to large azimuthal
distortions that will be accompanied by big $\theta$ variations both
on the random substrate and in the bulk. For intermediate pinning
strength, the system can perhaps even exhibit a random Fredericks-like
transition corresponding to a bulk escape from a planar surface
configuration, driven by a large $(\nabla\phi)^2 \gg 1/w^2$. We leave
the detailed study of the associated subtleties for this system, which
distinguish it from the simple $xy$ model to a future research.

\subsection{Smectic liquid crystal phase}

Another important realization of a random surface-pinning problem is
that of a smectic liquid crystal on a heterogeneous substrate, as for
example realized in recent experiments \cite{ClarkSmC} mentioned in the
Sec.~\ref{sec:intro}. We will focus on the experimentally and theoretically
more interesting case of the bookshelf geometry, illustrated in
Fig.~\ref{fig:LCcell}, where layers and director are, respectively,
perpendicular and parallel to the substrate.  Choosing the coordinate
system as indicated in Fig.~\ref{fig:smecticCartoon}, so that the 
smectic layers lie parallel to the $(x,y)$ plane,
with the average layer normal along the $z$ axis and the random
substrate located at $y=0$ and running perpendicular the $y$ axis, the
total energy is given by

\bea H_{sm}&=&\int d^{d-1}x\int_0^w dy
\left[\frac{K}{2}(\nabla^2_\perp u)^2 +\frac{B}{2}(\partial_z
  u)^2\right]
\nonumber\\
&& + H_{pin},
\label{Hsmectic}
\eea
where the first two terms describe the usual smectic elasticity, with
bending and compressional elastic constants $K$ and $B$ (for
simplicity taken to be harmonic \cite{commentHarmonic}), and the last
term is a surface-pinning energy given by
\bea
\hspace{-0.2cm}
H_{pin}&=&\int
d^{d-1}x dy \delta(y)\bigg[\frac{W}{2}(\nh\cdot\yh)^2
-(\nh\cdot{\bf g}(\rv))^2-V[u,\rv]\bigg],\nonumber\\
&\approx&\int
d^{d-1}x dy \delta(y)\bigg[\frac{W}{2}(\delta n_y)^2
- h(\xv)\delta n_x-V[u,\xv]\bigg],\nonumber\\
&\approx&\int
d^{d-1}x dy \delta(y)\bigg[\frac{W}{2}(\partial_y u)^2
-h(\xv)\partial_x u-V[u,\xv]\bigg],\nonumber\\
\label{Hsm_pin}
\eea
where for convenience we defined $\rv=(z,x,y)\equiv(\xv,y)$ and
extended its $\xv$ to $d-1$ dimensions transverse to $y$.

\begin{figure}
\includegraphics[height=5 cm]{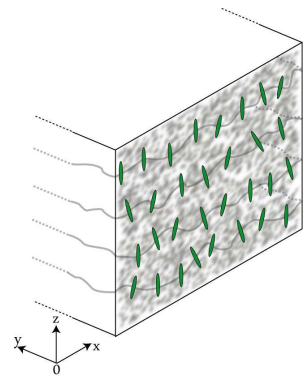}\\    
\caption{(Color online)
  Cartoon of smectic liquid crystal cell with random substrate at 
  $y=0$. The substrate fixes large orientation transversely, modeled by 
  $W(\delta n_y)^2\sim W(\partial_y u)^2$, and the surface disorder randomly 
  pins layer positions through $V[u,\xv]\delta(y)$ and layer orientations 
  through $(\hat{n}\cdot{\bf g}(\rv))^2\sim h(\xv)\partial_x u$.}
\label{fig:smecticCartoon}
\end{figure}

In above, the $W$ and $g(\xv)$ terms are the homogeneous and random
components of the orientational pinning [$W$ inducing a homogeneous
parallel to the surface director alignment and $g(\xv)$ capturing
random azimuthal director pinning within the heterogeneous substrate
plane] and $V[u,\xv]$ is the positional pinning of surface layers
with the random substrate \cite{RTaerogelPRL,RTaerogelPRB}. In getting
to the final form we expanded $\nh$ about its pinning-free orientation
along $\zh$ and used the smectic ``Higgs
mechanism'' \cite{deGennes,RTaerogelPRB} to make a replacement
$\delta{\nh}_\perp\rightarrow\nabla_\perp u$, valid inside a smectic
phase.

Following procedure used for the $xy$ model, this randomly
surface-pinned smectic model can be used to analyze the phenomenology
of a smectic liquid crystal cell with the hopes of understanding
long-scale random textures observed in Ref.~[\onlinecite{ClarkSmC}].
Because (as illustrated above) the smectic elasticity and pinning
differ qualitatively from that of an $xy$ model, we expect a
phenomenology that is qualitatively distinct from that found for the
simple $xy$ model and the nematic phase found above. We leave the
interesting and nontrivial study of a random smectic cell to a future
research.

\subsection{Experimental observables}

One attractive feature of liquid crystals is that their orientational
order can be readily studied via light microscopy.  In its simplest
form, the technique utilizes a crossed polarizer-analyzer pair on the
front and back of a cell, typically transversely oriented.  In this
geometry, the spatial (within the $xy$ plane) distribution of the
transmitted light intensity through the cell is sensitive to the
azimuthal variation of the local optic axis and therefore measures the
director's planar spatial distribution.  

For a fully ordered planar nematic state, with a {\em uniform}
director orientation at an azimuthal angle $\varphi$ with respect to
the polarizer (or analyzer), the transmitted light intensity $I$
through a uniaxial cell of thickness $w$ is given by \cite{deGennes}
\be
  I_w=I_0\sin^2(2\varphi)\sin^2(\delta/2).
\label{Iuniform}
\ee
In above, $\delta=\chi_o-\chi_e=2\pi (n_o-n_e) w/\lambda$ is the phase
difference between ordinary and extraordinary components of light with 
wavelength $\lambda$, respectively, characterized by $n_o$, $n_e$ indices of
refraction.  In the simplest case of the director uniformly aligned
along the polarizer ($\varphi=0$) or the analyzer ($\varphi=\pi/2$),
this leads to a uniformly vanishing transmitted light
intensity. Conversely, the maximum light transmission is produced for
a $\pi/4$ uniform director orientation relative to the polarizer (or
equivalently, transversely-crossed analyzer). The last factor in
\rfs{Iuniform} leads to transmission color selectivity with optical
anisotropy and cell thickness.

For a spatially {\em nonuniform} director variation, the analysis of
the transmitted light intensity is more complicated. However, for a
nematic variation on a scale longer than light's transverse coherent
length (a typical situation for illumination with incoherent source),
the output intensity simply images the transverse ($xy$-) optic axis
variation, with each coherence region treatable as an independent
column of depth $w$.  Furthermore, in the limit that spatial variation
along $z$ is also smooth on the scale of light's wavelength, the
transmission through each column can be treated in Mauguin limit,
where light components along and perpendicular to the optic axes
simply adiabatically follow the local director orientation
$\nh(\xv,z)$. 

For a $\pi/2$-crossed polarizer-analyzer pair, standard analysis in
this Mauguin limit then gives the output light intensity (behind the
analyzer, at $z = w$)
\bea
\hspace{-0.5 cm}
I_w(\xv) &=& I_0|\cos\varphi(\xv,0)\sin\varphi(\xv,w)e^{i\chi_e}\nonumber \\
&&-\sin\varphi(\xv,0)\cos\varphi(\xv,w)e^{i\chi_o}|^2\nonumber\\
&=&I_0\sin^2[\varphi(\xv,w)-\varphi(\xv,0)]\cos^2(\delta/2)\nonumber\\
&&+I_0\sin^2[\varphi(\xv,w)+\varphi(\xv,0)]\sin^2(\delta/2).
\eea
For the director on the back substrate aligned (by the Dirichlet
boundary conditions) with the polarizer
axis (and perpendicular to the analyzer), 
i.e., $\varphi(\xv,w)=0$, the output signal simplifies
considerably to
\bea
I_w(\xv)&=&I_0\sin^2[\varphi(\xv,0)]
\eea
and is thus directly related to the local surface orientational order
parameter, $\psi(\xv,0)=e^{i2\varphi(\xv,0)}\equiv e^{i\phi(\xv,0)}$
studied in this paper.  For example, a spatially averaged transmission
through a Dirichlet cell is given by
\bea
I_w&=&\overline{\langle I_w(\xv)\rangle} \nonumber\\
&=&\oh I_0\left(1-\overline{\langle e^{i\phi_0(\xv)}\rangle}\right)
\nonumber\\
&=&\oh I_0\left(1-\overline{\psi}(w,0)\right),
\eea
in which $\overline{\psi}(w,0)$ is computed in Eqs.~\rf{psi_d} and \rf{psi3D}. 
Thus, for this choice of
geometry, a thin Dirichlet cell has the expected vanishing
transmission, that grows with cell thickness to its maximum value of
$1/2$. More stringent tests of our predictions can further be made by
comparing ($xy$-) spatial correlations of light transmission,
$\overline{\langle I_w(\xv)I_w(\xv')\rangle}$, with orientational
correlation functions to which these are clearly directly related
according to 
\bea 
\overline{\langle I_w(\xv)I_w(\xv')\rangle} &=& I_0^2
\overline{\langle \sin^2{\varphi(\xv,0)}
  \sin^2{\varphi(\xv',0)} \rangle} \nonumber\\
&\approx&I_0^2\bigg[\frac{1}{4}-\half e^{-\overline{\langle
      \phi_0^2\rangle}/2}\nonumber\\
&&+\frac{1}{8}e^{-\overline{\langle(\phi_0(\xv)-\phi_0(\xv'))^2\rangle}/2}
\bigg].\nonumber\\
&&\label{IIcorr} 
\eea 
An even more direct probe of director correlations is possible through
the polarized confocal microscopy \cite{SmalyukhConfocal}, where an
image of the local director orientation at each depth $z$ can be
produced. A numerical computation of thereby measured director
correlation functions therefore allows a detailed comparison to
results predicted here.

\section{Summary and Conclusion}
\label{sec:conclusion}
In this paper, we have studied the stability of random distortions in
an $xy$ model perturbed by a random surface pinning and discussed our
findings in the context of nematic liquid crystal cell with a
dirty non-rubbed substrate. We found that for a thick 3D cell, at
long scales, the disordering effects of the random substrate always
marginally dominate over the bulk nematic order. Thus, a 3D nematic
order is marginally unstable with orientational ``roughness'' growing
as $\ln[\ln(x/a)]$ on long scales. We have also extended these results
to a finite-thickness cell, with a second homogeneous substrate with
parallel Dirichlet and Neumann boundary conditions. Not surprisingly,
in the former case, the nematic order is stabilized to arbitrary long
scales, but with the nematic order parameter (and the corresponding
birefringence) exhibiting a crossover from a large value for a thin
(weakly heterogeneous) cell to a small value for a thick (strongly
heterogeneous) cell at a characteristic cell thickness set by the
Larkin length, $\xi_L$.  We expect our predictions to be
experimentally testable via a polarizer-analyzer transmission
microscopy and by studying how the nematic order is recovered in
response to a tunable in-plane aligning electric or magnetic field. We
propose that the predicted statistical properties (correlation
functions) of the random substrate-induced director textures can be
quantitatively tested with the polarized confocal
microscopy \cite{SmalyukhConfocal}.

\section{Acknowledgments}

We thank N. Clark, V. Gurarie, M. Hermele, I. Smalyukh, and S. Todari
for discussions and acknowledge financial support by the National
Science Foundation through Grants No. DMR-0321848 and No. MRSEC
DMR-0820579 (L.R., Q.Z.) and the Berkeley Miller and the University of
Colorado Faculty Fellowships (L.R.).  L.R. thanks Berkeley Physics
Department for its hospitality during part of this work.

\appendix

\section{Larkin lengths analysis}
\label{app:LarkinLengths}
In this appendix, we provide the details for the analysis of the Larkin
length \cite{Larkin} in finite thickness, two- and three-dimensional
cells.  As derived in Sec.~\ref{sec:Larkin}, the Larkin length is
defined in the standard way, given by

\be
\overline{\langle \phi_0^2(\bfx)\rangle}=(2\pi)^2
=\int\frac{d^{d-1} q\Delta_f}{(2\pi)^{d-1}[\Gamma_q^{(a)}]^2}
\ee 
with the $\Gamma_q^{(a)}$'s given by
Eqs.~\rf{GammaI}-\rf{GammaN} and the lower momentum cutoff
of the integration given by $1/\xi_L$. In the limit of an infinitely
thick ($w=\infty$) cell above integral is straightforwardly computed,
in 2D ($d=2$) giving

\bea
\overline{\langle \phi_0^2(x)\rangle}&=&(2\pi)^2 \nonumber\\
&=&\frac{\Delta_f}{\pi K^2}\int_{1/\xi_L^{\infty}}^{\infty}\frac{dq}{q^2} \nonumber\\
&=&\frac{\Delta_f}{\pi K^2}\xi_L^{\infty},
\eea
which leads to a Larkin length
\be
 \xi_{L,2D}^{\infty}=4\pi^3K^2/\Delta_f,
\label{xiL2dinfty}
\ee 
with the superscript $\infty$ denoting the result of an infinitely
thick cell (that for simplicity of notation we will drop). 

In 3D, with the two-dimensional random substrate, we obtain
\bea
\overline{\langle \phi_0^2(\bfx) \rangle}&=&(2\pi)^2\nonumber\\
&=&\frac{\Delta_f}{K^2}\int\frac{d^2q}{(2\pi)^2}\frac{1}{q^2} \nonumber\\
&=&\frac{\Delta_f}{2\pi K^2}\int_{1/\xi_L}^{1/a}\frac{dq}{q} \nonumber\\
&=&\frac{\Delta_f}{2\pi K^2}\ln{(\frac{\xi_L}{a})},
\eea
which gives the Larkin length as
\be
\xi_{L,3D}=a e^{(2\pi)^3\frac{K^2}{\Delta_f}}.
\ee

For a general dimension $d<3$, we have 
\bea
\overline{\langle \phi_0^2(\bfx)\rangle}&=&(2\pi)^2\\
&=&\frac{\Delta_f}{K^2}\int\frac{d^{d-1}q}{(2\pi)^{d-1}}\frac{1}{q^2}
\nonumber \\
&=&\frac{\Delta_f}{K^2}\frac{2^{2-d}\pi^{\frac{1-d}{2}}}{\Gamma (\frac{d-1}{2})}\int_{1/\xi_L}^{1/a}\frac{dq}{q^{4-d}} \nonumber\\
&=&\frac{\Delta_f}{K^2}\frac{2^{2-d}\pi^{\frac{1-d}{2}}}{\Gamma (\frac{d-1}{2})}\frac{1}{3-d}(\xi_L^{3-d}-a^{3-d}),\nonumber
\eea
giving
\be
\xi_{L,d}=\left[\frac{4(3-d)\pi^2K^2}{\Delta_f}\frac{\Gamma(\frac{d-1}{2})}{2^{2-d}\pi^{\frac{1-d}{2}}}\right]^{\frac{1}{3-d}},
\ee
where we ignored the strongly subdominant (for $d<3$) $a$ term.

\subsection{Finite thickness in two dimensions}

For a finite-thickness ($w$) 2D Dirichlet cell, the surface variance
determines the Larkin length $\xi_L^{\D}$ according to
\bea
\overline{\langle \phi_0^2(x)\rangle}&=&(2\pi)^2\nonumber\\
&=&\frac{\Delta_f w}{\pi
  K^2}\int_{w/\xi_L^{\D}}^{\infty}\frac{dy}{y^2\coth^2{(y)}}.
\label{dirichlet_2_integral}
\eea
By scaling variables, this defines an implicit expression
\be
\int_{(\hat{\xi}_L^{\D})^{-1}}^\infty\frac{1}{y^2\coth^2{(y)}}dy=\hat{\xi}_L,
\label{dirichlet_2}
\ee 
for the Dirichlet Larkin length $\xi_L^{\D}(w,\xi_L) =
w\hat{\xi}_L^{\D}(\xi_L/w)$ in terms of the infinite cell's Larkin scale
$\xi_L$, latter simply a characterization of disorder given by
\rfs{xiL2dinfty}.

Similarly for a 2D Neumann cell, we have
\bea
\overline{\langle \phi_0^2(x) \rangle}&=&(2\pi)^2\nonumber\\
&=&\frac{\Delta_f w}{\pi
  K^2}\int_{w/\xi_L^{\N}}^{\infty}\frac{dy}{y^2\tanh^2{(y)}},
\label{neumann_2_integral}
\eea
which gives
\be
\int_{(\hat{\xi}_L^{\N})^{-1}}^{\infty}\frac{1}{y^2\tanh^2{(y)}}dy
=\hat{\xi}_L,
\label{neumann_2}
\ee
both evaluated numerically and plotted in Fig.~\ref{fig:xiDN}.

\subsection{Finite thickness in three dimensions}

Repeating the analysis in 3D for the Dirichlet cell, with scaled
ultraviolet cutoff $\hat{a}=a/w$ and 
\bea
\overline{\langle\phi_0^2(\bfx)\rangle}&=&(2\pi)^2 \nonumber\\
&=&\frac{\Delta_f}{2\pi
  K^2}\int_{w/\xi_L^{\D}}^{w/a}\frac{dy}{y\coth^2{(y)}}, \eea gives
\bea
\int_{(\hat{\xi}_L^{\D})^{-1}}^{\hat{a}^{-1}}\frac{dy}{y\coth^2{y}}=
\ln{(\xi_L/a)},\nonumber\\
\int_{(\hat{\xi}_L^{\D})^{-1}}^1\frac{dy}{y\coth^2{y}}
+\int_1^{\hat{a}^{-1}}\left(\frac{1}{y\coth^2{y}}-\frac{1}{y}\right)dy
=\ln{\hat{\xi}_L},\nonumber\\
\label{dirichlet_3}
\eea
which reduces to
\be
\int_{(\hat{\xi}_L^{\D})^{-1}}^1\frac{dy}{y\coth^2{y}}
=\ln{(1.18\hat{\xi}_L)}. \label{dirichlet_3_Larkin}
\ee
To get to this final result, we used the fact that the second integral
in \rfs{dirichlet_3} is finite in the ultraviolet, for $w\gg a$ giving
an $a$-independent constant about $-0.17$, thereby eliminating
dependence on $a$.

For the Neumann cell, we instead find
\bea
\int_{(\hat{\xi}_L^{\N})^{-1}}^1\frac{dy}{y\tanh^2{y}}
&=&\ln{(0.79\hat{\xi}_L)},
\label{neumann_3_Larkin}
\eea
where again a physically relevant limit $w\gg a$ was taken to eliminate
the $a$ dependence. Numerical evaluation of
Eqs.~\rf{dirichlet_3_Larkin} and \rf{neumann_3_Larkin} gives
the 3D results shown in Fig.~\ref{fig:xiDN}.

\subsection{Larkin length crossover}
\label{app:crossover}
As we can see from Eqs.~\rf{dirichlet_2}, \rf{neumann_2}, 
\rf{dirichlet_3_Larkin} and \rf{neumann_3_Larkin}, the Larkin length 
in a cell of thickness $w$ depends on a single dimensionless ratio,
$\xi_L/w$, of the infinite cell Larkin length (characterizing
pinning strength) to the cell thickness.  We expect that for a thick cell
($w\gg\xi_L$), the result of infinite thick cell should be
recovered. On the other hand, for thin cells ($w\ll\xi_L$),
we expect the homogeneous substrate boundary condition to play a
role. Namely, since the Dirichlet boundary condition on the top
homogeneous substrate explicitly orders the director, suppressing the
distortions of $\phi$, we expect $\xi_L$ to diverge for a thin
Dirichlet cell.  Furthermore, since the Neumann boundary condition
eliminates the stiffening by the bulk, in the thin Neumann cell we
expect $\xi_L$ to approach the value for a $(d-1)$-dimensional bulk
system with $(d-1)$-dimensional pinning. The expected crossover is
indeed confirmed by a numerical evaluation with the solution
illustrated in Fig.~\ref{fig:xiDN}, with the Dirichlet
$\xi_L$ diverging at $\xi_L/w\approx 1.71$ in 2D
and $\xi_L/w\approx 1.23$ in 3D.

The divergent asymptotic behavior can be obtained by expanding the
implicit expression for $\hat{\xi}_L$ in Eqs.~\rf{dirichlet_2} and
\rf{dirichlet_3_Larkin} around $1/\hat{\xi}_L=0$. For $d=2$, we
have

\be
\int_{0}^{\infty}\frac{dy}{y^2\coth^2{(y)}}
-\int^{(\hat{\xi}_L^{\D})^{-1}}_0\frac{dy}{y^2\coth^2{(y)}}=\hat{\xi}_L,
\ee 
which making use of $\frac{1}{y^2\coth^2{(y)}}\rightarrow 1$ as $y\rightarrow
0$ reduces to
 \be 
\hat{\xi}_L^{\D} = \frac{1}{\hat{\xi}_L^* - \hat{\xi}_L},
\ee
with
\be
\hat{\xi}_L^*=\int_{0}^{\infty}\frac{dy}{y^2\coth^2{(y)}}\simeq 1.705.
\ee

For $d=3$, using $\frac{1}{y\coth^2{(y)}}\rightarrow y$ as
$y\rightarrow 0$, we have 
\bea
\hspace{-1cm} \int_{0}^{1}\frac{dy}{y\coth^2{y}}
-\int_0^{(\hat{\xi}_L^{\D})^{-1}}\frac{dy}{y\coth^2{y}}
=\ln{(1.18\hat{\xi}_L)}, \nonumber\\
\ln{(1.18\hat{\xi}_L^*)}-\frac{(\hat{\xi}_L^{\D})^{-2}}{2}=\ln{(1.18
  \hat{\xi}_L)}, 
\eea 
with 
\bea
\ln{(1.18\hat{\xi}_L^*)}&=&\int_{0}^{1}\frac{dy}{y\coth^2{y}},
\nonumber\\
\hat{\xi}_L^*&\simeq& 1.233.  
\eea 
This leads to 
\be
\hat{\xi}_L^{\D}
=\frac{1}{\sqrt{2\ln{(\hat{\xi}_L^*/\hat{\xi}_L)}}}
\simeq\frac{\sqrt{\hat{\xi}_L^*/2}}
{\sqrt{\hat{\xi}_L^*-\hat{\xi}_L}}, 
\ee
quoted in the main text and consistent with the numerical evaluation of 
the integral solution.

\section{The cusp and the fixed point}
\label{app:cusp}
For completeness, we now fill in some of the details (previously
reviewed in Ref.~[\onlinecite{WieseDoussal}]) for the RG
evolution of the random potential variance $\hR(\phi)$ into its
universal cusped form.  To this end, by differentiating the flow
equation (\ref{FRGflowRT0}) for $\hR(\phi,\ell)$ with respect to
$\phi$, we obtain

\bea
\partial_{\ell}\hR''(\phi)&=&\epsilon \hR''(\phi)+\hR'''(\phi)^2+\hR''(\phi)\hR''''(\phi)\nonumber\\
&&-\hR''''(\phi)\hR''(0), \label{RGEFL2_S}\nonumber\\
\partial_{\ell}\hR''''(\phi)&=&\epsilon \hR''''(\phi)+3\hR''''(\phi)^2+4\hR'''(\phi)\hR^{(5)}(\phi)\nonumber\\
&&+\hR''(\phi)\hR^{(6)}(\phi)-\hR^{(6)}(\phi)\hR''(0). 
\eea
Setting $\phi$ to $0$, we obtain
\bea
\partial_{\ell}\hR''(0)&=&\epsilon \hR''(0)+\hR'''(0)^2\rightarrow \epsilon \hR''(0),\label{R2RGE_S}\nonumber\\
\partial_{\ell}\hR''''(0)&=&\epsilon \hR''''(0)+3\hR''''(0)^2+4\hR'''(0)\hR^{(5)}(0)\nonumber\\
&&\rightarrow \epsilon \hR''''(0)+3\hR''''(0)^2.\label{R4RGE_S}
\eea

We note that $R(\phi)$ is an even function and moreover (before the
cusp develops) is smooth in $\phi$, with  $\hR'''(0)=\hR^{(5)}(0)=0$.

Clearly, the flows of $\hR''(0)$ and $\hR''''(0)$ are exact, with the
later diverging in a finite RG time according to \cite{DSFisherFRG}

\be
\hR''''(0)|_{\ell}=\frac{ce^{\epsilon \ell}}{1-3c(e^{\epsilon
    \ell}-1)/\epsilon}, 
\ee 
with $c=\hR''''(0)|_{\ell=0}$. For a special case $\epsilon=0$
($d=3$ in our system with surface pinning), the cusp develops
according to

\be
\hR''''(0)|_{\ell}=\frac{c}{1-3c\ell}.  
\ee

\begin{figure}
\includegraphics[width=3 cm]{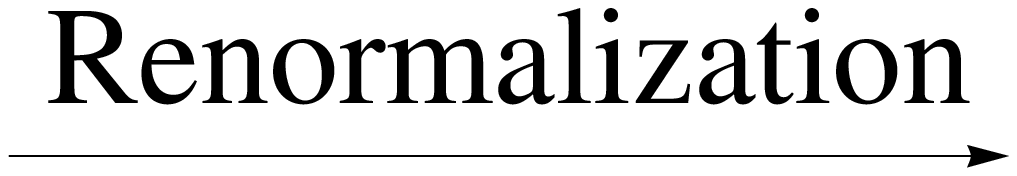}\\
\includegraphics[height=3.2 cm]{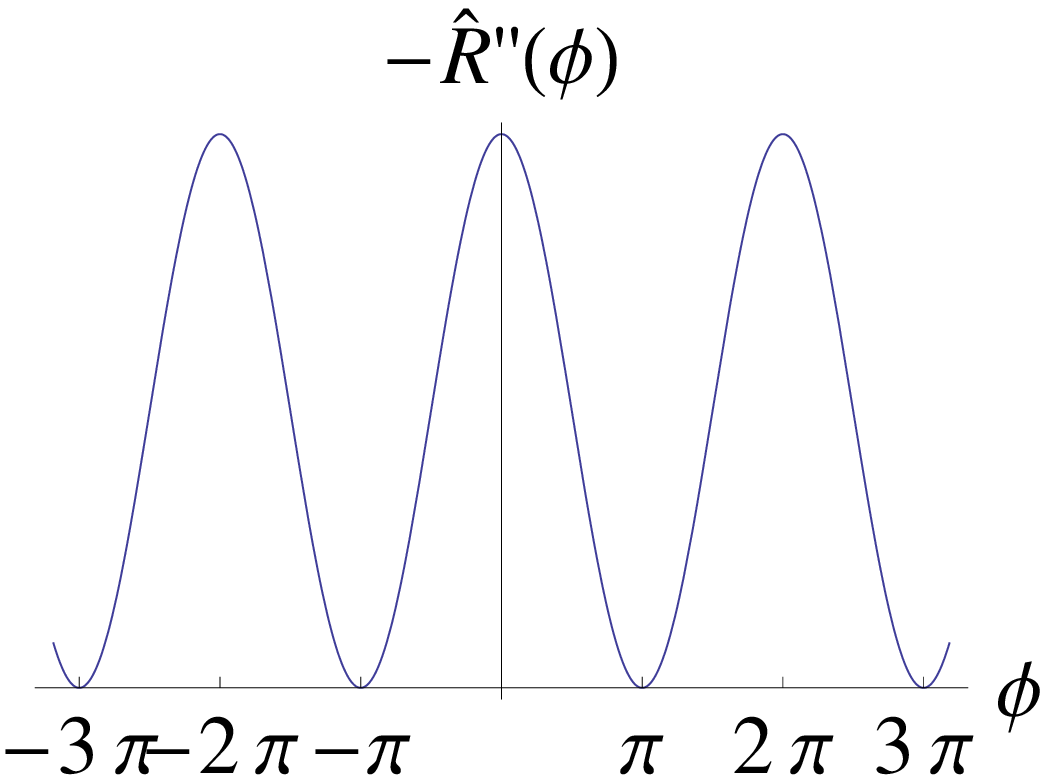}
\includegraphics[height=3.2 cm]{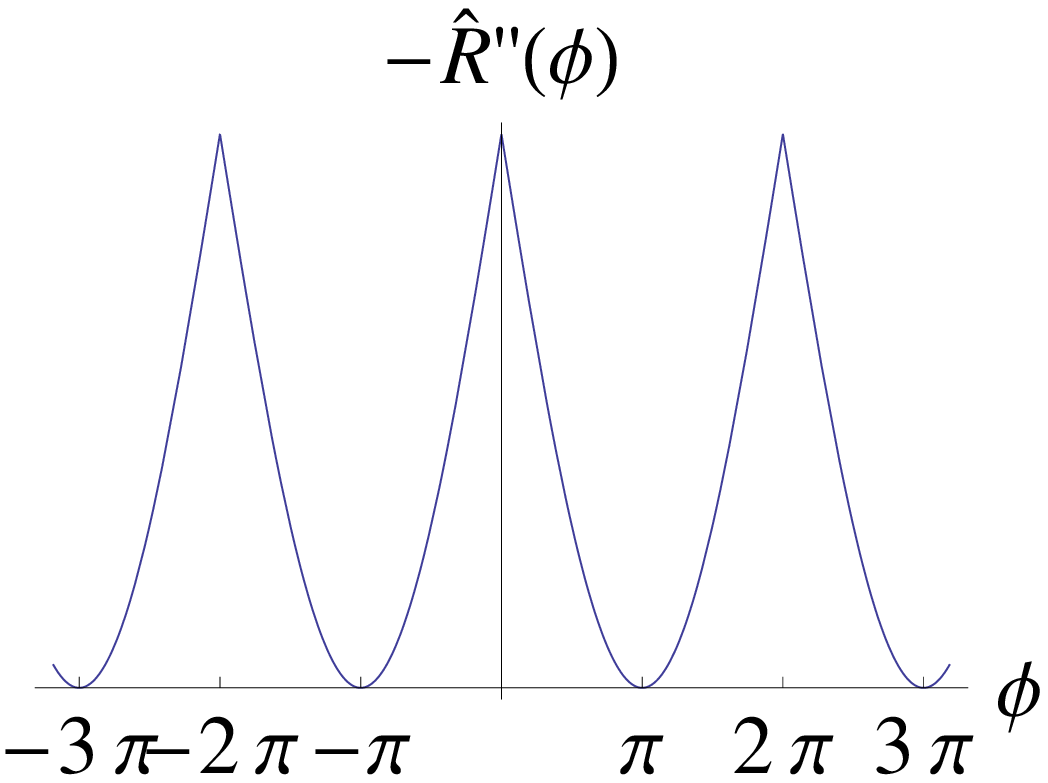}\\      
 \caption{(Color online) Evolution of $\hat{R}''(\phi)$ and 
  appearance of cusps under coarse-graining renormalization.}
\label{cusp}
\end{figure}
Thus after a finite RG time $\ell$, $\hR''''(0)$ diverges,
signaling the appearance of a cusp in $\hR''(\phi)$, as illustrated in
Fig.~\ref{cusp}. 

\section{Details of asymptotics of various correlation functions}
\label{app:asymptotics}
As derived in the main text, in an infinitely thick cell, the momentum
space correlation function has short- and long-scale
limits. The former one, computed in the random-torque (Larkin)
approximation (scale shorter than $\xi_L$) on the random substrate
($z=0$), is given by
\be
C(q)\approx\frac{\Delta_f}{K^2q^2},\ \mbox{$q>1/\xi_L$}.
\ee 
In the long-scale limit (small $q<1/\xi_L$), derived via FRG and
matching methods in Sec.\ref{sec:matching} it is instead given by 
\be
C(q)\approx\left\{\begin{array}{ll}
\frac{(3-d)\pi^2}{9C_{d-1}}\frac{1}{q^{d-1}},&\mbox{for $d < 3$},\\
-\frac{2\pi^3}{9}\frac{1}{q^{2}\ln(q a)},&\mbox{for $d = 3$},
\end{array}\right., \ \mbox{$q<1/\xi_L$}.
\ee
A Fourier transform of $C(q)$ [using above limits and generalized to
finite $z$; see Eqs.~\rf{phi}-\rf{phiN}] then gives this
correlation function in real space.

\subsection{Mean-squared distortion in Larkin approximation}
\label{app:msDistortion}
The mean-squared distortion of $\phi(\xv,z)$ can be given as
\be
\overline{\langle \phi^2(\xv,z)\rangle}\approx\int \frac{d^{d-1}q}{(2\pi)^{d-1}}
 \frac{\Delta_f}{K^2q^2}e^{-2qz}.
\ee

For $d=2$, we have (with $\hat{q}=2qz$)
\bea
\overline{\langle \phi^2(x,z)\rangle}
&\approx&\frac{\Delta_f}{\pi K^2}2z\int_{2z/\xi_L}^{\infty}
\frac{e^{-\hat{q}}}{\hat{q}^{2}}d\hat{q} \nonumber\\
&=&4\pi^2\Big(\frac{2z}{\xi_L}\Big)\Gamma(-1,2z/\xi_L),
\eea
in which the integral converges at large $q$ so we can ignore the 
upper cutoff $\Lambda=1/a$ of $q$ and the definition of $\xi_L$ 
as in Eq.~\rf{LarkinLengthl3} was used. Expanding this result at 
different range of $z$ values, we obtain
\be
\overline{\langle \phi^2(x,z)\rangle}\approx 4\pi^2\left\{\begin{array}{ll}
1-\frac{2z}{\xi_L}(\ln{\frac{\xi_L}{2z}}+1-\gamma),&a\ll 2z\ll \xi_L\\
\frac{\xi_L}{2z}e^{-2z/\xi_L},&2z\gg \xi_L,
\end{array}\right.
\ee
in which $\gamma\approx 0.58$ is the Euler's constant.

For $2<d<3$, we have
\bea
\hspace{-.5 cm}
\overline{\langle \phi^2(\xv,z)\rangle}
&\approx&\frac{C_{d-1}\Delta_f(2z)^{3-d}}{K^2}\int_{2z/\xi_L}^{\infty}
\frac{e^{-\hat{q}}}{\hat{q}^{4-d}}d\hat{q}, \nonumber\\
&=&4\pi^2(3-d)\Big(\frac{2z}{\xi_L}\Big)^{3-d}\Gamma(d-3,2z/\xi_L),
\eea
making use of $\hat{q}=2qz$ and the definition of $\xi_L$ 
as in Eq.~\rf{LarkinLengthl3}. For different range of $z$ values, we obtain
\be
\overline{\langle \phi^2(\xv,z)\rangle}\approx 4\pi^2\left\{\begin{array}{ll}
1-\Gamma(d-2)(\frac{2z}{\xi_L})^{3-d},&a\ll 2z\ll \xi_L\\
(3-d)\frac{\xi_L}{2z}e^{-2z/\xi_L},&2z\gg \xi_L,
\end{array}\right.
\ee

Similarly, for $d=3$, we need to consider the upper cutoff of $q$ and have
\bea
\overline{\langle \phi^2(\xv,z)\rangle}&\approx&\frac{\Delta_f}{2\pi K^2} 
 \int_{2z/\xi_L}^{2z/a}\frac{e^{-\hat{q}}}{\hat{q}}d\hat{q}
\nonumber\\
&=&\frac{4\pi^2}{\ln{(\xi_L/a)}}\left[\Gamma(0,\frac{2z}{\xi_L})
 -\Gamma(0,\frac{2z}{a})\right] \nonumber\\
&\approx&4\pi^2\left\{\begin{array}{ll}
1-\frac{\ln(2z/a)}{\ln{(\xi_L/a)}},&\hspace{-0.1 cm}a\ll 2z \ll \xi_L\\
\frac{\xi_L/2z}{\ln{(\xi_L/a)}}\ e^{-2z/\xi_L},&\hspace{-0.1 cm} 2z\gg\xi_L,
\end{array}\right.\hspace{-0.1 cm}\nonumber\\
\eea
where $\hat{q}=2qz$. The mean-squared 
distortions for $d=2$ and $d=3$ are plotted in Fig.~\ref{SelfCorrelation}.

\subsection{Correlation function in Larkin approximation}
\label{app:LarkinCorr}
The short scales (where Larkin approximation holds) contribution to
the real-space correlation is given by Eq.~\rf{CLxzz}. In an infinite
thick cell, we have
\be
C_L^{(\infty)}(\xv,z,z)\approx
\frac{2\Delta_f}{K^2}\int \frac{d^{d-1}q}{(2\pi)^{d-1}}
\frac{(1-\cos\qv\cdot\xv)e^{-2q z}}{q^2},
\ee
in which the integral of $q$ has lower cutoff $1/\xi_L$ and upper
cutoff $1/a$. For simplicity of notation, we will ignore the 
superscript $\infty$.

For $d=2$ and $x\ll \xi_L$, since the kernel is convergent at large 
$q$, we can extend the integral to infinity (requiring $x\gg a$) and obtain
\bea
C_L(x,z,z)&\approx&\frac{2\Delta_f}{\pi K^2}
  \int_{1/\xi_L}^{\infty}\frac{(1-\cos qx)e^{-2q z}}{q^2}dq
\nonumber\\
&=& 4\pi^2\Big[\frac{4z}{\xi_L}\Gamma(-1,\frac{2z}{\xi_L})
  -\frac{2z-ix}{\xi_L}\Gamma(-1,\frac{2z-ix}{\xi_L}) \nonumber \\
&&-\frac{2z+ix}{\xi_L}\Gamma(-1,\frac{2z+ix}{\xi_L})\Big].
\eea
On the heterogeneous substrate ($z=0$), we have
\bea
C_L(x,0,0)&\approx& 8\pi^2\left[1+\frac{\pi x}{2\xi_L}
  -\cos{\frac{x}{\xi_L}}-\frac{x}{\xi_L} 
  \text{Si}(\frac{x}{\xi_L})\right],\nonumber\\
&\approx& 8\pi^2\frac{\pi x}{2\xi_L},\ \ \mbox{for $x<<\xi_L$},
\eea
where $\text{Si}(z)=\int_0^z \frac{\sin{(t)}}{t} dt$ is the sine integral 
function.
Making use of the expansion of $\cos{(qx)}$ at $qx\ll 1$, 
for finite $z$ we obtain
\bea
&&\hspace{-0.6 cm}C_L(x,z,z)\nonumber\\
&\approx&\frac{8\pi^2}{\xi_L}\Big(\int_{1/\xi_L}^{1/x}
 + \int_{1/x}^{\infty}\Big)\frac{\left[1-\cos(qx)\right]}{q^2} e^{-2qz}dq \nonumber\\
&\approx&\frac{8\pi^2}{\xi_L}
  \Big(\frac{x^2}{2}\int_{1/\xi_L}^{1/x}e^{-2qz}dq
  +\int_{1/x}^{\infty}\frac{e^{-2qz}dq}{q^2}\Big) \nonumber\\
&\approx&4\pi^2\frac{x^2}{2z\xi_L}e^{-2z/\xi_L},
\eea
where the last approximation is taken for $z\gg x$ and the other terms 
are subdominant.

When $x\gg \xi_L$, the $\cos(qx)$ oscillates strongly giving
subdominant contribution. Then we have
\bea
&&\hspace{-.6 cm}C_L(x,z,z)\nonumber\\
&\approx& \frac{2\Delta_f}{\pi K^2}
  \int_{1/\xi_L}^{\infty}\frac{e^{-2q z}}{q^2}dq \nonumber\\
&=&8\pi^2\frac{2z}{\xi_L}\Gamma(-1,\frac{2z}{\xi_L},\frac{2z}{a}) 
\nonumber\\
&\approx&8\pi^2\left\{\begin{array}{ll}
1-\frac{2z}{\xi_L}(\ln{\frac{\xi_L}{2z}}+1-\gamma),&a\ll 2z\ll \xi_L\\
\frac{\xi_L}{2z}e^{-2z/\xi_L},&2z\gg \xi_L,
\end{array}\right. \nonumber\\
\eea
in which the generalized incomplete gamma function $\Gamma(p,z_1,z_2)
=\int_{z_1}^{z_2} t^{p-1} e^{-t} dt=\Gamma(p,z_1)-\Gamma(p,z_2)$ and
 we made use of expansions $\Gamma(-1,x)\approx \frac{1}{x}+
(\gamma-1+\ln{x})-x/2$ at small $x$ ($x\ll 1$) and $\Gamma(-1,x)\approx x^{-2}
e^{-x}$ at large $x$ ($x\gg 1$).

For $2<d<3$ and $x\ll \xi_L$, we can extend the integral to infinity
(requiring $x\gg a$) and obtain
\bea
&&\hspace{-.6 cm} C_L(\xv,z,z)\nonumber\\
&\approx&\frac{2\Delta_f}{K^2}\int \frac{d^{d-1}q}{(2\pi)^{d-1}}
\frac{\left[1-\cos{(\qv\cdot\xv)}\right]e^{-2q z}}{q^2} \nonumber\\
&=&\frac{8(3-d)\pi^2}{\xi_L^{3-d}}
\int_{1/\xi_L}^{\infty}\frac{\left[1-f_d{( qx)}\right]e^{-2q z}}{q^{4-d}}dq,
\eea
in which $f_d(qx)$ is the average of $\cos{(\qv\cdot\xv)}$ over a surface of 
a $(d-1)$-dimension unit sphere. The correlation function could be evaluated
numerically and on the heterogeneous substrate this result approaches
(with $\hat{q}=qx$)
\bea
C_L(\xv,0,0)&\approx& \frac{8(3-d)\pi^2}{\xi_L^{3-d}}
x^{3-d}\int_{x/\xi_L}^{\infty}\frac{1-f_d{(\hat{q})}}{\hat{q}^{4-d}}d\hat{q}
\nonumber\\
&\approx&8\pi^2\Big(\frac{x}{\xi_L}\Big)^{3-d},\ \mbox{when $d\lesssim 3$}, 
\eea
as given in Sec.~\ref{sec:Larkin}.
Making use of the expansion $1-f_d(qx)\approx 
\frac{(qx)^2}{2(d-1)}$ at $qx\ll 1$, 
we can evaluate the correlation function at finite $z$ as
\bea
&&\hspace{-0.6 cm}C_L(\xv,z,z)\nonumber\\
&\approx&\frac{8(3-d)\pi^2}{\xi_L^{3-d}}\Big(\int_{1/\xi_L}^{1/x}
 + \int_{1/x}^{1/a}\Big)\frac{(1-f_d(qx))}{q^{4-d}} e^{-2qz}dq \nonumber\\
&\approx&\frac{8(3-d)\pi^2}{\xi_L^{3-d}}
  \Big[\frac{x^2}{2(d-1)}\int_{1/\xi_L}^{1/x}e^{-2qz}q^{d-2}dq\nonumber\\
&&  +\int_{1/x}^{1/a}\frac{e^{-2qz}dq}{q^{4-d}}\Big] \nonumber\\
&\approx&\frac{4(3-d)\pi^2}{d-1}\frac{x^2(2z)^{1-d}}{\xi_L^{3-d}}\Gamma(d-1,\frac{2z}{\xi_L})
\eea
in which the last approximation is taken for $z\gg x$ and the other subdominant  
terms are ignored.

When $x\gg \xi_L$, the $\cos{(\qv\cdot \xv)}$ again oscillates at the 
full range of $q$ and obtain
\bea
&&\hspace{-0.5 cm}C_L(\xv,z,z) \nonumber\\
&\approx&\frac{8(3-d)\pi^2}{\xi_L^{3-d}}
  \int_{1/\xi_L}^{1/a}\frac{e^{-2q z}}{q^{4-d}}dq \nonumber\\
&=&(3-d)8\pi^2\Big(\frac{2z}{\xi_L}\Big)^{3-d}\Gamma(d-3,\frac{2z}{\xi_L},\frac{2z}{a}) \nonumber\\
&\approx&8\pi^2\left\{\begin{array}{ll}
1-\Gamma(d-2)(\frac{2z}{\xi_L})^{3-d},&a\ll 2z\ll \xi_L \\
(3-d)\frac{\xi_L}{2z}e^{-2z/\xi_L},&2z\gg\xi_L,
\end{array}\right.\nonumber\\
\eea
where we made use of expansions $\Gamma(d-3,x)\approx 
\Gamma(d-3)+\frac{x^{d-3}}{3-d}$ at small $x$ ($x\ll 1$) and $\Gamma(d-3,x)\approx x^{d-4}e^{-x}$ at large $x$ ($x\gg 1$).

For $d=3$, the variables $\qv$ and $\xv$ are two dimensional, thus
we have
\bea
C_L(\xv,z,z)&\approx&
\frac{2\Delta_f}{K^2}\int \frac{d^{d-1}q}{(2\pi)^{d-1}}
\frac{(1-\cos\qv\cdot\xv)e^{-2q z}}{q^2}, \nonumber\\
&=&\frac{\Delta_f}{\pi K^2}\int_{1/\xi_L}^{1/a} 
\frac{(1-J_0(qx))}{q}e^{-2q z} dq,
\eea
in which $J_0(qx)$ is the Bessel function of the first kind. 
The behavior of its correlation could be evaluated numerically and 
the asymptotics can be obtained approximately for different regions.

On the heterogeneous surface ($z=0$), for small $x$ ($a\ll x\ll \xi_L$),
we have $1-J_0(qx)\approx 1$ with $q\gg 1/x$ and $1-J_0(qx)
\approx (qx)^2/4$ with $q\ll 1/x$, so in this region we have
\bea
C_L(\xv,0,0)&\approx&\frac{8\pi^2}{\ln{(\xi_L/a)}}\Big(\int_{1/\xi_L}^{1/x}
 + \int_{1/x}^{1/a}\Big)\frac{\left[1-J_0(qx)\right]}{q} dq \nonumber\\
&\approx&\frac{8\pi^2}{\ln{(\xi_L/a)}}\Big(\frac{x^2}{4}\int_{1/\xi_L}^{1/x}qdq
 +\int_{1/x}^{1/a}\frac{dq}{q}\Big) \nonumber\\
&=&\frac{8\pi^2}{\ln{(\xi_L/a)}}\left[\frac{1}{8}-\frac{x^2}{8\xi_L^2}+\ln{(x/a)}\right] \nonumber\\
&\approx&\frac{8\pi^2}{\ln{(\xi_L/a)}}\ln{(x/a)}.
\eea
Making use of the expansion of $J_0(qx)$, we can evaluate 
the correlation function at finite $z$ as
\bea
&&\hspace{-0.6 cm}C_L(\xv,z,z)\nonumber\\
&\approx&\frac{8\pi^2}{\ln{(\xi_L/a)}}\Big(\int_{1/\xi_L}^{1/x}
 + \int_{1/x}^{1/a}\Big)\frac{(1-J_0(qx))}{q} e^{-2qz}dq \nonumber\\
&\approx&\frac{8\pi^2}{\ln{(\xi_L/a)}}
  \Big(\frac{x^2}{4}\int_{1/\xi_L}^{1/x}e^{-2qz}qdq
  +\int_{1/x}^{1/a}\frac{e^{-2qz}dq}{q}\Big) \nonumber\\
&\approx&\frac{2\pi^2}{\ln{(\xi_L/a)}}\frac{x^2}{(2z)^2}
  (1+\frac{2z}{\xi_L}) e^{-2z/\xi_L},
\eea
where in the last approximation we kept the leading term for $z\gg x$.

When $x\gg \xi_L$, the expression simplifies to 
\bea
&&\hspace{-.6 cm}C_L(\xv,z,z)\nonumber\\
&\approx& \frac{\Delta_f}{\pi K^2}\int_{1/\xi_L}^{1/a} 
\frac{e^{-2q z}}{q} dq \nonumber\\
&=&\frac{8\pi^2}{\ln{(\xi_L/a)}}\Gamma(0,2z/\xi_L,2z/a) \nonumber\\
&\approx&8\pi^2\left\{\begin{array}{ll}
1-\frac{\ln(2z/a)}{\ln{(\xi_L/a)}},&a\ll 2z\ll \xi_L\\
\frac{\xi_L/2z }{\ln(\xi_L/a)}e^{-2z/\xi_L},&2z\gg \xi_L,
\end{array}\right.
\eea
in which we made use of expansions $\Gamma(0,x)\approx 
(-\gamma-\ln{x})+x$ at small $x$ ($x\ll 1$) and $\Gamma(0,x)\approx x^{-1}
e^{-x}$ at large $x$ ($x\gg 1$).

\subsection{Universal (long-scales) part of correlation function}
\label{app:matching}
The correlation function in momentum space at small $q$ ($q<1/\xi_L$)
is obtained by FRG and matching methods in Sec.~\ref{sec:matching}.
Here we calculate the corresponding real-space correlation
functions. By construction, this form of $C(q)$ only holds at 
$0<q<\xi_L^{-1}$, with $\xi_L^{-1}$ therefore entering as the 
upper (UV) cutoff on all $q$ integrals done here.

For $d<3$, the FRG derived correlation function rated at $q<\xi_L^{-1}$
is given by
\bea
C_*(\xv,z,z)\approx \int \frac{d^{d-1}q}{(2\pi)^{d-1}}\frac{(3-d)\pi^2}{9C_{d-1}}
\frac{[1-\cos{(\qv\cdot\xv)}]e^{-2qz}}{q^{d-1}}.\nonumber\\
\eea
A better approximation is obtained by using a ``soft'' upper cutoff
by inserting a factor of $e^{-q\xi_L}$ inside above integrand.

For $d=2$, this correlation function is given by
\bea
C_*(x,z,z)&\approx& \frac{2\pi^2}{9}\int_{1/L}^{1/\xi_L}\frac{1-\cos{(qx)}}{q}e^{-2qz}dq \nonumber\\
&\approx&\frac{2\pi^2}{9}\int_{1/L}^{\infty}\frac{1-\cos{(qx)}}{q}e^{-2qz-q\xi_L}dq \nonumber\\
&=&\frac{\pi^2}{9}\Big[2\Gamma(0,\frac{2z+\xi_L}{L})-\Gamma(0,\frac{2z+\xi_L-ix}{L})\nonumber\\ 
  &&-\Gamma(0,\frac{2z+\xi_L+ix}{L})\Big],
\eea
where $L$ is system size to be taken to $\infty$ at the end of 
calculation. Using the gamma function expansion 
$\Gamma(0,x)\approx (-\gamma-\ln{x})+x$, we obtain
\bea
C_*(x,z,z)&\approx& \frac{\pi^2}{9}\big(-2\ln{\frac{2z+\xi_L}{L}}+\ln{\frac{2z+\xi_L-ix}{L}}\nonumber\\
  &&+\ln{\frac{2z+\xi_L+ix}{L}} \big) \nonumber\\
&\approx&\frac{\pi^2}{9}\ln{\Big[1+\frac{x^2}{(2z+\xi_L)^2}\Big]}.
\eea
Except for approximations associating with the matching method, 
this real-space result is an excellent approximation to a numerical
integration of $C(q)$, as shown in Fig.~\ref{fig:C_s2d}.
On the heterogeneous substrate and for
$x\gg\xi_L$, it reduces to $C_*(x,z,z)\approx
 \frac{2\pi^2}{9}\ln{(x/\xi_L)}$.

For $d=3$, the correlation function is given by
\bea
\hspace{-0.8 cm}
C_*(\xv,z,z)&\approx& -\frac{2\pi^2}{9C_2}\int\frac{d^2q}{(2\pi)^2}
  \frac{1-\cos{(\qv\cdot\xv)}}{q^2\ln{qa}}e^{-2qz} \nonumber\\
&=&-\frac{2\pi^2}{9}\int_{0}^{1/\xi_L}\frac{1-J_0{(qx)}}{q\ln{qa}}e^{-2qz}dq,
\eea
which we evaluated numerically with a soft cutoff $\frac{1}{1+(q\xi_L)^2}$ and plotted in Fig.~\ref{fig:C_s3d}.

Approximate asymptotic behavior of this correlation function can 
be obtained analytically
\bea
&&\hspace{-.6 cm}
C_*(\xv,z,z)\nonumber\\
&\approx&-\frac{2\pi^2}{9}\Big(\int_{0}^{1/x}
  +\int_{1/x}^{1/\xi_L}\Big)
  \frac{1-J_0{(qx)}}{q\ln{qa}}e^{-2qz}dq \nonumber\\
&=&-\frac{2\pi^2}{9}\Big(\frac{x^2}{4}\int_{0}^{1/x}\frac{qe^{-2qz}}{\ln{qa}}dq
   + \int_{1/x}^{1/\xi_L}\frac{e^{-2qz}}{q\ln{qa}}dq\Big) \nonumber\\
&=&-\frac{2\pi^2}{9}\left[
  \frac{1}{4}\int_0^1 dk\frac{ke^{-2 k z/x}}{\ln(k a/x)}
  + \int_{1}^{x/\xi_L} dk\frac{e^{-2 k z/x}}{k\ln(k a/x)}
 \right],\nonumber\\
\eea
in which $k=qx$.

On the heterogeneous substrate ($z=0$), the first integral is negligible,
thus for $x\gg \xi_L$ we have 
\bea
C_*(\xv,z,z)&\approx&-\frac{2\pi^2}{9}\int_{1/x}^{1/\xi_L}\frac{1}{q\ln{qa}}dq \nonumber\\
&=&\frac{2\pi^2}{9}\ln{\Big[\frac{\ln{(x/a)}}{\ln{(\xi_L/a)}}\Big]}.
\eea

Further away from the substrate such that $\xi_L\ll 2z\ll x$, 
the $e^{-2qz}$ acts like upper cutoff at $q\sim 1/2z$, giving
\bea
C_*(\xv,z,z)
&\approx&-\frac{2\pi^2}{9}\int_{1/x}^{1/2z}\frac{1}{q\ln{qa}}dq
\nonumber\\
&=&\frac{2\pi^2}{9}\ln{\Big[\frac{\ln{(x/a)}}{\ln{(2z/a)}}\Big]},
\eea

Finally, for $\xi_L\ll x\ll 2z$, we find
\bea
\hspace{-0.6 cm}
C_*(\xv,z,z)&\approx&-\frac{2\pi^2}{9}
  \frac{x^2}{4}\int_{0}^{1/2z}\frac{q}{\ln{qa}}dq \nonumber\\
&\approx&\frac{2\pi^2}{9}\frac{x^2}{16z^2}\frac{1}{2\ln{(2z/a)}},
\eea
with numerical prefactor that is off by a factor of $\half$ 
relative to the numerical integration that gives $C_*(\xv,z,z)
\approx \frac{2\pi^2}{9}\frac{x^2}{16z^2}\frac{1}{\ln{(2z/a)}}$, as given in 
Sec.~\ref{sec:matching}.


\begin{thebibliography}{99}
\bibitem{equivalentComment} This is certainly the case for the
  long-scale asymptotics of correlation functions, dominated by the
  $T=0$ ground-state distortion induced by the quenched random
  heterogeneity.
\bibitem{linearComment} Because we are perturbing about an ordered
  state $\phi_0(\xv)=0$, which is not a minimum of the random
  potential, the lowest nontrivial term in the Taylor expansion of the
  random potential is {\em linear} in $\phi_0$.
\bibitem{Delta0Comment} An additional operator
  $(\nabla\phi_{0<}^\alpha - \nabla\phi_{0<}^\beta)^2$ generated by
  the coarse-graining of $H_p$, is irrelevant in the RG sense for $d >
  1$, in contrast to its familiar bulk disorder counter-part.
\end{thebibliography}
\end{document}